\documentstyle[pictex,12pt]{article}
\newcommand{\Eq}[1]{Eq. (\ref{#1})}
\newcommand{\ben}{\begin{equation}}
\newcommand{\een}{\end{equation}}
\newcommand{\bea}{\begin{eqnarray}}
\newcommand{\eea}{\end{eqnarray}}
\newcommand{\bear}{\begin{array}}
\newcommand{\enar}{\end{array}}
\newcommand{\bdm}{\begin{displaymath}}
\newcommand{\edm}{\end{displaymath}}
\newcommand{\nn}{\nonumber \\ }
 
\newcommand{\hf}{\frac{1}{2}}

\newcommand{\br}{\langle}
\newcommand{\kt}{\rangle}

\newcommand{\vs}{\vspace}

\newcommand{\NP}[1]{Nucl.\ Phys.\ {\bf #1}}
\newcommand{\PL}[1]{Phys.\ Lett.\ {\bf #1}}
\newcommand{\CMP}[1]{Commun.\ Math.\ Phys.\ {\bf #1}}

\newcommand{\MPL}[1]{Mod.\ Phys.\ Lett.\ {\bf #1}}

\newcommand{\SJNP}[1]{Sov. J. Nucl. Phys.\ {\bf #1}}
                                                                   
\begin{document}

\topmargin -5mm
\oddsidemargin 5mm

\begin{titlepage}
\setcounter{page}{0}
\begin{flushright}
NBI-HE-96-18\\
AS-ITP-96-13\\
July 1996\\
hep-th/yymmxxx
\end{flushright}

\vs{8mm}
\begin{center}
{\Large FUSION, CROSSING AND MONODROMY}\\[.2cm]
{\Large IN CONFORMAL FIELD THEORY}\\[.2cm]
{\Large BASED ON $SL(2)$ CURRENT ALGEBRA}\\[.2cm]
{\Large WITH FRACTIONAL LEVEL}

\vs{8mm}
{\large Jens Lyng Petersen}\footnote{e-mail address:
 jenslyng@nbi.dk},
{\large J{\o}rgen Rasmussen}
\footnote{e-mail address: jrasmussen@nbi.dk},\\[.2cm]
{\em  The Niels Bohr Institute, Blegdamsvej 17, DK-2100 Copenhagen \O,
Denmark}\\[.5cm]
{\large Ming Yu}
\footnote{e-mail address:
 yum@itp.ac.cn}\\[.2cm]
{\em  Institute of Theoretical Physics, Academia Sinica, P.O.Box 2735, 
Beijing 100080, Peoples Republic of China}

\end{center}

\vs{8mm}
\centerline{{\bf{Abstract}}}

Based on our earlier work on free field realizations of conformal blocks for 
conformal field theories with $SL(2)$ current algebra and with fractional 
level and spins, we discuss in some detail the fusion rules which arise. 
By a careful analysis of the 4-point functions, we find that both the fusion 
rules previously found in the literature are realized in our formulation. 
Since this is somewhat contrary to our expectations in our first work based 
on 3-point
functions, we reanalyse the 3-point functions and come to the same conclusion.
We compare our results on 4-point conformal blocks in particular with a 
different realization of these
found by O. Andreev, and we argue for the equivalence. We describe in  
detail how integration contours have to be chosen to obtain convenient bases 
for conformal blocks, both in his and in our own formulation. 
We then carry out the rather lengthy calculation to 
obtain the crossing matrix between s- and t-channel blocks, and we use 
that to determine the monodromy invariant 4-point greens functions. We use the
monodromy coefficients to obtain the operator algebra coefficients for 
theories based on admissible representations. 

\end{titlepage}
\newpage
\renewcommand{\thefootnote}{\arabic{footnote}}
\setcounter{footnote}{0}
\section{Introduction}
There are several reasons why conformal field theories based on affine 
$\widehat{SL}(2)_k$ are interesting to study, not only for $k$ positive integer 
and for unitary, integrable representations based on usual integer and and half
integer spins, but also for fractional levels and for fractional spins, in
particular for admissible representations \cite{KK,MFF}. 
Thus for example it was
shown in Refs. \cite{HY,AGSY} how 2-d quantum gravity coupled to minimal 
conformal matter could in principle be described by a topological $G/G$ model 
with the $G$'s being affine $SL(2)$'s with the same levels. This possibility is
closely related to the realization that minimal conformal field theory is 
obtained via hamiltonian reduction of $SL(2)$ \cite{Bel,Pol,BO}. 
However, there a minimal theory labelled in the
standard way by $(p,q)$ ($p$ and $q$ co-prime integers) is related to an 
$SL(2)$ theory with level $k$ given by
\ben
t\equiv k+2=p/q
\een
and where admissible representations with fractional spins
of $\widehat{SL}(2)_k$ have to be used.
This approach appears potentially very interesting since a proper understanding
of it would seem to lead to obvious possibilities for generalizations based on
different groups and super-groups. 

Other completely different applications may be envisaged. For example
very interesting string backgrounds describing black holes may be obtained 
from $SL(2)$ theories with fractional levels \cite{bars}. 

To set up in detail conformal field theory based on $\widehat{SL}(2)_k$ 
with fractional levels and spins, one must first understand how to write down 
conformal blocks. In Refs. \cite{PRY} we have given a general description of
how this can be done based on the free field Wakimoto realization \cite{Wak}.
A number of 
technical obstacles arising from the occurrence of fractional powers
of free fields had to be overcome. Several other groups have also studied 
the conformal blocks from several different points of view with more or less
complete results \cite{FGPP,FF,An,AY,FM,FM95}. 
In this paper we have compared in 
detail to the approach by Andreev \cite{An} although so far only 3- and 4-point
functions have been written down there, but the 4-point function 
turns out to be very convenient for
calculation of the crossing matrix between s-and t-channel blocks \cite{DF}.

However, once the conformal blocks are obtained the next step in the program 
is to 
determine the monodromy invariant greens functions; these are the ones for
which physical applications can be made and they are necessary before for 
example an application to 2-d quantum gravity can be made. It is the principal 
goal of the present paper to obtain these monodromy invariant combinations,
and from the ensuing monodromy coefficients to determine the operator algebra
coefficients.
As described by Dotsenko and Fateev \cite{DF} this problem is conveniently 
solved by means of the crossing matrix relating the conformal blocks in the
s- and t-channels (in fact, just a particular row and column of that matrix).
A central portion of this paper is to show how to generalize the treatment of 
Dotsenko and Fateev \cite{DF} from minimal models to the problem at hand.

The conformal blocks for 4-point functions are characterized conveniently in 
terms of couplings to intermediate states. These in turn are determined by the
fusion rules of the theory. Fusion rules for $\widehat{SL}(2)_k$ theories based
on admissible representations have been obtained in 
Refs. \cite{AY,FM}, who agreed
that two different fusion rules were operating, only the first of which 
generalize in an obvious way the fusion rules found in integrable 
representations. In our first work on the subject \cite{PRY} we presented a 
calculation of the 3-point function, which appeared to give rise only to the 
first of the two fusion rules. However, we shall show explicitly in this paper,
that in fact the 4-point functions (also written down in \cite{PRY}) 
clearly imply both fusion rules. In addition we shall show in section 2 that
an analysis of the 3-point function based on the idea of 
over-screening will provide the same result.

The fusion rules provide a neat starting point for giving convenient bases for 
the conformal blocks in the s- and t-channels, namely by demanding 
that some of the 
monodromies become trivial. One must then understand how the general integral
representations can reproduce these bases. Here we shall use either the ones 
provided by us \cite{PRY} (to be referred to as PRY), 
or the one for the 4-point 
function obtained by Andreev \cite{An}. In either case it was not previously
specified how integration contours should be chosen in order to generate 
specific members 
of s- or t-channel bases. In the present paper we discuss that.
In the case of our own integral representation, we show how to obtain conformal
blocks in the s-channel corresponding to fusion rule I, and how 
to obtain conformal
blocks in the t-channel corresponding to fusion rule II,
using contours where the integration of the auxiliary variable, $u$,
(introduced in \cite{PRY}) is carried out first. We also show how to obtain
conformal blocks in the s-channel corresponding to fusion rule II, and how
to obtain conformal blocks in the t-channel corresponding to fusion rule I, 
using contours where the $u$-integration is done last. In the integral 
representation of Andreev \cite{An} there is no $u$-variable to worry about
and the contours we find are more tractable.
That his 4-point blocks are equivalent to ours is a priori
rather clear since both he and we have checked that the blocks we write down
satisfy the Knizhnik-Zamolodchikov equations \cite{KZ}. Nevertheless we find it
very instructive to attempt a direct analytic proof of how the equivalence may 
be obtained, in particular we shall see from the proof how our auxiliary 
integration may be gotten rid of in the process. The situation turns out to 
have a rather remarkable counterpart within the context of minimal models: in 
addition to the standard well known integral representation written down in 
Ref. \cite{DF}, an alternative form also exists, as mentioned in 
Ref. \cite{An}. We show in section 5 that
our integral representation and the one of \cite{An}, are related in a closely
analogous way. 

Having written down the full s- and t-channel bases for conformal blocks and
understood the corresponding integration contours, we go on to calculate the
relevant parts of the crossing matrix in section 6. 
It turns out that only a moderate
generalization will be needed compared to the similar calculation in minimal 
models \cite{DF}. In both cases the calculations are rather lengthy, however. 

In section 7 we use our results to calculate the monodromy invariant
greens functions for 4-point functions.

In section 8 we use the mondoromy coefficients to obtain the operator algebra 
coefficients, in particular for fusion rule I.

In section 9 we show how to generalize the previous treatment in 
which it was 
assumed that both vertices in the 4-point blocks pertain to the same fusion 
rule (I or II). The idea of over-screening used in section 2 
for the 3 point function may
be employed to obtain additional 4-point blocks in which there are different 
fusion rules (I and II) operating at the two vertices of the block. These
new 4-point blocks correspond to different sets of external spins from the
ones previously considered, and so they do not mix with these under crossing.
Based on the calculations carried out in the previous parts of the paper it is
fairly easy to read off what the new monodromy coefficients should be, and in
particular we obtain in this way new expressions for the operator algebra 
coefficients in the case of fusion rule II. Now these are parametrized in a way
quite different from what was obtained in section 7, but in a way which renders
the comparison with the coefficients pertaining to fusion rule I much more 
natural. In this parametrization we find identical functional forms for the 
operator algebra coefficients and we explain some differences in the 
parameters.

Finally, section 10 contains some concluding remarks.

\section{Notations}
We shall 
be interested in N-point functions (in this paper mostly 3- and 4-point
functions) of primary fields. These are taken to depend on a spin label, $j$,
a position variable $z$ (we only need specify the chiral dependencies for 
now), and one more variable, $x$ \cite{FZ,FGPP,PRY,An} which represents an 
equivalent but more convenient way of keeping track of the $SL(2)$ weight 
dependence on a weight, $m$. More precisely, if the affine currents are denoted 
by $J^a(z)$, $a=1,2,3$ or $a=+,-,3$, and if the primary field (chiral vertex 
operator) for short is denoted $\phi_j(w,x)$, then the OPE's take the form
\bea
J^a(z)\phi_j(w,x)&\sim&\frac{1}{z-w}D^a_x\phi_j(w,x)\nn
D^+_x&=&-x^2\partial_x+2xj\nn
D^3_x&=&-x\partial_x +j\nn
D^-_x&=&\partial_x
\label{OPE}
\eea
Correlators (conformal blocks) transform covariantly with respect to 
projective transformations of both $z$ and $x$ variables. Thus in a 4-point 
function 
$$\br \phi_{j_4}(z_4,x_4)\phi_{j_3}(z_3,x_3)\phi_{j_2}(z_2,x_2)
\phi_{j_1}(z_1,x_1)\kt$$
we consider as usual the limits
\bea
&\ \ z_4\rightarrow \infty, \ \ & x_4\rightarrow \infty\nn
&z_3\rightarrow  1, \ \ & x_3\rightarrow 1\nn
&z_2\rightarrow  z, \ \ & x_2\rightarrow  x\nn
&z_1\rightarrow  0, \ \ & x_1\rightarrow  0
\eea
so that the 4-point conformal blocks will be 
(in general multi-valued) functions of $(z,x)$. We 
label s- and t-channel conformal blocks by tree graphs, the meaning of which is
that in the limit $z\rightarrow 0$ followed by $x\rightarrow 0$ the s-channel
block corresponding to Fig. 1 has the behaviour following from the OPE's
\ben
S(z,x)\sim z^{h-h_1-h_2}(-x)^{j_1+j_2-j}(\mbox{const.}+{\cal O}(z,-x))
\label{schannel}
\een
whereas for the t-channel block we have in the limit $z\rightarrow 1$ 
followed by $x\rightarrow 1$
\ben
T(z,x)\sim (1-z)^{h-h_2-h_3}(x-1)^{j_2+j_3-j}(\mbox{const.}+{\cal O}(1-z,x-1))
\label{tchannel}
\een
\begin{figure}
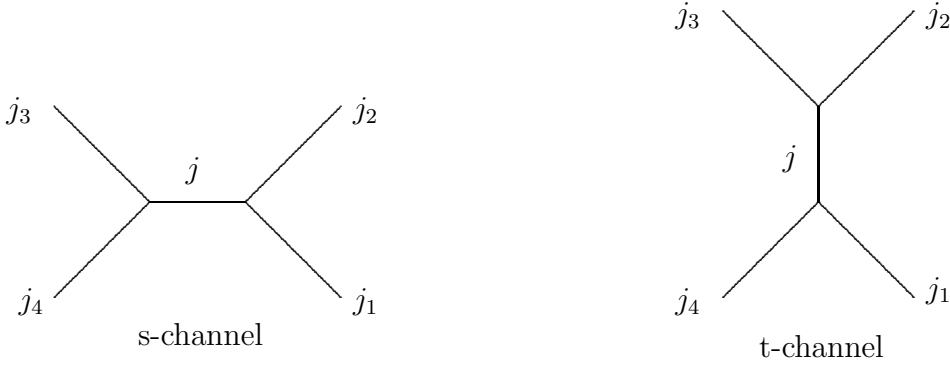

\font\thinlinefont=cmr5
\begingroup\makeatletter\ifx\SetFigFont\undefined
\def\x#1#2#3#4#5#6#7\relax{\def\x{#1#2#3#4#5#6}}%
\expandafter\x\fmtname xxxxxx\relax \def\y{splain}%
\ifx\x\y   
\gdef\SetFigFont#1#2#3{%
  \ifnum #1<17\tiny\else \ifnum #1<20\small\else
  \ifnum #1<24\normalsize\else \ifnum #1<29\large\else
  \ifnum #1<34\Large\else \ifnum #1<41\LARGE\else
     \huge\fi\fi\fi\fi\fi\fi
  \csname #3\endcsname}%
\else
\gdef\SetFigFont#1#2#3{\begingroup
  \count@#1\relax \ifnum 25<\count@\count@25\fi
  \def\x{\endgroup\@setsize\SetFigFont{#2pt}}%
  \expandafter\x
    \csname \romannumeral\the\count@ pt\expandafter\endcsname
    \csname @\romannumeral\the\count@ pt\endcsname
  \csname #3\endcsname}%
\fi
\fi\endgroup
\mbox{\beginpicture
\setcoordinatesystem units <1.00000cm,1.00000cm>
\unitlength=1.00000cm
\linethickness=1pt
\setplotsymbol ({\makebox(0,0)[l]{\tencirc\symbol{'160}}})
\setshadesymbol ({\thinlinefont .})
\setlinear
%
%
\linethickness= 0.500pt
\setplotsymbol ({\thinlinefont .})
\putrule from  2.540 22.860 to  3.810 22.860
%
%
\linethickness= 0.500pt
\setplotsymbol ({\thinlinefont .})
\plot  3.810 22.860  5.080 24.130 /
%
%
\linethickness= 0.500pt
\setplotsymbol ({\thinlinefont .})
\plot  2.540 22.860  1.270 21.590 /
%
%
\linethickness= 0.500pt
\setplotsymbol ({\thinlinefont .})
\plot  3.810 22.860  5.080 21.590 /
%
%
\linethickness= 0.500pt
\setplotsymbol ({\thinlinefont .})
\plot 10.160 25.400 11.430 24.130 /
%
%
\linethickness= 0.500pt
\setplotsymbol ({\thinlinefont .})
\plot 11.430 24.130 12.700 25.400 /
%
%
\linethickness= 0.500pt
\setplotsymbol ({\thinlinefont .})
\putrule from 11.430 24.130 to 11.430 22.860
%
%
\linethickness= 0.500pt
\setplotsymbol ({\thinlinefont .})
\plot 11.430 22.860 10.160 21.590 /
%
%
\linethickness= 0.500pt
\setplotsymbol ({\thinlinefont .})
\plot 11.430 22.860 12.700 21.590 /
%
%
\put{\SetFigFont{12}{14.4}{rm}s-channel} [lB] at  2.381 20.955
%
%
\linethickness= 0.500pt
\setplotsymbol ({\thinlinefont .})
\plot  1.270 24.130  2.540 22.860 /
%
%
\put{$j_1$} [lB] at  5.239 21.431
%
%
\put{$j$} [lB] at 10.954 23.336
%
%
\put{$j_2$} [lB] at  5.239 23.971
%
%
\put{$j_3$} [lB] at  0.635 23.971
%
%
\put{$j_4$} [lB] at  0.794 21.431
%
%
\put{$j$} [lB] at  3.016 23.178
%
%
\put{\SetFigFont{12}{14.4}{rm}t-channel} [lB] at 10.636 20.796
%
%
\put{$j_1$} [lB] at 12.859 21.590
%
%
\put{$j_2$} [lB] at 12.859 25.241
%
%
\put{$j_3$} [lB] at  9.525 25.241
%
%
\put{$j_4$} [lB] at  9.525 21.431
\linethickness=0pt
\putrectangle corners at  0.635 25.546 and 12.859 20.739
\endpicture}

\caption{Graphs for s- and t-channel blocks}
\label{fig1}
\end{figure}
Here the conformal weights are given by the standard expression
\ben
h_i=\frac{j_i(j_i+1)}{t}
\een
with $t=k+2$, where $k$ is the level. Admissible representations exist for
$t=p/q$ with $p$ and $q$ co-prime integers. Then the allowed values for the
spins are given by ($r,s$ integers) \cite{KK,MFF}
\bea
2j^+_{r,s}+1&=&r-st \ \ \ \ (r,s)\geq (1,0)\nn
2j^-_{r,s}+1&=&-r+st \ \ \ (r,s) \geq (1,1)
\label{jpmrs}
\eea
and we have the translation symmetry
\ben
j^\pm_{r+np,s+nq}=j^\pm_{r,s}
\een
Any $j^-$ may be written in terms of $j^+$
\bea
 j_{r,s}^-&=&-j_{r,s}^+-1\nn
 &=&j_{p-r,q-s}^+
\eea
so we may choose to work with the latter. Then 
for a coupling between 3 spins, $j_1,j_2,j_3$, labelled accordingly 
by $r_i,s_i$, the fusion rules are \cite{AY,FM}\\
{\bf Fusion rule I}
\bea
1+|r_1-r_2|\leq&r_3&\leq p-1-|r_1+r_2-p|\nn
|s_1-s_2|\leq &s_3&\leq q-1-|s_1+s_2-q+1|
\label{FI}
\eea
{\bf Fusion rule II}
\bea
1+|p-r_1-r_2|\leq &r_3&\leq p-1-|r_1-r_2|\nn
1+|q-s_1-s_2-1|\leq &s_3&\leq q-2-|s_1-s_2|
\label{FII}
\eea
In both cases, $r_3$ and $s_3$ jump in steps of 2. 
It is easily checked that both sets of fusion rules cannot be 
satisfied simultaneously.
In the next subsection we will discuss how fusion rule II
arises from our 3-point function \cite{PRY}.

A very convenient way to think 
about the fusion rules in our case consists in the following. Consider the 
s-channel coupling of $j_1,j_2$ to a $j$. When we parametrize 
$j=j_I$ for fusion 
rule I as $j_1+j_2-j_I=r-st$ 
we shall see that the integers $r,s$ are related to
the number of screenings of the first and second kinds \cite{BO,PRY} around the
$j_1j_2j_I$ vertex. The singular behaviour of the s-channel block in the limit
$z\rightarrow 0, x\rightarrow 0$ is then
\ben
z^{h-h_1-h_2}(-x)^{j_1+j_2-j_I}=z^{h-h_1-h_2}(-x)^{r-st}
\label{singFI}
\een 
with $h=j_I(j_I+1)/t$. For fusion rule II we may then parametrize the internal
$j$ as
\ben
j_{II}\equiv -j_I-1
\een
Of course the conformal dimensions for $j_I$ and $j_{II}$ are the same, but we
find the singular behaviour of the s-channel block to be
\ben
z^{h-h_1-h_2}(-x)^{j_1+j_2-j_{II}}=z^{h-h_1-h_2}(-x)^{2j_1+2j_2-r+st+1}
\label{singFII}
\een
All these statements follow by analysing the fusion rules \Eq{FI}, \Eq{FII}.
By analysing the s-channel 4-point blocks in the limit 
$z\rightarrow 0, x\rightarrow 0$ we indeed find both of these singular 
behaviours and hence verify that the blocks realize both fusion rules I and II.
In the t-channel the discussion is analogous, with $j_1\leftrightarrow j_3$,
$z\rightarrow 1-z$ and $x\rightarrow 1-x$, so that we consider the limits
$z\rightarrow 1$ followed by $x\rightarrow 1$.

\subsection{Fusion rule II and the 3-point function}
In \cite{PRY} we found using the Felder contours \cite{F} the following
expression for the 3-point function (here corrected for minor misprints)
\bea
 W_{F}^{r,s}&=&e^{i\pi r(r+1-2r_1)/t}e^{i\pi ts(s-1-2s_1)} \nn
 &\cdot&\prod_{j=1}^r\frac{(1-e^{2\pi i(r_1-j)/t})(1-e^{2\pi i j/t})}
  {1-e^{2\pi i/t}}
  \prod_{j=1}^s\frac{(1-e^{2\pi it(s_1+1-j)})(1-e^{2\pi itj})}
  {1-e^{2\pi it}}\nn
 &\cdot&\frac{\Gamma(2j_2+1)}{\Gamma(j_2+j_3-j_1+1)}
  t^{2rs}\prod_{i=1}^r\frac{\Gamma(i/t)}{\Gamma(1/t)}\prod_{i=1}^s
  \frac{\Gamma(it-r)}{\Gamma (t)}\nn
 &\cdot&\prod_{i=0}^{r-1}
  \frac{\Gamma(s_1+1+(1-r_1+i)/t)\Gamma(s_2+(1-r_2+i)/t)}
  {\Gamma(s_1+s_2+1-2s+(r-r_1-r_2+i+1)/t)}\nn
 &\cdot&\prod_{i=0}^{s-1}\frac{\Gamma(r_1-r+(i-s_1)t)\Gamma(r_2-r+(1-s_2+i)t)}
  {\Gamma(r_1-r+r_2+(s-s_1-s_2+i)t)}
\label{3-point}
\eea
It turns out that the Felder contours alone cannot produce a well-defined
and non-vanishing 3-point function corresponding to fusion rule II. We
need the combination that the $r$ screening variables of the first kind are 
integrated along Dotsenko-Fateev contours, while the $s$ screenings
of the second kind are taken along Felder contours (or vice versa). 
This leads to
\bea
 W_{DFF}^{r,s}&=&\lambda_r(1/t)\chi_s^{(2)}(s_1;t)\nn
 &\cdot&\frac{\Gamma(2j_2+1)}{\Gamma(j_2+j_3-j_1+1)}
  t^{2rs}\prod_{i=1}^r\frac{\Gamma(i/t)}{\Gamma(1/t)}\prod_{i=1}^s
  \frac{\Gamma(it-r)}{\Gamma (t)}\nn
 &\cdot&\prod_{i=1}^{r}
  \frac{\Gamma(s_1+1+(-r_1+i)/t)\Gamma(s_2+(-r_2+i)/t)}
  {\Gamma(s_1+s_2+1-2s+(r-r_1-r_2+i)/t)}\nn
 &\cdot&\prod_{i=1}^{s}\frac{\Gamma(r_1-r+(i-1-s_1)t)\Gamma(r_2-r+(-s_2+i)t)}
  {\Gamma(r_1-r+r_2+(s-s_1-s_2-1+i)t)}
\label{WDFF}
\eea
Here we have introduced the functions similar to Ref. \cite{DF}
\ben
\lambda_r(1/t)=\prod_{j=1}^re^{-i\pi(j-1)/t}\frac{s(j/t)}{s(1/t)}
\label{lambda}
\een
with
\ben
s(x)\equiv \sin(\pi x)
\een
and the functions
\bea
 \chi_s^{(2)}(s_1;t)&=&e^{i\pi ts(s-1-2s_1)}
  \prod_{j=1}^s\frac{(1-e^{2\pi it(s_1+1-j)})(1-e^{2\pi itj})}
  {1-e^{2\pi it}}\nn           
 &=&(2i)^se^{i\pi ts(s-1-s_1)}\prod_{j=1}^s\frac{s((j-s_1-1)t)s(jt)}{s(t)}
\label{chis}
\eea
If one chooses the alternative combination where the $r$ screenings of the 
first kind are integrated along Felder contours, while the $s$ screenings 
of the second kind are taken along Dotsenko-Fateev contours, the pre-factor
in (\ref{WDFF}) would be $\chi_r^{(1)}(r_1;1/t)\lambda_s(t)$ where
\bea
 \chi_r^{(1)}(r_1;1/t)&=&e^{i\pi r(r+1-2r_1)/t}
  \prod_{j=1}^r\frac{(1-e^{2\pi i(r_1-j)/t})(1-e^{2\pi i j/t})}{1-e^{2\pi i/t}}
  \nn
 &=&(2i)^re^{i\pi r(r-r_1)/t}\prod_{j=1}^r\frac{s((j-r_1)/t)s(jt)}{s(t)}
\label{chir}
\eea 
There is a considerable freedom in choosing the numbers of screenings
subject to the charge conservation since $p-qt=0$. 
We will denote the following choice
\bea
 2r&=&r_1+r_2-r_3-1+p\nn
 2s&=&s_1+s_2-s_3+q
\label{overscreen}
\eea
as over-screening due to the addition of $p,q$. The analysis of (\ref{WDFF})
in terms of fusion rules is standard and using (\ref{overscreen}) one finds
(\ref{FII}), fusion rule II. 
In the process we encounter the cancellation $\Gamma(0)/\Gamma(0)=1$.
It should be mentioned that for fusion rule I
the choice of contours only affects the normalization.
\section{Conformal blocks for fusion rules I and II according to PRY}

The new feature discussed here compared to our discussion in PRY, 
Ref. \cite{PRY}, is 
the precise specification of integration contours for the various variables in 
the integral representation. Using the results of PRY for the 4-point 
function, we want to show here that the integration contours we
indicate will give rise first to a set of 
s-channel conformal blocks corresponding to the intermediate state ($j$)
in Fig. 1 
being given by fusion rule I, and second to a set of conformal blocks in the
t-channel corresponding to the intermediate $j$ being given by fusion rule II.

\begin{figure}
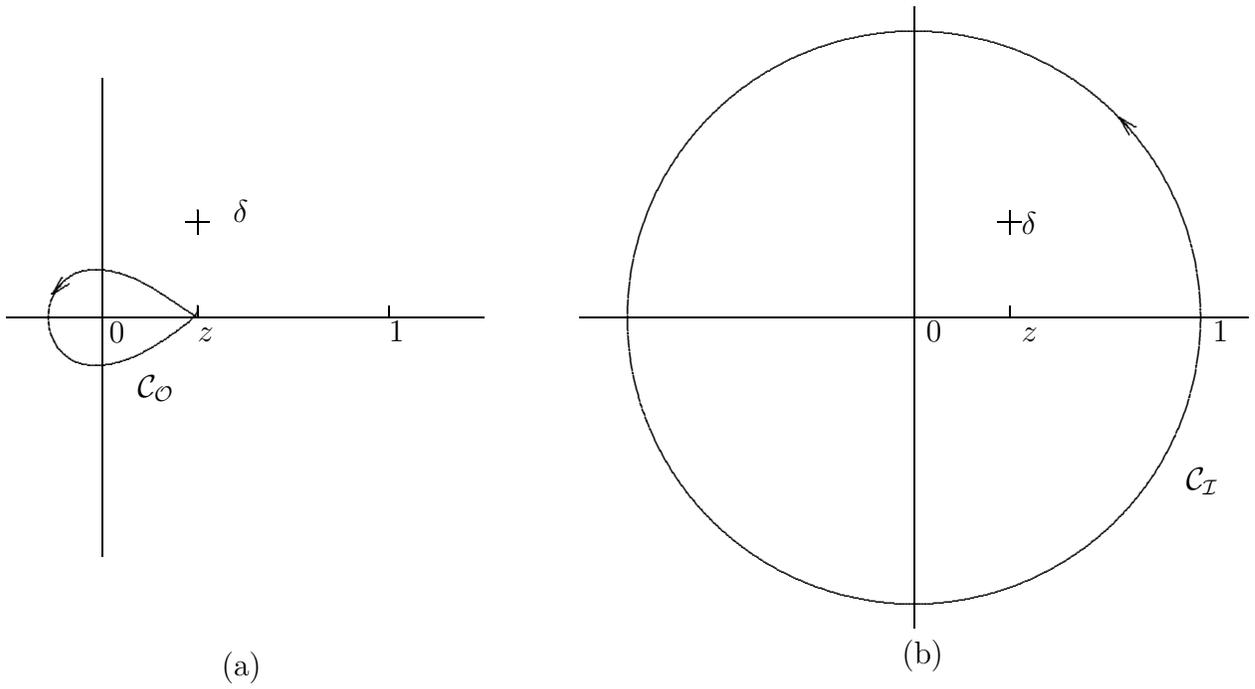

\font\thinlinefont=cmr5
\begingroup\makeatletter\ifx\SetFigFont\undefined
\def\x#1#2#3#4#5#6#7\relax{\def\x{#1#2#3#4#5#6}}%
\expandafter\x\fmtname xxxxxx\relax \def\y{splain}%
\ifx\x\y   
\gdef\SetFigFont#1#2#3{%
  \ifnum #1<17\tiny\else \ifnum #1<20\small\else
  \ifnum #1<24\normalsize\else \ifnum #1<29\large\else
  \ifnum #1<34\Large\else \ifnum #1<41\LARGE\else
     \huge\fi\fi\fi\fi\fi\fi
  \csname #3\endcsname}%
\else
\gdef\SetFigFont#1#2#3{\begingroup
  \count@#1\relax \ifnum 25<\count@\count@25\fi
  \def\x{\endgroup\@setsize\SetFigFont{#2pt}}%
  \expandafter\x
    \csname \romannumeral\the\count@ pt\expandafter\endcsname
    \csname @\romannumeral\the\count@ pt\endcsname
  \csname #3\endcsname}%
\fi
\fi\endgroup
\mbox{\beginpicture
\setcoordinatesystem units <1.00000cm,1.00000cm>
\unitlength=1.00000cm
\linethickness=1pt
\setplotsymbol ({\makebox(0,0)[l]{\tencirc\symbol{'160}}})
\setshadesymbol ({\thinlinefont .})
\setlinear
%
%
\linethickness= 0.500pt
\setplotsymbol ({\thinlinefont .})
\putrule from  1.270 21.590 to  7.620 21.590
%
%
\linethickness= 0.500pt
\setplotsymbol ({\thinlinefont .})
\putrule from  2.540 24.765 to  2.540 18.415
%
%
\linethickness= 0.500pt
\setplotsymbol ({\thinlinefont .})
\putrule from  3.810 21.749 to  3.810 21.590
%
%
\linethickness= 0.500pt
\setplotsymbol ({\thinlinefont .})
\putrule from  6.350 21.749 to  6.350 21.590
%
%
\linethickness= 0.500pt
\setplotsymbol ({\thinlinefont .})
\plot  3.810 23.019  3.810 23.019 /
%
%
\linethickness= 0.500pt
\setplotsymbol ({\thinlinefont .})
\putrule from  3.810 23.019 to  3.810 22.701
%
%
\linethickness= 0.500pt
\setplotsymbol ({\thinlinefont .})
\putrule from  3.651 22.860 to  3.969 22.860
%
%
\linethickness= 0.500pt
\setplotsymbol ({\thinlinefont .})
\putrule from  8.890 21.590 to 17.780 21.590
%
%
\linethickness= 0.500pt
\setplotsymbol ({\thinlinefont .})
\putrule from 13.335 25.718 to 13.335 17.462
%
%
\linethickness= 0.500pt
\setplotsymbol ({\thinlinefont .})
\putrule from 14.605 21.749 to 14.605 21.590
%
%
\linethickness= 0.500pt
\setplotsymbol ({\thinlinefont .})
\putrule from 14.605 23.019 to 14.605 22.701
%
%
\linethickness= 0.500pt
\setplotsymbol ({\thinlinefont .})
\putrule from 14.446 22.860 to 14.764 22.860
%
%
\linethickness= 0.500pt
\setplotsymbol ({\thinlinefont .})
\plot 16.173 24.145 16.061 24.257 /
%
%
\plot 16.286 24.122 16.061 24.257 16.196 24.032 /
%
%
%
\linethickness= 0.500pt
\setplotsymbol ({\thinlinefont .})
\ellipticalarc axes ratio  3.810:3.810  360 degrees 
	from 17.145 21.590 center at 13.335 21.590
%
%
\linethickness= 0.500pt
\setplotsymbol ({\thinlinefont .})
\plot  1.990 22.013  1.877 21.901 /
%
%
\plot  2.012 22.126  1.877 21.901  2.102 22.036 /
%
%
%
\put{${\cal C}_{\cal I}$} [lB] at 16.954 19.304
\linethickness= 0.500pt
\setplotsymbol ({\thinlinefont .})
%
%
%
\plot	 3.810 21.590  3.334 21.907
 	 3.216 21.982
	 3.101 22.046
	 2.988 22.101
	 2.877 22.146
	 2.769 22.180
	 2.664 22.205
	 2.561 22.220
	 2.461 22.225
	 2.365 22.220
	 2.277 22.205
	 2.196 22.180
	 2.123 22.146
	 1.999 22.046
	 1.948 21.982
	 1.905 21.907
	 1.870 21.828
	 1.845 21.749
	 1.831 21.669
	 1.826 21.590
	 1.831 21.511
	 1.845 21.431
	 1.870 21.352
	 1.905 21.273
	 1.948 21.198
	 1.999 21.134
	 2.123 21.034
	 2.196 21.000
	 2.277 20.975
	 2.365 20.960
	 2.461 20.955
	 2.561 20.960
	 2.664 20.975
	 2.769 21.000
	 2.877 21.034
	 2.988 21.079
	 3.101 21.134
	 3.216 21.198
	 3.334 21.273
	 3.445 21.348
	 3.542 21.416
	 3.624 21.477
	 3.691 21.530
	 3.780 21.615
	 3.810 21.669
	 /
\plot  3.810 21.669  3.810 21.749 /
%
%
\put{$\delta$} [lB] at  4.286 22.860
%
%
\put{$z$} [lB] at  3.810 21.273
%
%
\put{$1$} [lB] at  6.350 21.273
%
%
\put{$1$} [lB] at 17.304 21.273
%
%
\put{\SetFigFont{12}{14.4}{rm}(a)} [lB] at  4.128 16.828
%
%
\put{$z$} [lB] at 14.764 21.273
%
%
\put{$\delta$} [lB] at 14.764 22.701
%
%
\put{$0$} [lB] at 13.494 21.273
%
%
\put{\SetFigFont{12}{14.4}{rm}(b)} [lB] at 13.176 16.986
%
%
\put{$0$} [lB] at  2.633 21.258
%
%
\put{${\cal C}_{\cal O}$} [lB] at  3.016 20.542
\linethickness=0pt
\putrectangle corners at  1.245 25.743 and 17.805 16.751
\endpicture}

\caption{The integration contours ${\cal C}_{\cal O}$ (a) and
${\cal C}_{\cal I}$ (b) for an s-channel block corresponding to fusion rule I.}
\label{figpryws}
\end{figure}

We first describe the situation in the s-channel corresponding to fusion rule
I. Here the $u$-integration is carried out first. We write for the conformal
block (cf. \cite{PRY})
\bea
&&W^{(R,S)}_{(r,s)}(j_1,j_2,j_3,j_4;z,x)\nn
&=&z^{2j_1j_2/t}(1-z)^{2j_2j_3/t}
\oint_{{\cal C}_{\cal I}} \prod_{i\in {\cal I}}\frac{dw_i}{2\pi i}
 \oint_{{\cal C}_{\cal O}}\prod_{j\in {\cal O}}\frac{dw_j}{2\pi i}
 \oint_{{\cal C}_u}\frac{du}{2\pi i}\nn
&&w_i^{2k_ij_1/t}(w_i-z)^{2k_ij_2/t}(w_i-1)^{2k_ij_3/t}
 \prod_{\stackrel{i,j\in{\cal A}}{i<j}}(w_i-w_j)^{2k_ik_j/t}\nn
&&\prod_{i\in{\cal A}}\left ( -\frac{u}{w_i-1}+\frac{x}{w_i-z}
 \right )^{-k_i}
 (1-u)^{2j_2+2j_3-R +S t}u^{-2j_3-1}
\label{pryblock}
\eea
Here we are considering an integral representation of the 4-point conformal 
block with a total of $R$ screening operators of the first kind 
(\cite{PRY}) and a total of $S$ screening operators of the second kind. 
The $w_i$'s are the positions of the screening operators, and $k_i=-1$ for 
screenings of the first kind and $k_i=t$ for screenings of the second kind. 
When $i\in{\cal O}$ the corresponding $w_i$ is integrated along the contour of 
Fig. \ref{figpryws}(a) (whether it is of the first or second kind) 
corresponding to a screening of the vertex, $j_1j_2j$. Different 
$i,i'\in{\cal O}$ 
are taken along slightly different contours in order to avoid 
the singularity coming from $(w_i-w_{i'})^{2k_ik_{i'}/t}$. Similarly, the 
$w_j$'s for $j\in{\cal I}$ are integrated along the contour, 
Fig. \ref{figpryws}(b),
corresponding to a screening of the vertex, $jj_3j_4$. 
${\cal A}={\cal O}\cup{\cal I}$ is simply the combined index set. We denote the
numbers of screenings of the first kind at the $j_1j_2j$ and the $jj_3j_4$ 
vertices respectively as $r$ and $R-r$. Similarly the corresponding 
numbers of screenings of the second kinds at the two vertices are denoted 
$s$ and $S-s$. In the product of factors $(w_i-w_j)^{2k_ik_j/t}$
an arbitrary ordering of the indices is implied.

For fixed $w_i$'s the integrand has singularities in the $u$-plane at 
$u=0,1,\Delta_i$, where
\ben
\Delta_i=\frac{w_i-1}{w_i-z}x
\een
The integration contour for $u$ is to divide the singularities, $\Delta_i$,
so that the ones for $i\in{\cal O}$ lie outside ${\cal C}_u$ and the ones for
$i\in{\cal I}$ lie inside ${\cal C}_u$, and ${\cal C}_u$ should pass through 
$u=1$. For $z$ and $x$ sufficiently small,
we may take ${\cal C}_u$ to be the unit circle, Fig. \ref{figpryus}. 
Remember that in order to 
identify the nature of the block and the value of the intermediate $j$, we are 
going to investigate the limit $z\rightarrow 0$ followed by $x\rightarrow 0$. 
\begin{figure}

\font\thinlinefont=cmr5
\begingroup\makeatletter\ifx\SetFigFont\undefined
\def\x#1#2#3#4#5#6#7\relax{\def\x{#1#2#3#4#5#6}}%
\expandafter\x\fmtname xxxxxx\relax \def\y{splain}%
\ifx\x\y   
\gdef\SetFigFont#1#2#3{%
  \ifnum #1<17\tiny\else \ifnum #1<20\small\else
  \ifnum #1<24\normalsize\else \ifnum #1<29\large\else
  \ifnum #1<34\Large\else \ifnum #1<41\LARGE\else
     \huge\fi\fi\fi\fi\fi\fi
  \csname #3\endcsname}%
\else
\gdef\SetFigFont#1#2#3{\begingroup
  \count@#1\relax \ifnum 25<\count@\count@25\fi
  \def\x{\endgroup\@setsize\SetFigFont{#2pt}}%
  \expandafter\x
    \csname \romannumeral\the\count@ pt\expandafter\endcsname
    \csname @\romannumeral\the\count@ pt\endcsname
  \csname #3\endcsname}%
\fi
\fi\endgroup
\mbox{\beginpicture
\setcoordinatesystem units <1.00000cm,1.00000cm>
\unitlength=1.00000cm
\linethickness=1pt
\setplotsymbol ({\makebox(0,0)[l]{\tencirc\symbol{'160}}})
\setshadesymbol ({\thinlinefont .})
\setlinear
%
%
\linethickness= 0.500pt
\setplotsymbol ({\thinlinefont .})
\putrule from  1.270 19.050 to  8.890 19.050
%
%
\linethickness= 0.500pt
\setplotsymbol ({\thinlinefont .})
\putrule from  5.080 22.860 to  5.080 15.240
%
%
\linethickness= 0.500pt
\setplotsymbol ({\thinlinefont .})
\plot  7.097 20.684  6.985 20.796 /
%
%
\plot  7.210 20.662  6.985 20.796  7.120 20.572 /
%
%
%
\put{$\Delta_{i\in{\cal O}}$} [lB] at  7.461 21.114
%
%
\linethickness= 0.500pt
\setplotsymbol ({\thinlinefont .})
\ellipticalarc axes ratio  2.544:2.544  360 degrees 
        from  7.654 19.078 center at  5.110 19.078
%
%
\put{$\Delta_{i\in{\cal I}}$} [lB] at  5.556 19.685
%
%
\put{$0$} [lB] at  5.239 18.733
%
%
\put{${\cal C}_u$} [lB] at  6.985 16.669
%
%
\put{\SetFigFont{12}{14.4}{rm}$u$-plane} [lB] at  7.938 22.225
%
%
\put{$1$} [lB] at  7.779 18.733
\linethickness=0pt
\putrectangle corners at  1.245 22.885 and  8.915 15.215
\endpicture}

\caption{The integration contour ${\cal C}_u$ for an s-channel block 
corresponding to fusion rule I.}
\label{figpryus}
\end{figure}
\begin{figure}
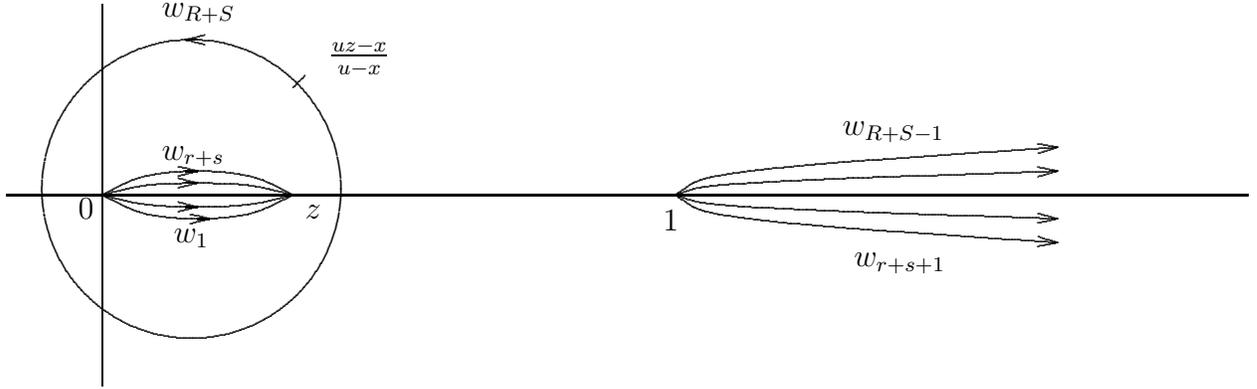

\font\thinlinefont=cmr5
\begingroup\makeatletter\ifx\SetFigFont\undefined
\def\x#1#2#3#4#5#6#7\relax{\def\x{#1#2#3#4#5#6}}%
\expandafter\x\fmtname xxxxxx\relax \def\y{splain}%
\ifx\x\y   
\gdef\SetFigFont#1#2#3{%
  \ifnum #1<17\tiny\else \ifnum #1<20\small\else
  \ifnum #1<24\normalsize\else \ifnum #1<29\large\else
  \ifnum #1<34\Large\else \ifnum #1<41\LARGE\else
     \huge\fi\fi\fi\fi\fi\fi
  \csname #3\endcsname}%
\else
\gdef\SetFigFont#1#2#3{\begingroup
  \count@#1\relax \ifnum 25<\count@\count@25\fi
  \def\x{\endgroup\@setsize\SetFigFont{#2pt}}%
  \expandafter\x
    \csname \romannumeral\the\count@ pt\expandafter\endcsname
    \csname @\romannumeral\the\count@ pt\endcsname
  \csname #3\endcsname}%
\fi
\fi\endgroup
\mbox{\beginpicture
\setcoordinatesystem units <1.00000cm,1.00000cm>
\unitlength=1.00000cm
\linethickness=1pt
\setplotsymbol ({\makebox(0,0)[l]{\tencirc\symbol{'160}}})
\setshadesymbol ({\thinlinefont .})
\setlinear
%
%
\linethickness= 0.500pt
\setplotsymbol ({\thinlinefont .})
\putrule from  1.270 21.590 to 17.780 21.590
%
%
\linethickness= 0.500pt
\setplotsymbol ({\thinlinefont .})
\putrule from  2.540 24.130 to  2.540 19.050
%
%
\linethickness= 0.500pt
\setplotsymbol ({\thinlinefont .})
\putrule from  5.239 23.019 to  5.239 23.019
%
%
\linethickness= 0.500pt
\setplotsymbol ({\thinlinefont .})
\plot  5.080 23.019  5.239 23.178 /
%
%
\linethickness= 0.500pt
\setplotsymbol ({\thinlinefont .})
\putrule from  3.651 21.907 to  3.810 21.907
%
%
\plot  3.556 21.844  3.810 21.907  3.556 21.971 /
%
%
%
\linethickness= 0.500pt
\setplotsymbol ({\thinlinefont .})
\putrule from  3.651 21.749 to  3.810 21.749
%
%
\plot  3.556 21.685  3.810 21.749  3.556 21.812 /
%
%
%
\linethickness= 0.500pt
\setplotsymbol ({\thinlinefont .})
\putrule from  3.651 21.431 to  3.810 21.431
%
%
\plot  3.556 21.368  3.810 21.431  3.556 21.495 /
%
%
%
\linethickness= 0.500pt
\setplotsymbol ({\thinlinefont .})
\putrule from  3.810 21.273 to  3.969 21.273
%
%
\plot  3.715 21.209  3.969 21.273  3.715 21.336 /
%
%
%
\linethickness= 0.500pt
\setplotsymbol ({\thinlinefont .})
\putrule from  3.810 23.654 to  3.651 23.654
%
%
\plot  3.905 23.717  3.651 23.654  3.905 23.590 /
%
%
%
\linethickness= 0.500pt
\setplotsymbol ({\thinlinefont .})
\plot 11.589 26.035 11.589 26.035 /
\linethickness= 0.500pt
\setplotsymbol ({\thinlinefont .})
%
%
%
\plot	 2.540 21.590  2.857 21.669
 	 2.942 21.688
	 3.036 21.704
	 3.140 21.718
	 3.254 21.729
	 3.378 21.738
	 3.444 21.741
	 3.512 21.744
	 3.583 21.746
	 3.656 21.748
	 3.732 21.748
	 3.810 21.749
	 3.888 21.748
	 3.964 21.748
	 4.037 21.746
	 4.108 21.744
	 4.176 21.741
	 4.242 21.738
	 4.366 21.729
	 4.480 21.718
	 4.584 21.704
	 4.678 21.688
	 4.763 21.669
	 /
\plot  4.763 21.669  5.080 21.590 /
\linethickness= 0.500pt
\setplotsymbol ({\thinlinefont .})
%
%
%
\plot	 5.080 21.590  4.763 21.511
 	 4.678 21.492
	 4.584 21.476
	 4.480 21.462
	 4.366 21.451
	 4.242 21.442
	 4.176 21.439
	 4.108 21.436
	 4.037 21.434
	 3.964 21.432
	 3.888 21.432
	 3.810 21.431
	 3.732 21.432
	 3.656 21.432
	 3.583 21.434
	 3.512 21.436
	 3.444 21.439
	 3.378 21.442
	 3.254 21.451
	 3.140 21.462
	 3.036 21.476
	 2.942 21.492
	 2.857 21.511
	 /
\plot  2.857 21.511  2.540 21.590 /
\linethickness= 0.500pt
\setplotsymbol ({\thinlinefont .})
%
%
%
\plot	 2.540 21.590  2.857 21.749
 	 2.942 21.786
	 3.036 21.818
	 3.140 21.845
	 3.254 21.868
	 3.378 21.885
	 3.444 21.892
	 3.512 21.898
	 3.583 21.902
	 3.656 21.905
	 3.732 21.907
	 3.810 21.907
	 3.888 21.907
	 3.964 21.905
	 4.037 21.902
	 4.108 21.898
	 4.176 21.892
	 4.242 21.885
	 4.366 21.868
	 4.480 21.845
	 4.584 21.818
	 4.678 21.786
	 4.763 21.749
	 /
\plot  4.763 21.749  5.080 21.590 /
%
%
\linethickness= 0.500pt
\setplotsymbol ({\thinlinefont .})
\ellipticalarc axes ratio  1.988:1.988  360 degrees 
	from  5.717 21.670 center at  3.730 21.670
\linethickness= 0.500pt
\setplotsymbol ({\thinlinefont .})
%
%
%
\plot	 2.540 21.590  2.857 21.431
 	 2.942 21.394
	 3.036 21.362
	 3.140 21.335
	 3.254 21.312
	 3.378 21.295
	 3.444 21.288
	 3.512 21.282
	 3.583 21.278
	 3.656 21.275
	 3.732 21.273
	 3.810 21.273
	 3.888 21.273
	 3.964 21.275
	 4.037 21.278
	 4.108 21.282
	 4.176 21.288
	 4.242 21.295
	 4.366 21.312
	 4.480 21.335
	 4.584 21.362
	 4.678 21.394
	 4.763 21.431
	 /
\plot  4.763 21.431  5.080 21.590 /
%
%
\put{$w_{R+S}$} [lB] at  3.334 23.971
\linethickness= 0.500pt
\setplotsymbol ({\thinlinefont .})
%
%
%
\plot	10.160 21.590 10.398 21.669
 	10.490 21.689
	10.560 21.699
	10.646 21.709
	10.748 21.719
	10.867 21.729
	10.932 21.734
	11.001 21.739
	11.075 21.744
	11.152 21.749
	11.234 21.754
	11.319 21.759
	11.408 21.764
	11.502 21.769
	11.599 21.774
	11.701 21.779
	11.807 21.783
	11.916 21.788
	12.030 21.793
	12.147 21.798
	12.269 21.803
	12.395 21.808
	12.459 21.811
	12.525 21.813
	12.591 21.816
	12.658 21.818
	12.727 21.821
	12.796 21.823
	12.867 21.826
	12.938 21.828
	 /
\plot 12.938 21.828 15.240 21.907 /
%
%
\plot 14.988 21.835 15.240 21.907 14.984 21.962 /
\linethickness= 0.500pt
\setplotsymbol ({\thinlinefont .})
%
%
%
\plot	10.160 21.590 10.398 21.749
 	10.490 21.788
	10.560 21.808
	10.646 21.828
	10.748 21.848
	10.867 21.868
	10.932 21.878
	11.001 21.888
	11.075 21.898
	11.152 21.907
	11.234 21.917
	11.319 21.927
	11.408 21.937
	11.502 21.947
	11.599 21.957
	11.701 21.967
	11.807 21.977
	11.916 21.987
	12.030 21.997
	12.147 22.007
	12.269 22.017
	12.395 22.027
	12.459 22.032
	12.525 22.036
	12.591 22.041
	12.658 22.046
	12.727 22.051
	12.796 22.056
	12.867 22.061
	12.938 22.066
	 /
\plot 12.938 22.066 15.240 22.225 /
%
%
\plot 14.991 22.144 15.240 22.225 14.982 22.271 /
\linethickness= 0.500pt
\setplotsymbol ({\thinlinefont .})
%
%
%
\plot	10.160 21.590 10.398 21.511
 	10.490 21.491
	10.560 21.481
	10.646 21.471
	10.748 21.461
	10.867 21.451
	10.932 21.446
	11.001 21.441
	11.075 21.436
	11.152 21.431
	11.234 21.426
	11.319 21.421
	11.408 21.416
	11.502 21.411
	11.599 21.406
	11.701 21.401
	11.807 21.397
	11.916 21.392
	12.030 21.387
	12.147 21.382
	12.269 21.377
	12.395 21.372
	12.459 21.369
	12.525 21.367
	12.591 21.364
	12.658 21.362
	12.727 21.359
	12.796 21.357
	12.867 21.354
	12.938 21.352
	 /
\plot 12.938 21.352 15.240 21.273 /
%
%
\plot 14.984 21.218 15.240 21.273 14.988 21.345 /
\linethickness= 0.500pt
\setplotsymbol ({\thinlinefont .})
%
%
%
\plot	10.160 21.590 10.398 21.431
 	10.490 21.392
	10.560 21.372
	10.646 21.352
	10.748 21.332
	10.867 21.312
	10.932 21.302
	11.001 21.292
	11.075 21.282
	11.152 21.273
	11.234 21.263
	11.319 21.253
	11.408 21.243
	11.502 21.233
	11.599 21.223
	11.701 21.213
	11.807 21.203
	11.916 21.193
	12.030 21.183
	12.147 21.173
	12.269 21.163
	12.395 21.153
	12.459 21.148
	12.525 21.144
	12.591 21.139
	12.658 21.134
	12.727 21.129
	12.796 21.124
	12.867 21.119
	12.938 21.114
	 /
\plot 12.938 21.114 15.240 20.955 /
%
%
\plot 14.982 20.909 15.240 20.955 14.991 21.036 /
%
%
%
\put{$0$} [lB] at  2.223 21.273
%
%
\put{$z$} [lB] at  5.239 21.273
%
%
\put{$w_{r+s}$} [lB] at  3.334 22.066
%
%
\put{$w_1$} [lB] at  3.493 20.955
%
%
\put{$\frac{uz-x}{u-x}$} [lB] at  5.556 23.336
%
%
\put{$1$} [lB] at 10.001 21.114
%
%
\put{$w_{R+S-1}$} [lB] at 12.383 22.384
%
%
\put{$w_{r+s+1}$} [lB] at 12.541 20.637
\linethickness=0pt
\putrectangle corners at  1.245 26.060 and 17.805 19.025
\endpicture}

\caption{Integration contours for an s-channel block corresponding to fusion
rule II. The corresponding $u$-integration is simply along the
unit circle.}
\label{figprywsII}
\end{figure}
The different positions of the singularities, $\Delta_i$ in $u$ for 
$i\in{\cal O}$ and $i\in{\cal I}$ mean that they give rise to different 
singularities in the corresponding $w_i$ planes after the $u$-integration 
has been performed. 
In fact, for $i\in {\cal O}$ there occurs a pinching of singularities
when $\Delta_i$ collides with either $0$ of $1$ (the additional singularities 
in $u$). This happens for $w_i$ equal to $1$, and for $w_i$ equal to 
$$\delta =\frac{x-z}{x-1}$$
respectively. In particular no extra singularity is generated at $w_i=z$. 
This is why we may take the contour in $w_i$ to start from $z$ as indicated, 
since the
singularity for $w_i$ is what we term ``pure'', meaning that it is of the form
$$(w_i-z)^a(1+{\cal O}(w-z))$$
One can easily check that this is enough to ensure that the corresponding block
will satisfy the 
Knizhnik-Zamolodchikov equation, going over the proof presented
in PRY \cite{PRY}. In contrast, the singularity at $w_i=0$ is ``non-pure'': it 
is a mixture of different powers of $w_i$. Hence we cannot allow the contour to
end in $w_i=0$, it has to surround that point as indicated. 

Turning to the singularities in $w_i$ for $i\in{\cal I}$, we see that pinching
occurs only when $\Delta_i=1$, so that there is no extra singularity produced
at $w_i=1$: it remains pure, and we may take the integration contour to start 
in $w_i=1$ as indicated. If more convenient, one may take the contour to wrap
around the real axis form $1$ to $\infty$, which is a form closer to the one
used by Dotsenko and Fateev \cite{DF}.

Having established that the choice of contours 
indicated is allowed in the sense
that the conformal block will satisfy the Knizhnik-Zamolodchikov equations, it
is a relatively simple matter to find the leading singularity in the limit
$z\rightarrow 0$ followed by $x\rightarrow 0$. In fact, we may scale all the 
$w_i$'s with $i\in{\cal O}$ as
$$w_i\rightarrow zw_i$$
In the limit $z\rightarrow 0$ this is easily seen to result in a leading $z$ 
behaviour of the form
\ben
W^{(R,S)}_{(r,s)}(z,x)\sim z^{-h(j_1)-h(j_2)+h(j_I)}
\een
where
$$h(j)=j(j+1)/t$$
and where 
\ben
j_I=j_1+j_2-r+s t
\een
This is not enough to prove that indeed the intermediate state corresponds 
to a primary field with that value of $j$, since
\ben
h(j)=h(-j-1)
\een
In fact, according to our earlier discussion, the difference between fusion 
rules I and II is exactly that for fusion rule I we should obtain $j=j_I$ 
whereas for fusion rule II we should obtain $j=j_{II}=-j_I-1$. In other words
the $z$ behaviour is precisely unable to distinguish between the two fusion 
rules. To distinguish we must investigate the leading $x$ behaviour 
in the limit
$x\rightarrow 0$ after we have taken $z\rightarrow 0$. However, it is an 
easy matter to do so and to find the behaviour
\ben
W^{(R,S)}_{(r,s)}\sim z^{-h(j_1)-h(j_2)+h(j_I)}(-x)^{r-s t}
\een
This is the proof that the conformal block we have constructed corresponds to
fusion rule I, since $r-s t=j_1+j_2-j_I$.

We next describe how contours have to be chosen in order to produce a 
t-channel block corresponding to fusion rule II. This situation can occur
only provided there is at least one screening operator of the second kind 
\cite{PRY}. We use the same defining 
equation as in \Eq{pryblock}, but the sets of indices as well as $r$ 
and $s$ have different meanings. Again we have a total of $R$ 
and $S$ screenings of the first and second kinds. There are $r$ and
$s$ screenings associated with the upper vertex, and the corresponding
index set for the $w_i$'s is ${\cal O}$. There are $R-r$ and $S-s$ 
screening operators of the two kinds associated with the lower vertex and the 
corresponding index set for the $w_i$'s is ${\cal I}$.
The integration contour,
${\cal C}_u$, is indicated in Fig. \ref{figpryut}. One checks that in the limit
$z\rightarrow 1$ followed by $x\rightarrow 1$ the two sets of singularities,
$\Delta_i$ for $i\in{\cal O}$ and $i\in{\cal I}$ respectively are well 
separated, so that the contour may be taken to separate them as indicated.
The contours for the two
sets of $w_i$ contours, ${\cal C}_{\cal O}$ and ${\cal C}_{\cal I}$ are shown 
in Fig. \ref{figprywt} (a) and (b) respectively.
\begin{figure}
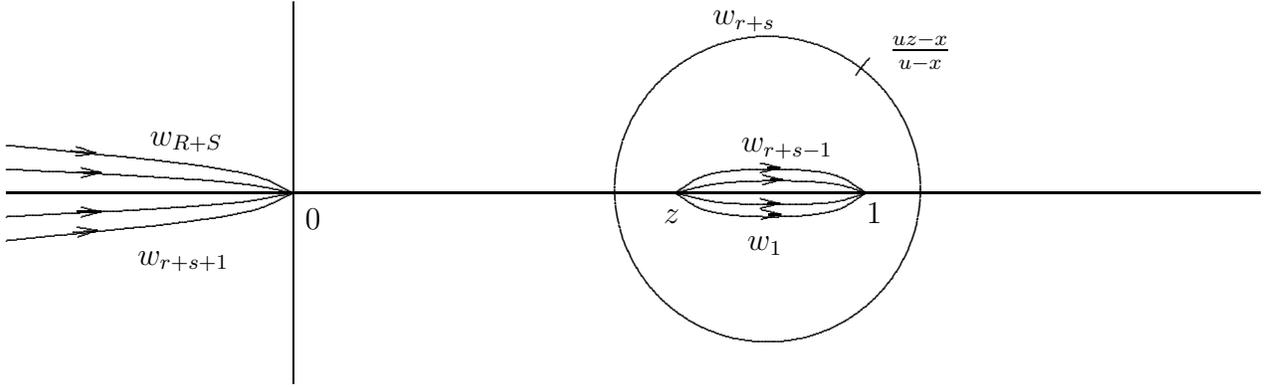

\font\thinlinefont=cmr5
\begingroup\makeatletter\ifx\SetFigFont\undefined
\def\x#1#2#3#4#5#6#7\relax{\def\x{#1#2#3#4#5#6}}%
\expandafter\x\fmtname xxxxxx\relax \def\y{splain}%
\ifx\x\y   
\gdef\SetFigFont#1#2#3{%
  \ifnum #1<17\tiny\else \ifnum #1<20\small\else
  \ifnum #1<24\normalsize\else \ifnum #1<29\large\else
  \ifnum #1<34\Large\else \ifnum #1<41\LARGE\else
     \huge\fi\fi\fi\fi\fi\fi
  \csname #3\endcsname}%
\else
\gdef\SetFigFont#1#2#3{\begingroup
  \count@#1\relax \ifnum 25<\count@\count@25\fi
  \def\x{\endgroup\@setsize\SetFigFont{#2pt}}%
  \expandafter\x
    \csname \romannumeral\the\count@ pt\expandafter\endcsname
    \csname @\romannumeral\the\count@ pt\endcsname
  \csname #3\endcsname}%
\fi
\fi\endgroup
\mbox{\beginpicture
\setcoordinatesystem units <1.00000cm,1.00000cm>
\unitlength=1.00000cm
\linethickness=1pt
\setplotsymbol ({\makebox(0,0)[l]{\tencirc\symbol{'160}}})
\setshadesymbol ({\thinlinefont .})
\setlinear
%
%
\linethickness= 0.500pt
\setplotsymbol ({\thinlinefont .})
\putrule from  1.270 21.590 to 17.939 21.590
%
%
\linethickness= 0.500pt
\setplotsymbol ({\thinlinefont .})
\putrule from  5.080 24.130 to  5.080 19.050
%
%
\linethickness= 0.500pt
\setplotsymbol ({\thinlinefont .})
\plot  2.127 22.147  2.451 22.121 /
%
%
\plot  2.193 22.078  2.451 22.121  2.203 22.204 /
%
%
%
\linethickness= 0.500pt
\setplotsymbol ({\thinlinefont .})
\putrule from  2.206 21.861 to  2.523 21.861
%
%
\plot  2.269 21.797  2.523 21.861  2.269 21.924 /
%
%
%
\linethickness= 0.500pt
\setplotsymbol ({\thinlinefont .})
\putrule from  2.206 21.336 to  2.523 21.336
%
%
\plot  2.269 21.273  2.523 21.336  2.269 21.400 /
%
%
%
\linethickness= 0.500pt
\setplotsymbol ({\thinlinefont .})
\plot  2.155 21.063  2.470 21.090 /
%
%
\plot  2.223 21.005  2.470 21.090  2.212 21.132 /
%
%
%
\linethickness= 0.500pt
\setplotsymbol ({\thinlinefont .})
\putrule from 11.390 21.924 to 11.549 21.924
%
%
\plot 11.295 21.861 11.549 21.924 11.295 21.988 /
%
%
%
\linethickness= 0.500pt
\setplotsymbol ({\thinlinefont .})
\putrule from 11.422 21.766 to 11.580 21.766
%
%
\plot 11.326 21.702 11.580 21.766 11.326 21.829 /
%
%
%
\linethickness= 0.500pt
\setplotsymbol ({\thinlinefont .})
\putrule from 11.390 21.448 to 11.549 21.448
%
%
\plot 11.295 21.385 11.549 21.448 11.295 21.512 /
%
%
%
\linethickness= 0.500pt
\setplotsymbol ({\thinlinefont .})
\putrule from 11.405 21.304 to 11.563 21.304
%
%
\plot 11.309 21.241 11.563 21.304 11.309 21.368 /
%
%
%
\linethickness= 0.500pt
\setplotsymbol ({\thinlinefont .})
\plot 12.565 23.163 12.738 23.385 /
\linethickness= 0.500pt
\setplotsymbol ({\thinlinefont .})
%
%
%
\plot	 5.080 21.590  4.683 21.669
 	 4.567 21.689
	 4.495 21.699
	 4.415 21.709
	 4.327 21.719
	 4.229 21.729
	 4.123 21.739
	 4.008 21.749
	 3.885 21.759
	 3.820 21.764
	 3.753 21.769
	 3.684 21.774
	 3.612 21.779
	 3.539 21.783
	 3.463 21.788
	 3.385 21.793
	 3.305 21.798
	 3.222 21.803
	 3.138 21.808
	 3.051 21.813
	 2.962 21.818
	 2.871 21.823
	 2.778 21.828
	 /
\plot  2.778 21.828  1.270 21.907 /
\linethickness= 0.500pt
\setplotsymbol ({\thinlinefont .})
%
%
%
\plot	 5.080 21.590  4.763 21.749
 	 4.663 21.788
	 4.599 21.808
	 4.524 21.828
	 4.440 21.848
	 4.346 21.868
	 4.242 21.888
	 4.128 21.907
	 4.003 21.927
	 3.938 21.937
	 3.870 21.947
	 3.799 21.957
	 3.726 21.967
	 3.650 21.977
	 3.572 21.987
	 3.491 21.997
	 3.408 22.007
	 3.323 22.017
	 3.235 22.027
	 3.144 22.036
	 3.051 22.046
	 2.955 22.056
	 2.857 22.066
	 /
\plot  2.857 22.066  1.270 22.225 /
%
%
\linethickness= 0.500pt
\setplotsymbol ({\thinlinefont .})
\ellipticalarc axes ratio  2.032:2.032  360 degrees 
	from 13.422 21.639 center at 11.390 21.639
\linethickness= 0.500pt
\setplotsymbol ({\thinlinefont .})
%
%
%
\plot	 5.080 21.590  4.763 21.431
 	 4.663 21.392
	 4.599 21.372
	 4.524 21.352
	 4.440 21.332
	 4.346 21.312
	 4.242 21.292
	 4.128 21.273
	 4.003 21.253
	 3.938 21.243
	 3.870 21.233
	 3.799 21.223
	 3.726 21.213
	 3.650 21.203
	 3.572 21.193
	 3.491 21.183
	 3.408 21.173
	 3.323 21.163
	 3.235 21.153
	 3.144 21.144
	 3.051 21.134
	 2.955 21.124
	 2.857 21.114
	 /
\plot  2.857 21.114  1.270 20.955 /
%
%
\put{$w_{r+s}$} [lB] at 10.660 23.829
\linethickness= 0.500pt
\setplotsymbol ({\thinlinefont .})
%
%
%
\plot	 5.080 21.590  4.683 21.511
 	 4.567 21.491
	 4.495 21.481
	 4.415 21.471
	 4.327 21.461
	 4.229 21.451
	 4.123 21.441
	 4.008 21.431
	 3.885 21.421
	 3.820 21.416
	 3.753 21.411
	 3.684 21.406
	 3.612 21.401
	 3.539 21.397
	 3.463 21.392
	 3.385 21.387
	 3.305 21.382
	 3.222 21.377
	 3.138 21.372
	 3.051 21.367
	 2.962 21.362
	 2.871 21.357
	 2.778 21.352
	 /
\plot  2.778 21.352  1.270 21.273 /
\linethickness= 0.500pt
\setplotsymbol ({\thinlinefont .})
%
%
%
\plot	10.160 21.590 10.478 21.669
 	10.563 21.688
	10.661 21.704
	10.771 21.718
	10.894 21.729
	10.960 21.734
	11.029 21.738
	11.102 21.741
	11.177 21.744
	11.255 21.746
	11.337 21.748
	11.422 21.748
	11.509 21.749
	11.597 21.748
	11.681 21.748
	11.761 21.746
	11.837 21.744
	11.909 21.741
	11.978 21.738
	12.105 21.729
	12.216 21.718
	12.313 21.704
	12.395 21.688
	12.462 21.669
	 /
\plot 12.462 21.669 12.700 21.590 /
\linethickness= 0.500pt
\setplotsymbol ({\thinlinefont .})
%
%
%
\plot	10.160 21.590 10.398 21.749
 	10.466 21.786
	10.552 21.818
	10.655 21.845
	10.775 21.868
	10.842 21.877
	10.913 21.885
	10.988 21.892
	11.068 21.898
	11.152 21.902
	11.240 21.905
	11.333 21.907
	11.430 21.907
	11.527 21.907
	11.620 21.905
	11.708 21.902
	11.792 21.898
	11.872 21.892
	11.947 21.885
	12.018 21.877
	12.085 21.868
	12.205 21.845
	12.308 21.818
	12.394 21.786
	12.462 21.749
	 /
\plot 12.462 21.749 12.700 21.590 /
\linethickness= 0.500pt
\setplotsymbol ({\thinlinefont .})
%
%
%
\plot	10.160 21.590 10.398 21.431
 	10.466 21.394
	10.552 21.362
	10.655 21.335
	10.775 21.312
	10.842 21.303
	10.913 21.295
	10.988 21.288
	11.068 21.282
	11.152 21.278
	11.240 21.275
	11.333 21.273
	11.430 21.273
	11.527 21.273
	11.620 21.275
	11.708 21.278
	11.792 21.282
	11.872 21.288
	11.947 21.295
	12.018 21.303
	12.085 21.312
	12.205 21.335
	12.308 21.362
	12.394 21.394
	12.462 21.431
	 /
\plot 12.462 21.431 12.700 21.590 /
\linethickness= 0.500pt
\setplotsymbol ({\thinlinefont .})
%
%
%
\plot	10.160 21.590 10.478 21.511
 	10.563 21.492
	10.661 21.476
	10.771 21.462
	10.894 21.451
	10.960 21.446
	11.029 21.442
	11.102 21.439
	11.177 21.436
	11.255 21.434
	11.337 21.432
	11.422 21.432
	11.509 21.431
	11.597 21.432
	11.681 21.432
	11.761 21.434
	11.837 21.436
	11.909 21.439
	11.978 21.442
	12.105 21.451
	12.216 21.462
	12.313 21.476
	12.395 21.492
	12.462 21.511
	 /
\plot 12.462 21.511 12.700 21.590 /
%
%
\put{$w_{R+S}$} [lB] at  3.175 22.225
%
%
\put{$w_{r+s+1}$} [lB] at  3.016 20.637
%
%
\put{$0$} [lB] at  5.239 21.114
%
%
\put{$w_{r+s-1}$} [lB] at 11.041 22.147
%
%
\put{$w_1$} [lB] at 11.119 20.828
%
%
\put{$z$} [lB] at 10.008 21.194
%
%
\put{$1$} [lB] at 12.706 21.194
%
%
\put{$\frac{uz-x}{u-x}$} [lB] at 13.009 23.385
\linethickness=0pt
\putrectangle corners at  1.245 24.155 and 17.964 19.025
\endpicture}

\caption{Integration contours for the screening charges in the case of a 
t-channel block corresponding to fusion rule I. The $u$-integration is along 
a closed contour starting in 1 and surrounding $x$.}
\label{figprywtI}
\end{figure}
\begin{figure} 

\font\thinlinefont=cmr5
\begingroup\makeatletter\ifx\SetFigFont\undefined
\def\x#1#2#3#4#5#6#7\relax{\def\x{#1#2#3#4#5#6}}%
\expandafter\x\fmtname xxxxxx\relax \def\y{splain}%
\ifx\x\y   
\gdef\SetFigFont#1#2#3{%
  \ifnum #1<17\tiny\else \ifnum #1<20\small\else
  \ifnum #1<24\normalsize\else \ifnum #1<29\large\else
  \ifnum #1<34\Large\else \ifnum #1<41\LARGE\else
     \huge\fi\fi\fi\fi\fi\fi
  \csname #3\endcsname}%
\else
\gdef\SetFigFont#1#2#3{\begingroup
  \count@#1\relax \ifnum 25<\count@\count@25\fi
  \def\x{\endgroup\@setsize\SetFigFont{#2pt}}%
  \expandafter\x
    \csname \romannumeral\the\count@ pt\expandafter\endcsname
    \csname @\romannumeral\the\count@ pt\endcsname
  \csname #3\endcsname}%
\fi
\fi\endgroup
\mbox{\beginpicture
\setcoordinatesystem units <1.00000cm,1.00000cm>
\unitlength=1.00000cm
\linethickness=1pt
\setplotsymbol ({\makebox(0,0)[l]{\tencirc\symbol{'160}}})
\setshadesymbol ({\thinlinefont .})
\setlinear
%
%
\linethickness= 0.500pt
\setplotsymbol ({\thinlinefont .})
\putrule from  3.810 21.590 to 12.700 21.590
%
%
\linethickness= 0.500pt
\setplotsymbol ({\thinlinefont .})
\putrule from 10.160 23.495 to 10.160 23.654
%
%
\plot 10.224 23.400 10.160 23.654 10.097 23.400 /
\linethickness= 0.500pt
\setplotsymbol ({\thinlinefont .})
%
%
%
\plot    8.890 21.590  8.255 22.225
         8.178 22.304
         8.106 22.384
         8.039 22.463
         7.977 22.543
         7.920 22.622
         7.868 22.701
         7.821 22.781
         7.779 22.860
         7.742 22.939
         7.709 23.019
         7.682 23.098
         7.660 23.178
         7.642 23.257
         7.630 23.336
         7.622 23.416
         7.620 23.495
         7.622 23.574
         7.630 23.654
         7.642 23.733
         7.660 23.813
         7.682 23.892
         7.709 23.971
         7.742 24.051
         7.779 24.130
         7.821 24.209
         7.868 24.289
         7.920 24.368
         7.977 24.448
         8.039 24.527
         8.106 24.606
         8.178 24.686
         8.255 24.765
         8.334 24.839
         8.414 24.904
         8.493 24.958
         8.572 25.003
         8.652 25.038
         8.731 25.063
         8.811 25.078
         8.890 25.082
         8.969 25.078
         9.049 25.063
         9.128 25.038
         9.207 25.003
         9.287 24.958
         9.366 24.904
         9.446 24.839
         9.525 24.765
         9.602 24.686
         9.674 24.606
         9.741 24.527
         9.803 24.448
         9.860 24.368
         9.912 24.289
         9.959 24.209
        10.001 24.130
        10.038 24.051
        10.071 23.971
        10.098 23.892
        10.120 23.813
        10.138 23.733
        10.150 23.654
        10.158 23.574
        10.160 23.495
        10.158 23.416
        10.150 23.336
        10.138 23.257
        10.120 23.178
        10.098 23.098
        10.071 23.019
        10.038 22.939
        10.001 22.860
         9.959 22.781
         9.912 22.701
         9.860 22.622
         9.803 22.543
         9.741 22.463
         9.674 22.384
         9.602 22.304
         9.525 22.225
         /
\plot  9.525 22.225  8.890 21.590 /
%
%
\linethickness= 0.500pt
\setplotsymbol ({\thinlinefont .})
\putrule from  5.080 24.130 to  5.080 19.050
%
%
\put{\SetFigFont{12}{14.4}{rm}$u$-plane} [lB] at 10.795 25.082
%
%
\put{${\cal C}_u$} [lB] at  9.842 22.066
%
%
\put{$\Delta_{j\in{\cal I}}$} [lB] at  8.255 23.336
%
%
\put{$\Delta_{i\in{\cal O}}$} [lB] at  5.715 24.606
%
%
\put{$1$} [lB] at  8.890 21.273
\linethickness=0pt
\putrectangle corners at  3.785 25.387 and 12.725 19.025
\endpicture}

\caption{The integration contour ${\cal C}_u$ for a t-channel block 
corresponding to fusion rule II.}
\label{figpryut}
\end{figure}
\begin{figure}
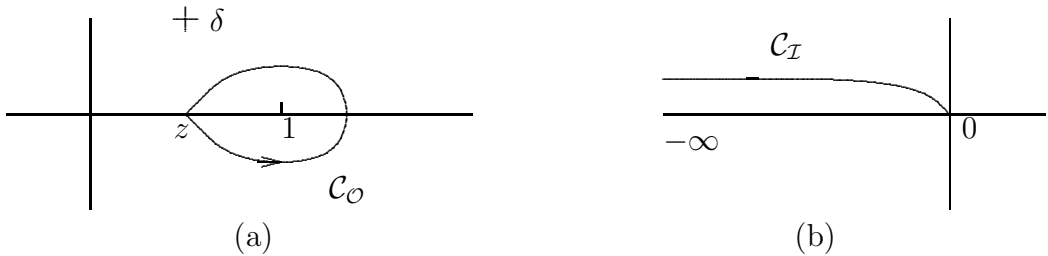

\font\thinlinefont=cmr5
\begingroup\makeatletter\ifx\SetFigFont\undefined
\def\x#1#2#3#4#5#6#7\relax{\def\x{#1#2#3#4#5#6}}%
\expandafter\x\fmtname xxxxxx\relax \def\y{splain}%
\ifx\x\y   
\gdef\SetFigFont#1#2#3{%
  \ifnum #1<17\tiny\else \ifnum #1<20\small\else
  \ifnum #1<24\normalsize\else \ifnum #1<29\large\else
  \ifnum #1<34\Large\else \ifnum #1<41\LARGE\else
     \huge\fi\fi\fi\fi\fi\fi
  \csname #3\endcsname}%
\else
\gdef\SetFigFont#1#2#3{\begingroup
  \count@#1\relax \ifnum 25<\count@\count@25\fi
  \def\x{\endgroup\@setsize\SetFigFont{#2pt}}%
  \expandafter\x
    \csname \romannumeral\the\count@ pt\expandafter\endcsname
    \csname @\romannumeral\the\count@ pt\endcsname
  \csname #3\endcsname}%
\fi
\fi\endgroup
\mbox{\beginpicture
\setcoordinatesystem units <1.00000cm,1.00000cm>
\unitlength=1.00000cm
\linethickness=1pt
\setplotsymbol ({\makebox(0,0)[l]{\tencirc\symbol{'160}}})
\setshadesymbol ({\thinlinefont .})
\setlinear
%
%
\linethickness= 0.500pt
\setplotsymbol ({\thinlinefont .})
\putrule from  1.429 22.860 to  7.620 22.860
%
%
\linethickness= 0.500pt
\setplotsymbol ({\thinlinefont .})
\putrule from  5.080 23.019 to  5.080 22.860
%
%
\linethickness= 0.500pt
\setplotsymbol ({\thinlinefont .})
\putrule from 10.160 22.860 to 15.240 22.860
%
%
\linethickness= 0.500pt
\setplotsymbol ({\thinlinefont .})
\putrule from 13.970 24.130 to 13.970 21.590
%
%
\linethickness= 0.500pt
\setplotsymbol ({\thinlinefont .})
\putrule from 11.271 23.336 to 11.430 23.336
\putrule from 11.430 23.336 to 11.430 23.336
%
%
\linethickness= 0.500pt
\setplotsymbol ({\thinlinefont .})
\putrule from  4.763 22.225 to  5.080 22.225
%
%
\plot  4.826 22.162  5.080 22.225  4.826 22.289 /
\linethickness= 0.500pt
\setplotsymbol ({\thinlinefont .})
%
%
%
\plot	 3.810 22.860  4.128 23.178
 	 4.212 23.252
	 4.306 23.316
	 4.410 23.371
	 4.524 23.416
	 4.648 23.450
	 4.714 23.464
	 4.782 23.475
	 4.853 23.484
	 4.926 23.490
	 5.002 23.494
	 5.080 23.495
	 5.158 23.494
	 5.231 23.490
	 5.301 23.484
	 5.368 23.475
	 5.489 23.450
	 5.596 23.416
	 5.688 23.371
	 5.765 23.316
	 5.827 23.252
	 5.874 23.178
	 5.908 23.098
	 5.933 23.019
	 5.948 22.939
	 5.953 22.860
	 5.948 22.781
	 5.933 22.701
	 5.908 22.622
	 5.874 22.543
	 5.827 22.468
	 5.765 22.404
	 5.688 22.349
	 5.596 22.304
	 5.489 22.270
	 5.368 22.245
	 5.301 22.236
	 5.231 22.230
	 5.158 22.226
	 5.080 22.225
	 5.002 22.226
	 4.926 22.230
	 4.853 22.236
	 4.782 22.245
	 4.714 22.256
	 4.648 22.270
	 4.524 22.304
	 4.410 22.349
	 4.306 22.404
	 4.212 22.468
	 4.128 22.543
	 /
\plot  4.128 22.543  3.810 22.860 /
\linethickness= 0.500pt
\setplotsymbol ({\thinlinefont .})
%
%
%
\plot	 3.810 23.971  3.810 24.130
 	 /
\plot  3.810 24.130  3.810 24.289 /
\linethickness= 0.500pt
\setplotsymbol ({\thinlinefont .})
%
%
%
\plot	 3.651 24.130  3.810 24.130
 	 /
\plot  3.810 24.130  3.969 24.130 /
%
%
\linethickness= 0.500pt
\setplotsymbol ({\thinlinefont .})
\putrule from  2.540 24.130 to  2.540 21.590
\linethickness= 0.500pt
\setplotsymbol ({\thinlinefont .})
%
%
%
\plot	10.160 23.336 11.430 23.336
 	11.509 23.336
	11.586 23.336
	11.661 23.336
	11.735 23.335
	11.807 23.334
	11.878 23.333
	11.948 23.332
	12.015 23.331
	12.082 23.330
	12.146 23.328
	12.271 23.325
	12.389 23.321
	12.502 23.316
	12.608 23.311
	12.707 23.305
	12.801 23.299
	12.889 23.292
	12.970 23.284
	13.045 23.275
	13.114 23.266
	13.176 23.257
	13.290 23.236
	13.395 23.212
	13.489 23.186
	13.573 23.158
	13.648 23.127
	13.712 23.093
	13.811 23.019
	 /
\plot 13.811 23.019 13.970 22.860 /
%
%
\put{${\cal C}_{\cal I}$} [lB] at 11.589 23.654
%
%
\put{$z$} [lB] at  3.651 22.543
%
%
\put{$1$} [lB] at  5.080 22.543
%
%
\put{$\delta$} [lB] at  4.128 23.971
%
%
\put{$0$} [lB] at 14.129 22.543
%
%
\put{$-\infty$} [lB] at 10.160 22.384
%
%
\put{\SetFigFont{12}{14.4}{rm}(b)} [lB] at 11.906 21.114
%
%
\put{\SetFigFont{12}{14.4}{rm}(a)} [lB] at  4.445 21.114
%
%
\put{${\cal C}_{\cal O}$} [lB] at  5.715 21.749
\linethickness=0pt
\putrectangle corners at  1.403 24.306 and 15.265 21.038
\endpicture}

\caption{The integration contours for the $w_i$'s: ${\cal C}_{\cal O}$ (a) and
${\cal C}_{\cal I}$ (b) for a t-channel block corresponding to fusion rule II.}
\label{figprywt}
\end{figure}
In all cases one checks as for the s-channel block that the nature of 
singularities is such that the contours may be chosen as indicated, and that
the block thus defined will satisfy the Knizhnik-Zamolodchikov equations, 
following PRY \cite{PRY}. Then we investigate the combined behaviour 
$z\rightarrow 1$ followed by $x\rightarrow 1$. To this end we perform the 
following scalings of the integration variables:
\bea
w_i&\rightarrow&\frac{w_i-1}{z-1}, \ \ i\in{\cal O}\nn
w_i&\rightarrow&\frac{w_i}{w_i-1}, \ \ i\in{\cal I}\nn
u&\rightarrow&\frac{u-\Delta_{j_0}}{1-\Delta_{j_0}}
\eea
where $j_0$ is an arbitrary index in the set ${\cal I}$, however with the 
restriction that $w_{j_0}$ is the position of a screening operator of the 
second kind. It is rather straightforward 
to check that this gives rise to the combined singular behaviour
\ben
W^{(R,S)}_{(r,s)}(z,x)\sim (1-z)^{-h(j_2)-h(j_3)+h(j_{II})}
(x-1)^{2j_2+2j_3-r+s t+1}
\label{pryII}
\een
where 
\bea
j_{II}&=&-j_I-1\nn
j_I&=&j_2+j_3-r+s t
\eea
so that \Eq{pryII} exactly demonstrates that we have fusion rule II, since
$j_2+j_3-j_{II}=2j_2+2j_3-r+s t+1$.

It follows that the conformal blocks defined on the basis of the free field 
realization elaborated in PRY \cite{PRY}, indeed do give rise to both the
fusion rules previously found in the literature \cite{AY,FM}.

We have also 
found realizations of conformal blocks corresponding to fusion rule
II in the s-channel and of ones corresponding to fusion rule I in the 
t-channel. However, these are not given by quite as 
simple contours as above. It is the appearance of non-pure
singularities which has prevented us from finding such simple contours. 
The new idea is to carry out first the integrations of
the screening operators letting the contours depend on $u$. Then there are
only pure singularities in the $w_i$ planes and there will be no problems
caused by non-pure singularities. It turns out that it is possible to
find contours like that leaving, upon integration, a simple $u$ integral.
Let us first consider the conformal blocks in the s-channel corresponding
to fusion rule II, where the contours are depicted in Fig. \ref{figprywsII}
\bea
 &&W^{(R,S)}_{(r,s)}(j_1,j_2,j_3,j_4;z,x)\nn
 &=&z^{2j_1j_2/t}(1-z)^{2j_2j_3/t}
  \oint_{{\cal C}_u}\frac{du}{2\pi i}
  \oint_{{\cal C}_w}\frac{dw}{2\pi i}
  \int_1^\infty\prod_{i\in {\cal I}}dw_i
  \int_0^z\prod_{i\in {\cal O}} dw_i \nn
 &\cdot&w_i^{2k_ij_1/t}(w_i-z)^{2k_ij_2/t}(w_i-1)^{2k_ij_3/t}
  \prod_{\stackrel{i,j\in{\cal A}}{i<j}}(w_i-w_j)^{2k_ik_j/t}\nn
 &\cdot&\prod_{i\in{\cal A}}\left ( -\frac{u}{w_i-1}+
  \frac{x}{w_i-z}\right )^{-k_i}(1-u)^{2j_2+2j_3-R +S t}u^{-2j_3-1}
\label{pryb}
\eea
where 
\bea
{\cal A}&=&{\cal I}\cup {\cal O}\cup \{ R+S\} \nn
w &\equiv & w_{R+S} \nn
k_{R+S} &=& t
\eea
To see that the above formula produces the right singular behaviour 
in the limit
$z\rightarrow 0$ followed by $x\rightarrow 0$, we may scale all the 
$w_j$'s with $j\in{\cal O}$ as
$$w_j\rightarrow zw_j, \ \ \ \ \ \ \mbox{for all } \ \ j \in {\cal O}$$ 
and scale all the $w_i$'s with $i\in{\cal I}$ as
$$w_i\rightarrow 1/w_i, \ \ \ \ \ \ \mbox{for all } \ \ i \in {\cal I}$$
and also
$$w\rightarrow \frac{uz-x}{u-x} w$$
We can then show that
\ben
W^{(R,S)}_{(r,s)}(z,x)\sim z^{-h(j_1)-h(j_2)+h(j_{II})}(-x)^{j_1+j_2-j_{II}}
\een
where 
\ben
j_{II}=-j_1-j_2+r-s t-1
\een
This is precisely the expected singular behaviour.

Finally we consider the conformal blocks in the t-channel corresponding to
fusion rule I, see Fig. \ref{figprywtI}
\bea
 &&W^{(R,S)}_{(r,s)}(j_1,j_2,j_3,j_4;z,x)\nn
 &=&z^{2j_1j_2/t}(1-z)^{2j_2j_3/t}
  \oint_{{\cal C}_u} \frac{du}{2\pi i}
  \oint_{{\cal C}_w}\frac{dw}{2\pi i}
  \int_0^{-\infty}\prod_{i\in {\cal I}}\frac{dw_i}{2\pi i}
  \int_z^1\prod_{i\in {\cal O}}\frac{dw_i}{2\pi i}\nn
 &\cdot&
   w_i^{2k_ij_1/t}(w_i-z)^{2k_ij_2/t}(1-w_i)^{2k_ij_3/t}
  \prod_{\stackrel{i,j\in{\cal A}}{i<j}}(w_i-w_j)^{2k_ik_j/t}\nn
 &\cdot&
  \prod_{i\in{\cal A}}\left ( -\frac{u}{w_i-1}+\frac{x}{w_i-z}\right )^{-k_i} 
  (1-u)^{2j_2+2j_3-R +S t}u^{-2j_3-1}
\label{pryt1a}
\eea
where 
\bea
 {\cal A}&=&{\cal I}\cup {\cal O}\cup \{ r+s\} \nn
 w &\equiv & w_{r+s} \nn
 k_{r+s} &=& t
\eea
To see that the above formula produces the right singular behaviour 
in the limit
$z\rightarrow 1$ followed by $x\rightarrow 1$, we may scale all the 
$w_j$'s with $j\in{\cal O}$ as
$$w_j\rightarrow 1-(1-z)w_j, \ \ \ \ \ \ \mbox{for all  } j \in {\cal O}$$ 
and scale all $w_i$'s with $i\in{\cal I}$ as
$$w_i\rightarrow w_i/(w_i-1), \ \ \ \ \ \ \mbox{for all } i \in {\cal I}$$
and also
$$w\rightarrow 1-\frac{(1-z)}{u-x}u w $$
We also scale $u$ as
$$u \rightarrow x+(1-x)u $$
It should be noticed that the final $u$ contour starts at 1 and goes
along the unit circle such that it surrounds 0 and the other points
which are away from 0 by a distance of order $(1-z)/(1-x)$. This means that
we can not deform the $u$ contour to the form $\int_0^1du$, or in terms of the 
original $u$ variable, that cannot be deformed into $\int_x^1du$.
Using these scalings, we show that in the presence of at least one 
screening charge of the second kind in the scaling region (the region
close to $1$ and $z$) the singular behaviour is  
\ben
W^{(R,S)}_{(r,s)}(z,x)\sim (1-z)^{-h(j_2)-h(j_3)+h(j_I)}(x-1)^{j_2+j_3-j_I}
\een
where 
\ben
j_I=j_2+j_3-r+s t
\een
What happens if there is no screening charge of the second kind in the scaling
region? Then the above method does not apply, but in that case
$j_2+j_3-j_I$ is an integer, and 
\ben
W^{(R,S)}_{(r,0)}(z,x)\sim (1-z)^{-h(j_2)-h(j_3)+h(j_I)}(x-1)^{j_2+j_3-j_I}
\een
is a polynomial in $x$.
There will be no extra singularities present in $w$'s,
such as at $\delta = \frac{x-z}{x-1}$, if we integrate over $u$
first. Thus we could choose the following contours ($w=w_{r+s}=w_r\in
{\cal O}$)
\bea
 \int_z^1dw_j&,&j\in {\cal O}\nn
 \int_{-\infty}^0dw_i&,& i \in {\cal I}\nn
 &\int_0^1du&
\eea
These contours are effectively closed in the sense that a total derivative
integrated along them vanishes, such as is required for the 
Knizhnik-Zamolodchikov equations to be satisfied \cite{PRY}.
Notice, however, that these contours are not closed (in the same sense) 
when there is a screening
charge of the second kind in the scaling region. However,  
it is difficult to determine explicitly the $(1-x)$ behaviour for these 
contours, but since we know that our formula is both projective and $SL(2)$ 
invariant, we could express the above formula in terms of $x_3=0$ and
$x_1=1$, where the $(1-x)$ behaviour is manifest. 

It may seem 
surprising that one could not make use of the $j_1\leftrightarrow j_3$ symmetry
to obtain t-channel contours from s-channel ones and vice versa. The reason is 
that the simple form of the 4-point function we have given with only one 
auxiliary $u$ integration, breaks this symmetry, since not all 4 primary fields
are treated on the same footing. For a more symmetric treatment, more $u$ 
integrations have to be introduced, which is also inconvenient, however.

In the next 
section we start our detailed comparison between the 4-point functions written
down by us \cite{PRY} and by Andreev \cite{An}. For those latter ones, 
it turns out that simpler contours may be devised.

\section{Conformal blocks in Andreev's representation}

In this section we base our discussion on the integral realization of Andreev 
\cite{An}. In the next section we discuss 
the equivalence between that realization
and ours \cite{PRY}, described in the preceding section. In this section we 
show how to choose simple integration contours so that we produce both s-and 
t-channel blocks corresponding to both fusion rules I and II. It will turn out 
that the t-channel blocks are obtained in a very simple way from the s-channel 
blocks so we mostly concentrate on the latter. It is the specification of the 
integration contours which is our contribution here over Ref. \cite{An}. 
The advantage of the realization of \cite{An} is that contrary to the case
with ours, there is no auxiliary integration in addition to the integration 
over positions of screening charges. The disadvantage is that the 
representation (so far) has no underlying free field realization and therefore
only is known for 4-point blocks.
For our purpose later on it is convenient to have different names for s- and 
t-channel blocks. We denote them by letters ${\cal S}$ or $S$ and 
${\cal T}$ or $T$. The difference will be explained.

\subsection{Fusion Rule I}
\begin{figure}
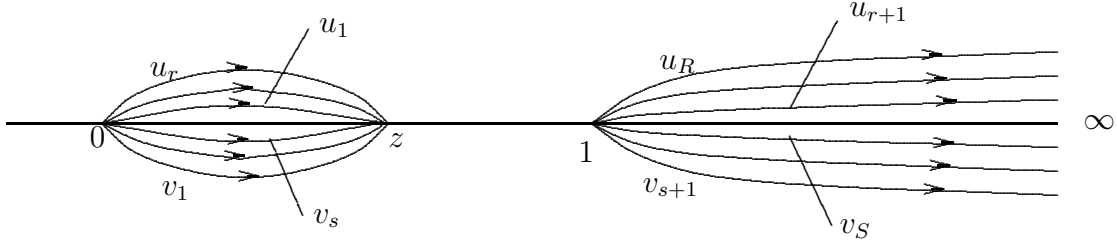

\font\thinlinefont=cmr5
\begingroup\makeatletter\ifx\SetFigFont\undefined
\def\x#1#2#3#4#5#6#7\relax{\def\x{#1#2#3#4#5#6}}%
\expandafter\x\fmtname xxxxxx\relax \def\y{splain}%
\ifx\x\y   
\gdef\SetFigFont#1#2#3{%
  \ifnum #1<17\tiny\else \ifnum #1<20\small\else
  \ifnum #1<24\normalsize\else \ifnum #1<29\large\else
  \ifnum #1<34\Large\else \ifnum #1<41\LARGE\else
     \huge\fi\fi\fi\fi\fi\fi
  \csname #3\endcsname}%
\else
\gdef\SetFigFont#1#2#3{\begingroup
  \count@#1\relax \ifnum 25<\count@\count@25\fi
  \def\x{\endgroup\@setsize\SetFigFont{#2pt}}%
  \expandafter\x
    \csname \romannumeral\the\count@ pt\expandafter\endcsname
    \csname @\romannumeral\the\count@ pt\endcsname
  \csname #3\endcsname}%
\fi
\fi\endgroup
\mbox{\beginpicture
\setcoordinatesystem units <1.00000cm,1.00000cm>
\unitlength=1.00000cm
\linethickness=1pt
\setplotsymbol ({\makebox(0,0)[l]{\tencirc\symbol{'160}}})
\setshadesymbol ({\thinlinefont .})
\setlinear
%
%
\linethickness= 0.500pt
\setplotsymbol ({\thinlinefont .})
\plot  4.714 21.861  5.285 22.845 /
%
%
\linethickness= 0.500pt
\setplotsymbol ({\thinlinefont .})
\plot  4.777 21.385  5.207 20.369 /
%
%
\linethickness= 0.500pt
\setplotsymbol ({\thinlinefont .})
\putrule from  4.477 20.892 to  4.619 20.892
%
%
\plot  4.365 20.828  4.619 20.892  4.365 20.955 /
%
%
%
\linethickness= 0.500pt
\setplotsymbol ({\thinlinefont .})
\putrule from  4.286 21.162 to  4.428 21.162
%
%
\plot  4.174 21.099  4.428 21.162  4.174 21.226 /
%
%
%
\linethickness= 0.500pt
\setplotsymbol ({\thinlinefont .})
\putrule from  4.333 21.385 to  4.475 21.385
%
%
\plot  4.221 21.321  4.475 21.385  4.221 21.448 /
%
%
%
\linethickness= 0.500pt
\setplotsymbol ({\thinlinefont .})
\putrule from  4.350 21.861 to  4.492 21.861
%
%
\plot  4.238 21.797  4.492 21.861  4.238 21.924 /
%
%
%
\linethickness= 0.500pt
\setplotsymbol ({\thinlinefont .})
\putrule from  4.396 22.066 to  4.538 22.066
%
%
\plot  4.284 22.003  4.538 22.066  4.284 22.130 /
%
%
%
\linethickness= 0.500pt
\setplotsymbol ({\thinlinefont .})
\putrule from  4.320 22.337 to  4.462 22.337
%
%
\plot  4.208 22.274  4.462 22.337  4.208 22.401 /
%
%
%
\linethickness= 0.500pt
\setplotsymbol ({\thinlinefont .})
\plot  9.207 19.050  9.207 19.050 /
%
%
\linethickness= 0.500pt
\setplotsymbol ({\thinlinefont .})
\putrule from 13.540 22.511 to 13.540 22.511
\putrule from 13.540 22.511 to 13.699 22.511
%
%
\plot 13.445 22.447 13.699 22.511 13.445 22.574 /
%
%
%
\linethickness= 0.500pt
\setplotsymbol ({\thinlinefont .})
\putrule from 13.604 22.193 to 13.604 22.193
\putrule from 13.604 22.193 to 13.763 22.193
%
%
\plot 13.509 22.130 13.763 22.193 13.509 22.257 /
%
%
%
\linethickness= 0.500pt
\setplotsymbol ({\thinlinefont .})
\putrule from 13.731 21.893 to 13.731 21.893
\putrule from 13.731 21.893 to 13.890 21.893
%
%
\plot 13.636 21.829 13.890 21.893 13.636 21.956 /
%
%
%
\linethickness= 0.500pt
\setplotsymbol ({\thinlinefont .})
\putrule from 13.699 21.336 to 13.699 21.336
\putrule from 13.699 21.336 to 13.858 21.336
%
%
\plot 13.604 21.273 13.858 21.336 13.604 21.400 /
%
%
%
\linethickness= 0.500pt
\setplotsymbol ({\thinlinefont .})
\putrule from 13.650 21.035 to 13.650 21.035
\putrule from 13.650 21.035 to 13.809 21.035
%
%
\plot 13.555 20.972 13.809 21.035 13.555 21.099 /
%
%
%
\linethickness= 0.500pt
\setplotsymbol ({\thinlinefont .})
\putrule from 13.572 20.718 to 13.572 20.718
\putrule from 13.572 20.718 to 13.731 20.718
%
%
\plot 13.477 20.654 13.731 20.718 13.477 20.781 /
%
%
%
\linethickness= 0.500pt
\setplotsymbol ({\thinlinefont .})
\plot 11.699 21.812 12.317 22.955 /
%
%
\linethickness= 0.500pt
\setplotsymbol ({\thinlinefont .})
\plot 11.682 21.416 12.239 20.242 /
\linethickness= 0.500pt
\setplotsymbol ({\thinlinefont .})
%
%
%
\plot	 2.540 21.590  3.413 21.749
 	 3.523 21.767
	 3.634 21.783
	 3.746 21.797
	 3.860 21.808
	 3.974 21.817
	 4.090 21.823
	 4.207 21.827
	 4.326 21.828
	 4.446 21.827
	 4.567 21.823
	 4.689 21.817
	 4.812 21.808
	 4.937 21.797
	 5.063 21.783
	 5.126 21.776
	 5.190 21.767
	 5.254 21.758
	 5.318 21.749
	 /
\plot  5.318 21.749  6.350 21.590 /
\linethickness= 0.500pt
\setplotsymbol ({\thinlinefont .})
%
%
%
\plot	 2.540 21.590  3.493 21.431
 	 3.611 21.413
	 3.729 21.397
	 3.847 21.383
	 3.964 21.372
	 4.080 21.363
	 4.196 21.357
	 4.311 21.353
	 4.425 21.352
	 4.539 21.353
	 4.652 21.357
	 4.765 21.363
	 4.877 21.372
	 4.988 21.383
	 5.099 21.397
	 5.209 21.413
	 5.318 21.431
	 /
\plot  5.318 21.431  6.191 21.590 /
\linethickness= 0.500pt
\setplotsymbol ({\thinlinefont .})
%
%
%
\plot	 2.540 21.590  2.857 21.749
 	 2.942 21.787
	 3.036 21.823
	 3.140 21.857
	 3.254 21.888
	 3.378 21.916
	 3.444 21.930
	 3.512 21.942
	 3.583 21.954
	 3.656 21.966
	 3.732 21.977
	 3.810 21.987
	 3.887 21.996
	 3.958 22.005
	 4.083 22.022
	 4.186 22.035
	 4.266 22.046
	 4.366 22.066
	 4.485 22.046
	 4.576 22.035
	 4.693 22.022
	 4.761 22.014
	 4.834 22.005
	 4.914 21.996
	 5.001 21.987
	 5.088 21.977
	 5.173 21.966
	 5.255 21.954
	 5.333 21.942
	 5.408 21.930
	 5.481 21.916
	 5.550 21.902
	 5.616 21.888
	 5.739 21.857
	 5.849 21.823
	 5.947 21.787
	 6.032 21.749
	 6.106 21.712
	 6.166 21.679
	 6.251 21.630
	 6.271 21.590
	 /
\plot  6.271 21.590  6.191 21.590 /
%
%
\linethickness= 0.500pt
\setplotsymbol ({\thinlinefont .})
\putrule from  1.270 21.590 to 15.240 21.590
\linethickness= 0.500pt
\setplotsymbol ({\thinlinefont .})
%
%
%
\plot	 2.540 21.590  2.857 21.431
 	 2.942 21.393
	 3.036 21.357
	 3.140 21.323
	 3.254 21.292
	 3.378 21.264
	 3.444 21.250
	 3.512 21.238
	 3.583 21.226
	 3.656 21.214
	 3.732 21.203
	 3.810 21.193
	 3.887 21.184
	 3.958 21.175
	 4.083 21.158
	 4.186 21.145
	 4.266 21.134
	 4.366 21.114
	 4.485 21.134
	 4.576 21.145
	 4.693 21.158
	 4.761 21.166
	 4.834 21.175
	 4.914 21.184
	 5.001 21.193
	 5.088 21.203
	 5.173 21.214
	 5.255 21.226
	 5.333 21.238
	 5.408 21.250
	 5.481 21.264
	 5.550 21.278
	 5.616 21.292
	 5.739 21.323
	 5.849 21.357
	 5.947 21.393
	 6.032 21.431
	 6.106 21.468
	 6.166 21.501
	 6.251 21.550
	 6.271 21.590
	 /
\plot  6.271 21.590  6.191 21.590 /
%
%
\put{$v_S$} [lB] at 12.349 20.130
\linethickness= 0.500pt
\setplotsymbol ({\thinlinefont .})
%
%
%
\plot	 2.540 21.590  2.778 21.828
 	 2.845 21.886
	 2.927 21.942
	 3.024 21.996
	 3.135 22.046
	 3.262 22.095
	 3.331 22.118
	 3.403 22.141
	 3.479 22.163
	 3.559 22.184
	 3.643 22.205
	 3.731 22.225
	 3.820 22.244
	 3.909 22.260
	 3.999 22.273
	 4.088 22.285
	 4.177 22.293
	 4.266 22.299
	 4.356 22.303
	 4.445 22.304
	 4.534 22.303
	 4.624 22.299
	 4.713 22.293
	 4.802 22.285
	 4.891 22.273
	 4.981 22.260
	 5.070 22.244
	 5.159 22.225
	 5.247 22.205
	 5.331 22.184
	 5.411 22.163
	 5.487 22.141
	 5.559 22.118
	 5.628 22.095
	 5.755 22.046
	 5.866 21.996
	 5.963 21.942
	 6.045 21.886
	 6.112 21.828
	 /
\plot  6.112 21.828  6.350 21.590 /
\linethickness= 0.500pt
\setplotsymbol ({\thinlinefont .})
%
%
%
\plot	 2.540 21.590  2.778 21.352
 	 2.845 21.294
	 2.927 21.238
	 3.024 21.184
	 3.135 21.134
	 3.262 21.085
	 3.331 21.062
	 3.403 21.039
	 3.479 21.017
	 3.559 20.996
	 3.643 20.975
	 3.731 20.955
	 3.820 20.936
	 3.909 20.920
	 3.999 20.907
	 4.088 20.895
	 4.177 20.887
	 4.266 20.881
	 4.356 20.877
	 4.445 20.876
	 4.534 20.877
	 4.624 20.881
	 4.713 20.887
	 4.802 20.895
	 4.891 20.907
	 4.981 20.920
	 5.070 20.936
	 5.159 20.955
	 5.247 20.975
	 5.331 20.996
	 5.411 21.017
	 5.487 21.039
	 5.559 21.062
	 5.628 21.085
	 5.755 21.134
	 5.866 21.184
	 5.963 21.238
	 6.045 21.294
	 6.112 21.352
	 /
\plot  6.112 21.352  6.350 21.590 /
\linethickness= 0.500pt
\setplotsymbol ({\thinlinefont .})
%
%
%
\plot	 9.049 21.590  9.446 21.669
 	 9.504 21.679
	 9.581 21.689
	 9.675 21.699
	 9.788 21.709
	 9.851 21.714
	 9.918 21.719
	 9.990 21.724
	10.067 21.729
	10.148 21.734
	10.233 21.739
	10.323 21.744
	10.418 21.749
	10.517 21.754
	10.620 21.759
	10.728 21.764
	10.841 21.769
	10.958 21.774
	11.079 21.779
	11.205 21.783
	11.270 21.786
	11.336 21.788
	11.403 21.791
	11.471 21.793
	11.540 21.796
	11.610 21.798
	11.682 21.801
	11.754 21.803
	11.828 21.806
	11.903 21.808
	11.978 21.811
	12.055 21.813
	12.134 21.816
	12.213 21.818
	12.293 21.821
	12.375 21.823
	12.457 21.826
	12.541 21.828
	 /
\plot 12.541 21.828 15.240 21.907 /
\linethickness= 0.500pt
\setplotsymbol ({\thinlinefont .})
%
%
%
\plot	 9.049 21.590  9.446 21.511
 	 9.504 21.501
	 9.581 21.491
	 9.675 21.481
	 9.788 21.471
	 9.851 21.466
	 9.918 21.461
	 9.990 21.456
	10.067 21.451
	10.148 21.446
	10.233 21.441
	10.323 21.436
	10.418 21.431
	10.517 21.426
	10.620 21.421
	10.728 21.416
	10.841 21.411
	10.958 21.406
	11.079 21.401
	11.205 21.397
	11.270 21.394
	11.336 21.392
	11.403 21.389
	11.471 21.387
	11.540 21.384
	11.610 21.382
	11.682 21.379
	11.754 21.377
	11.828 21.374
	11.903 21.372
	11.978 21.369
	12.055 21.367
	12.134 21.364
	12.213 21.362
	12.293 21.359
	12.375 21.357
	12.457 21.354
	12.541 21.352
	 /
\plot 12.541 21.352 15.240 21.273 /
\linethickness= 0.500pt
\setplotsymbol ({\thinlinefont .})
%
%
%
\plot	 9.049 21.590  9.366 21.749
 	 9.451 21.787
	 9.545 21.823
	 9.649 21.857
	 9.763 21.888
	 9.887 21.916
	 9.953 21.930
	10.021 21.942
	10.092 21.954
	10.165 21.966
	10.241 21.977
	10.319 21.987
	10.404 21.997
	10.501 22.007
	10.610 22.017
	10.731 22.027
	10.795 22.032
	10.863 22.036
	10.934 22.041
	11.007 22.046
	11.084 22.051
	11.163 22.056
	11.245 22.061
	11.331 22.066
	11.419 22.071
	11.510 22.076
	11.604 22.081
	11.702 22.086
	11.802 22.091
	11.905 22.096
	12.011 22.101
	12.120 22.106
	12.231 22.111
	12.346 22.116
	12.464 22.121
	12.585 22.126
	12.708 22.131
	12.835 22.136
	12.899 22.138
	12.964 22.141
	13.030 22.143
	13.097 22.146
	 /
\plot 13.097 22.146 15.240 22.225 /
\linethickness= 0.500pt
\setplotsymbol ({\thinlinefont .})
%
%
%
\plot	 9.049 21.590  9.366 21.431
 	 9.451 21.393
	 9.545 21.357
	 9.649 21.323
	 9.763 21.292
	 9.887 21.264
	 9.953 21.250
	10.021 21.238
	10.092 21.226
	10.165 21.214
	10.241 21.203
	10.319 21.193
	10.404 21.183
	10.501 21.173
	10.610 21.163
	10.731 21.153
	10.795 21.148
	10.863 21.144
	10.934 21.139
	11.007 21.134
	11.084 21.129
	11.163 21.124
	11.245 21.119
	11.331 21.114
	11.419 21.109
	11.510 21.104
	11.604 21.099
	11.702 21.094
	11.802 21.089
	11.905 21.084
	12.011 21.079
	12.120 21.074
	12.231 21.069
	12.346 21.064
	12.464 21.059
	12.585 21.054
	12.708 21.049
	12.835 21.044
	12.899 21.042
	12.964 21.039
	13.030 21.037
	13.097 21.034
	 /
\plot 13.097 21.034 15.240 20.955 /
\linethickness= 0.500pt
\setplotsymbol ({\thinlinefont .})
%
%
%
\plot	 9.049 21.590  9.366 21.828
 	 9.451 21.886
	 9.545 21.942
	 9.649 21.996
	 9.763 22.046
	 9.887 22.095
	 9.953 22.118
	10.021 22.141
	10.092 22.163
	10.165 22.184
	10.241 22.205
	10.319 22.225
	10.404 22.245
	10.501 22.263
	10.610 22.282
	10.731 22.299
	10.795 22.308
	10.863 22.316
	10.934 22.325
	11.007 22.333
	11.084 22.341
	11.163 22.349
	11.245 22.356
	11.331 22.364
	11.419 22.371
	11.510 22.378
	11.604 22.386
	11.702 22.392
	11.802 22.399
	11.905 22.406
	12.011 22.412
	12.120 22.418
	12.231 22.425
	12.346 22.431
	12.464 22.436
	12.585 22.442
	12.708 22.448
	12.835 22.453
	12.899 22.456
	12.964 22.458
	13.030 22.461
	13.097 22.463
	 /
\plot 13.097 22.463 15.240 22.543 /
\linethickness= 0.500pt
\setplotsymbol ({\thinlinefont .})
%
%
%
\plot	 9.049 21.590  9.366 21.352
 	 9.451 21.294
	 9.545 21.238
	 9.649 21.184
	 9.763 21.134
	 9.887 21.085
	 9.953 21.062
	10.021 21.039
	10.092 21.017
	10.165 20.996
	10.241 20.975
	10.319 20.955
	10.404 20.935
	10.501 20.917
	10.610 20.898
	10.731 20.881
	10.795 20.872
	10.863 20.864
	10.934 20.855
	11.007 20.847
	11.084 20.839
	11.163 20.831
	11.245 20.824
	11.331 20.816
	11.419 20.809
	11.510 20.802
	11.604 20.794
	11.702 20.788
	11.802 20.781
	11.905 20.774
	12.011 20.768
	12.120 20.762
	12.231 20.755
	12.346 20.749
	12.464 20.744
	12.585 20.738
	12.708 20.732
	12.835 20.727
	12.899 20.724
	12.964 20.722
	13.030 20.719
	13.097 20.717
	 /
\plot 13.097 20.717 15.240 20.637 /
%
%
\put{$0$} [lB] at  2.381 21.273
%
%
\put{$z$} [lB] at  6.350 21.273
%
%
\put{$v_1$} [lB] at  3.334 20.637
%
%
\put{$u_r$} [lB] at  3.175 22.225
%
%
\put{$u_1$} [lB] at  5.429 22.813
%
%
\put{$v_s$} [lB] at  5.349 20.273
%
%
\put{$1$} [lB] at  8.873 21.082
%
%
\put{$\infty$} [lB] at 15.587 21.526
%
%
\put{$u_R$} [lB] at  9.953 22.320
%
%
\put{$u_{r+1}$} [lB] at 12.476 23.019
%
%
\put{$v_{s+1}$} [lB] at  9.730 20.686
\linethickness=0pt
\putrectangle corners at  1.245 23.247 and 15.587 19.025
\endpicture}

\caption{Integration contours for $u$ and $v$ variables for an s-channel block 
for fusion rule I.}
\label{figanI}
\end{figure}

We define the complex block in the s-channel for fusion rule I with $r$ 
screenings of the first kind and $s$ screenings of the second kind at the
right vertex as follows:
\bea
 {\cal S}^{(R,S)}_{(r,s,0)}(z,x)&\equiv&z^{2j_1j_2/t}(1-z)^{2j_2j_3/t}
  \int_0^z\prod_{i\in I_1,k\in J_1}du_idv_k
  \int_1^\infty\prod_{j\in I_2,l\in J_2}du_jdv_l\nn
 &&u_i^{a'}(1-u_i)^{b'}(z-u_i)^{c'}\prod_{i<i',\in I_1}(u_i-u_{i'})^{2\rho'}
  u_j^{a'}(u_j-1)^{b'}(u_j-z)^{c'}\nn
 &&\prod_{j<j',\in I_2}(u_j-u_{j'})^{2\rho'} 
  \prod_{i\in I_1,j\in I_2}(u_j-u_i)^{2\rho'}\nn
 &&v_k^a(1-v_k)^b(z-v_k)^c\prod_{k<k', \in J_1}(v_k-v_{k'})^{2\rho}
  v_l^a(v_l-1)^b(v_l-z)^c\nn
 &&\prod_{l<l',\in J_2}(v_l-v_{l'})^{2\rho}
  \prod_{k\in J_1,l\in J_2}(v_l-v_k)^{2\rho}\nn
 &&\prod_{i,k}(u_i-v_k)^{-2}
  \prod_{i,l}(u_i-v_l)^{-2}
  \prod_{j,k}(u_j-v_k)^{-2}\prod_{j,l}(u_j-v_l)^{-2}\nn
 && \prod_{i,j,k,l}(u_i-x)(u_j-x)(v_k-x)^{-\rho}(v_l-x)^{-\rho}
\label{anblockI}
\eea
Here we have introduced the following index sets
\bea
I_1&=&\{1,...,r\}\nn
I_2&=&\{r+1,...,R\}\nn
J_1&=&\{1,...,s\}\nn
J_2&=&\{s+1,...,S\}
\eea
where $R$ and $S$ are the total numbers of screenings of the first and second 
kinds respectively. Variables $u$ and $v$ belong to screenings of the first and
second kind respectively, although this language is rather symbolic, since as 
yet there exists 
no known free field realization which directly gives this form.
Also the integrals are taken along complex Dotsenko-Fateev 
contours shown in Fig. \ref{figanI}. Notice that expressions of the form
$(u_i-u_{i'})^{2\rho'}$ have a phase defined by the fact that the first of the
two integration variables have a lower imaginary part than the last variable.
Finally
\bea
a&=&-2j_3+t+R-St-1\nn
b&=&-2j_1+t+R-St-1\nn
c&=&2j_1+2j_2+2j_3-R+St+1\nn
\rho&=&t, \ \ \rho'=1/t\nn
a'&=&-a/t,\ \ \ b'=-b/t,\ \ \ c'=-c/t
\label{abc}
\eea
The integrand of this expression is provided in a slightly different form in 
Ref. \cite{An}. 
In fact there, the $j$'s are replaced by their parametrizations 
\Eq{jpmrs} giving rise to 4 independent forms for the integrand depending on 
whether the $j_i^+$ or the $j_i^-$ form is used. The above form holds in 
general. By analysing the small $z$ and small $x$ behaviour of this form it
is easy to establish that this conformal block corresponds to the s-channel 
diagram Fig. \ref{fig1} 
with the intermediate $j$ given by fusion rule I. Indeed
by scaling $u_i\rightarrow zu_i,v_k\rightarrow zv_k, i\in I_1,k\in J_1$ we find
\ben
{\cal S}^{(R,S)}_{(r,s,0)}(z,x)\sim z^{-h(j_1)-h(j_2)+h(j_I)}(-x)^{j_1+j_2-j_I}
\een
with
\ben
j_I=j_1+j_2-r+st
\een

The contours in Fig. \ref{figanI} are essentially equal to the contours in Ref.
\cite{DF} for minimal models.

\subsection{Fusion rule II}
\begin{figure}
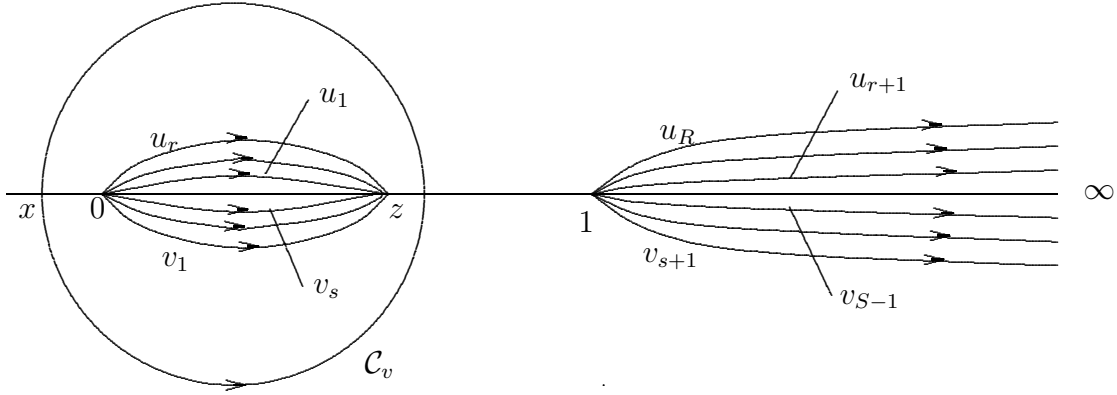

\font\thinlinefont=cmr5
\begingroup\makeatletter\ifx\SetFigFont\undefined
\def\x#1#2#3#4#5#6#7\relax{\def\x{#1#2#3#4#5#6}}%
\expandafter\x\fmtname xxxxxx\relax \def\y{splain}%
\ifx\x\y   
\gdef\SetFigFont#1#2#3{%
  \ifnum #1<17\tiny\else \ifnum #1<20\small\else
  \ifnum #1<24\normalsize\else \ifnum #1<29\large\else
  \ifnum #1<34\Large\else \ifnum #1<41\LARGE\else
     \huge\fi\fi\fi\fi\fi\fi
  \csname #3\endcsname}%
\else
\gdef\SetFigFont#1#2#3{\begingroup
  \count@#1\relax \ifnum 25<\count@\count@25\fi
  \def\x{\endgroup\@setsize\SetFigFont{#2pt}}%
  \expandafter\x
    \csname \romannumeral\the\count@ pt\expandafter\endcsname
    \csname @\romannumeral\the\count@ pt\endcsname
  \csname #3\endcsname}%
\fi
\fi\endgroup
\mbox{\beginpicture
\setcoordinatesystem units <1.00000cm,1.00000cm>
\unitlength=1.00000cm
\linethickness=1pt
\setplotsymbol ({\makebox(0,0)[l]{\tencirc\symbol{'160}}})
\setshadesymbol ({\thinlinefont .})
\setlinear
%
%
\linethickness= 0.500pt
\setplotsymbol ({\thinlinefont .})
\putrule from  1.270 21.590 to 15.240 21.590
%
%
\linethickness= 0.500pt
\setplotsymbol ({\thinlinefont .})
\plot  4.714 21.861  5.285 22.845 /
%
%
\linethickness= 0.500pt
\setplotsymbol ({\thinlinefont .})
\plot  4.777 21.385  5.207 20.369 /
%
%
\linethickness= 0.500pt
\setplotsymbol ({\thinlinefont .})
\putrule from  4.477 20.892 to  4.619 20.892
%
%
\plot  4.365 20.828  4.619 20.892  4.365 20.955 /
%
%
%
\linethickness= 0.500pt
\setplotsymbol ({\thinlinefont .})
\putrule from  4.286 21.162 to  4.428 21.162
%
%
\plot  4.174 21.099  4.428 21.162  4.174 21.226 /
%
%
%
\linethickness= 0.500pt
\setplotsymbol ({\thinlinefont .})
\putrule from  4.333 21.385 to  4.475 21.385
%
%
\plot  4.221 21.321  4.475 21.385  4.221 21.448 /
%
%
%
\linethickness= 0.500pt
\setplotsymbol ({\thinlinefont .})
\putrule from  4.350 21.861 to  4.492 21.861
%
%
\plot  4.238 21.797  4.492 21.861  4.238 21.924 /
%
%
%
\linethickness= 0.500pt
\setplotsymbol ({\thinlinefont .})
\putrule from  4.396 22.066 to  4.538 22.066
%
%
\plot  4.284 22.003  4.538 22.066  4.284 22.130 /
%
%
%
\linethickness= 0.500pt
\setplotsymbol ({\thinlinefont .})
\putrule from  4.320 22.337 to  4.462 22.337
%
%
\plot  4.208 22.274  4.462 22.337  4.208 22.401 /
%
%
%
\linethickness= 0.500pt
\setplotsymbol ({\thinlinefont .})
\plot  9.207 19.050  9.207 19.050 /
%
%
\linethickness= 0.500pt
\setplotsymbol ({\thinlinefont .})
\putrule from 13.540 22.511 to 13.540 22.511
\putrule from 13.540 22.511 to 13.699 22.511
%
%
\plot 13.445 22.447 13.699 22.511 13.445 22.574 /
%
%
%
\linethickness= 0.500pt
\setplotsymbol ({\thinlinefont .})
\putrule from 13.604 22.193 to 13.604 22.193
\putrule from 13.604 22.193 to 13.763 22.193
%
%
\plot 13.509 22.130 13.763 22.193 13.509 22.257 /
%
%
%
\linethickness= 0.500pt
\setplotsymbol ({\thinlinefont .})
\putrule from 13.731 21.893 to 13.731 21.893
\putrule from 13.731 21.893 to 13.890 21.893
%
%
\plot 13.636 21.829 13.890 21.893 13.636 21.956 /
%
%
%
\linethickness= 0.500pt
\setplotsymbol ({\thinlinefont .})
\putrule from 13.699 21.336 to 13.699 21.336
\putrule from 13.699 21.336 to 13.858 21.336
%
%
\plot 13.604 21.273 13.858 21.336 13.604 21.400 /
%
%
%
\linethickness= 0.500pt
\setplotsymbol ({\thinlinefont .})
\putrule from 13.650 21.035 to 13.650 21.035
\putrule from 13.650 21.035 to 13.809 21.035
%
%
\plot 13.555 20.972 13.809 21.035 13.555 21.099 /
%
%
%
\linethickness= 0.500pt
\setplotsymbol ({\thinlinefont .})
\putrule from 13.572 20.718 to 13.572 20.718
\putrule from 13.572 20.718 to 13.731 20.718
%
%
\plot 13.477 20.654 13.731 20.718 13.477 20.781 /
%
%
%
\linethickness= 0.500pt
\setplotsymbol ({\thinlinefont .})
\plot 11.699 21.812 12.317 22.955 /
%
%
\linethickness= 0.500pt
\setplotsymbol ({\thinlinefont .})
\plot 11.682 21.416 12.239 20.242 /
%
%
\linethickness= 0.500pt
\setplotsymbol ({\thinlinefont .})
\putrule from  4.286 19.050 to  4.445 19.050
%
%
\plot  4.191 18.986  4.445 19.050  4.191 19.114 /
\linethickness= 0.500pt
\setplotsymbol ({\thinlinefont .})
%
%
%
\plot	 2.540 21.590  3.413 21.749
 	 3.523 21.767
	 3.634 21.783
	 3.746 21.797
	 3.860 21.808
	 3.974 21.817
	 4.090 21.823
	 4.207 21.827
	 4.326 21.828
	 4.446 21.827
	 4.567 21.823
	 4.689 21.817
	 4.812 21.808
	 4.937 21.797
	 5.063 21.783
	 5.126 21.776
	 5.190 21.767
	 5.254 21.758
	 5.318 21.749
	 /
\plot  5.318 21.749  6.350 21.590 /
\linethickness= 0.500pt
\setplotsymbol ({\thinlinefont .})
%
%
%
\plot	 2.540 21.590  3.493 21.431
 	 3.611 21.413
	 3.729 21.397
	 3.847 21.383
	 3.964 21.372
	 4.080 21.363
	 4.196 21.357
	 4.311 21.353
	 4.425 21.352
	 4.539 21.353
	 4.652 21.357
	 4.765 21.363
	 4.877 21.372
	 4.988 21.383
	 5.099 21.397
	 5.209 21.413
	 5.318 21.431
	 /
\plot  5.318 21.431  6.191 21.590 /
\linethickness= 0.500pt
\setplotsymbol ({\thinlinefont .})
%
%
%
\plot	 2.540 21.590  2.857 21.749
 	 2.942 21.787
	 3.036 21.823
	 3.140 21.857
	 3.254 21.888
	 3.378 21.916
	 3.444 21.930
	 3.512 21.942
	 3.583 21.954
	 3.656 21.966
	 3.732 21.977
	 3.810 21.987
	 3.887 21.996
	 3.958 22.005
	 4.083 22.022
	 4.186 22.035
	 4.266 22.046
	 4.366 22.066
	 4.485 22.046
	 4.576 22.035
	 4.693 22.022
	 4.761 22.014
	 4.834 22.005
	 4.914 21.996
	 5.001 21.987
	 5.088 21.977
	 5.173 21.966
	 5.255 21.954
	 5.333 21.942
	 5.408 21.930
	 5.481 21.916
	 5.550 21.902
	 5.616 21.888
	 5.739 21.857
	 5.849 21.823
	 5.947 21.787
	 6.032 21.749
	 6.106 21.712
	 6.166 21.679
	 6.251 21.630
	 6.271 21.590
	 /
\plot  6.271 21.590  6.191 21.590 /
%
%
\linethickness= 0.500pt
\setplotsymbol ({\thinlinefont .})
\ellipticalarc axes ratio  2.540:2.540  360 degrees 
	from  6.826 21.590 center at  4.286 21.590
\linethickness= 0.500pt
\setplotsymbol ({\thinlinefont .})
%
%
%
\plot	 2.540 21.590  2.857 21.431
 	 2.942 21.393
	 3.036 21.357
	 3.140 21.323
	 3.254 21.292
	 3.378 21.264
	 3.444 21.250
	 3.512 21.238
	 3.583 21.226
	 3.656 21.214
	 3.732 21.203
	 3.810 21.193
	 3.887 21.184
	 3.958 21.175
	 4.083 21.158
	 4.186 21.145
	 4.266 21.134
	 4.366 21.114
	 4.485 21.134
	 4.576 21.145
	 4.693 21.158
	 4.761 21.166
	 4.834 21.175
	 4.914 21.184
	 5.001 21.193
	 5.088 21.203
	 5.173 21.214
	 5.255 21.226
	 5.333 21.238
	 5.408 21.250
	 5.481 21.264
	 5.550 21.278
	 5.616 21.292
	 5.739 21.323
	 5.849 21.357
	 5.947 21.393
	 6.032 21.431
	 6.106 21.468
	 6.166 21.501
	 6.251 21.550
	 6.271 21.590
	 /
\plot  6.271 21.590  6.191 21.590 /
%
%
\put{${\cal C}_v$} [lB] at  6.032 19.209
\linethickness= 0.500pt
\setplotsymbol ({\thinlinefont .})
%
%
%
\plot	 2.540 21.590  2.778 21.828
 	 2.845 21.886
	 2.927 21.942
	 3.024 21.996
	 3.135 22.046
	 3.262 22.095
	 3.331 22.118
	 3.403 22.141
	 3.479 22.163
	 3.559 22.184
	 3.643 22.205
	 3.731 22.225
	 3.820 22.244
	 3.909 22.260
	 3.999 22.273
	 4.088 22.285
	 4.177 22.293
	 4.266 22.299
	 4.356 22.303
	 4.445 22.304
	 4.534 22.303
	 4.624 22.299
	 4.713 22.293
	 4.802 22.285
	 4.891 22.273
	 4.981 22.260
	 5.070 22.244
	 5.159 22.225
	 5.247 22.205
	 5.331 22.184
	 5.411 22.163
	 5.487 22.141
	 5.559 22.118
	 5.628 22.095
	 5.755 22.046
	 5.866 21.996
	 5.963 21.942
	 6.045 21.886
	 6.112 21.828
	 /
\plot  6.112 21.828  6.350 21.590 /
\linethickness= 0.500pt
\setplotsymbol ({\thinlinefont .})
%
%
%
\plot	 2.540 21.590  2.778 21.352
 	 2.845 21.294
	 2.927 21.238
	 3.024 21.184
	 3.135 21.134
	 3.262 21.085
	 3.331 21.062
	 3.403 21.039
	 3.479 21.017
	 3.559 20.996
	 3.643 20.975
	 3.731 20.955
	 3.820 20.936
	 3.909 20.920
	 3.999 20.907
	 4.088 20.895
	 4.177 20.887
	 4.266 20.881
	 4.356 20.877
	 4.445 20.876
	 4.534 20.877
	 4.624 20.881
	 4.713 20.887
	 4.802 20.895
	 4.891 20.907
	 4.981 20.920
	 5.070 20.936
	 5.159 20.955
	 5.247 20.975
	 5.331 20.996
	 5.411 21.017
	 5.487 21.039
	 5.559 21.062
	 5.628 21.085
	 5.755 21.134
	 5.866 21.184
	 5.963 21.238
	 6.045 21.294
	 6.112 21.352
	 /
\plot  6.112 21.352  6.350 21.590 /
\linethickness= 0.500pt
\setplotsymbol ({\thinlinefont .})
%
%
%
\plot	 9.049 21.590  9.446 21.669
 	 9.504 21.679
	 9.581 21.689
	 9.675 21.699
	 9.788 21.709
	 9.851 21.714
	 9.918 21.719
	 9.990 21.724
	10.067 21.729
	10.148 21.734
	10.233 21.739
	10.323 21.744
	10.418 21.749
	10.517 21.754
	10.620 21.759
	10.728 21.764
	10.841 21.769
	10.958 21.774
	11.079 21.779
	11.205 21.783
	11.270 21.786
	11.336 21.788
	11.403 21.791
	11.471 21.793
	11.540 21.796
	11.610 21.798
	11.682 21.801
	11.754 21.803
	11.828 21.806
	11.903 21.808
	11.978 21.811
	12.055 21.813
	12.134 21.816
	12.213 21.818
	12.293 21.821
	12.375 21.823
	12.457 21.826
	12.541 21.828
	 /
\plot 12.541 21.828 15.240 21.907 /
\linethickness= 0.500pt
\setplotsymbol ({\thinlinefont .})
%
%
%
\plot	 9.049 21.590  9.446 21.511
 	 9.504 21.501
	 9.581 21.491
	 9.675 21.481
	 9.788 21.471
	 9.851 21.466
	 9.918 21.461
	 9.990 21.456
	10.067 21.451
	10.148 21.446
	10.233 21.441
	10.323 21.436
	10.418 21.431
	10.517 21.426
	10.620 21.421
	10.728 21.416
	10.841 21.411
	10.958 21.406
	11.079 21.401
	11.205 21.397
	11.270 21.394
	11.336 21.392
	11.403 21.389
	11.471 21.387
	11.540 21.384
	11.610 21.382
	11.682 21.379
	11.754 21.377
	11.828 21.374
	11.903 21.372
	11.978 21.369
	12.055 21.367
	12.134 21.364
	12.213 21.362
	12.293 21.359
	12.375 21.357
	12.457 21.354
	12.541 21.352
	 /
\plot 12.541 21.352 15.240 21.273 /
\linethickness= 0.500pt
\setplotsymbol ({\thinlinefont .})
%
%
%
\plot	 9.049 21.590  9.366 21.749
 	 9.451 21.787
	 9.545 21.823
	 9.649 21.857
	 9.763 21.888
	 9.887 21.916
	 9.953 21.930
	10.021 21.942
	10.092 21.954
	10.165 21.966
	10.241 21.977
	10.319 21.987
	10.404 21.997
	10.501 22.007
	10.610 22.017
	10.731 22.027
	10.795 22.032
	10.863 22.036
	10.934 22.041
	11.007 22.046
	11.084 22.051
	11.163 22.056
	11.245 22.061
	11.331 22.066
	11.419 22.071
	11.510 22.076
	11.604 22.081
	11.702 22.086
	11.802 22.091
	11.905 22.096
	12.011 22.101
	12.120 22.106
	12.231 22.111
	12.346 22.116
	12.464 22.121
	12.585 22.126
	12.708 22.131
	12.835 22.136
	12.899 22.138
	12.964 22.141
	13.030 22.143
	13.097 22.146
	 /
\plot 13.097 22.146 15.240 22.225 /
\linethickness= 0.500pt
\setplotsymbol ({\thinlinefont .})
%
%
%
\plot	 9.049 21.590  9.366 21.431
 	 9.451 21.393
	 9.545 21.357
	 9.649 21.323
	 9.763 21.292
	 9.887 21.264
	 9.953 21.250
	10.021 21.238
	10.092 21.226
	10.165 21.214
	10.241 21.203
	10.319 21.193
	10.404 21.183
	10.501 21.173
	10.610 21.163
	10.731 21.153
	10.795 21.148
	10.863 21.144
	10.934 21.139
	11.007 21.134
	11.084 21.129
	11.163 21.124
	11.245 21.119
	11.331 21.114
	11.419 21.109
	11.510 21.104
	11.604 21.099
	11.702 21.094
	11.802 21.089
	11.905 21.084
	12.011 21.079
	12.120 21.074
	12.231 21.069
	12.346 21.064
	12.464 21.059
	12.585 21.054
	12.708 21.049
	12.835 21.044
	12.899 21.042
	12.964 21.039
	13.030 21.037
	13.097 21.034
	 /
\plot 13.097 21.034 15.240 20.955 /
\linethickness= 0.500pt
\setplotsymbol ({\thinlinefont .})
%
%
%
\plot	 9.049 21.590  9.366 21.828
 	 9.451 21.886
	 9.545 21.942
	 9.649 21.996
	 9.763 22.046
	 9.887 22.095
	 9.953 22.118
	10.021 22.141
	10.092 22.163
	10.165 22.184
	10.241 22.205
	10.319 22.225
	10.404 22.245
	10.501 22.263
	10.610 22.282
	10.731 22.299
	10.795 22.308
	10.863 22.316
	10.934 22.325
	11.007 22.333
	11.084 22.341
	11.163 22.349
	11.245 22.356
	11.331 22.364
	11.419 22.371
	11.510 22.378
	11.604 22.386
	11.702 22.392
	11.802 22.399
	11.905 22.406
	12.011 22.412
	12.120 22.418
	12.231 22.425
	12.346 22.431
	12.464 22.436
	12.585 22.442
	12.708 22.448
	12.835 22.453
	12.899 22.456
	12.964 22.458
	13.030 22.461
	13.097 22.463
	 /
\plot 13.097 22.463 15.240 22.543 /
\linethickness= 0.500pt
\setplotsymbol ({\thinlinefont .})
%
%
%
\plot	 9.049 21.590  9.366 21.352
 	 9.451 21.294
	 9.545 21.238
	 9.649 21.184
	 9.763 21.134
	 9.887 21.085
	 9.953 21.062
	10.021 21.039
	10.092 21.017
	10.165 20.996
	10.241 20.975
	10.319 20.955
	10.404 20.935
	10.501 20.917
	10.610 20.898
	10.731 20.881
	10.795 20.872
	10.863 20.864
	10.934 20.855
	11.007 20.847
	11.084 20.839
	11.163 20.831
	11.245 20.824
	11.331 20.816
	11.419 20.809
	11.510 20.802
	11.604 20.794
	11.702 20.788
	11.802 20.781
	11.905 20.774
	12.011 20.768
	12.120 20.762
	12.231 20.755
	12.346 20.749
	12.464 20.744
	12.585 20.738
	12.708 20.732
	12.835 20.727
	12.899 20.724
	12.964 20.722
	13.030 20.719
	13.097 20.717
	 /
\plot 13.097 20.717 15.240 20.637 /
%
%
\put{$0$} [lB] at  2.381 21.273
%
%
\put{$z$} [lB] at  6.350 21.273
%
%
\put{$v_1$} [lB] at  3.334 20.637
%
%
\put{$u_r$} [lB] at  3.175 22.225
%
%
\put{$u_1$} [lB] at  5.429 22.813
%
%
\put{$v_s$} [lB] at  5.349 20.273
%
%
\put{$1$} [lB] at  8.873 21.082
%
%
\put{$\infty$} [lB] at 15.587 21.526
%
%
\put{$u_R$} [lB] at  9.953 22.320
%
%
\put{$v_{s+1}$} [lB] at  9.730 20.686
%
%
\put{$v_{S-1}$} [lB] at 12.349 20.130
%
%
\put{$u_{r+1}$} [lB] at 12.476 23.019
%
%
\put{$x$} [lB] at  1.429 21.273
\linethickness=0pt
\putrectangle corners at  1.245 24.145 and 15.587 19.025
\endpicture}

\caption{Integration contours for $u$ and $v$ variables for an s-channel block
for fusion rule II.}
\label{figanII}
\end{figure}

Fig. \ref{figanII} shows 
the integration contours.
The s-channel block for fusion rule II is given by
\bea
&&{\cal S}_{(r,s,1)}^{(R,S)}(z,x)\nn
&=&z^{2j_1j_2/t}(1-z)^{2j_2j_3/t}
\int_0^z du_idv_k\int_1^\infty du_jdv_l
\oint_{{\cal C}_v}\frac{dv}{2\pi i}\nn
&&u_i^{a'}(1-u_i)^{b'}(z-u_i)^{c'}\prod_{i<i'}(u_i-u_{i'})^{2\rho'}
u_j^{a'}(u_j-1)^{b'}(u_j-z)^{c'}\nn
&&\prod_{j<j'}(u_j-u_{j'})^{2\rho'}
\prod_{i,j}(u_j-u_i)^{2\rho'}\nn
&&v_k^a(1-v_k)^b(z-v_k)^c\prod_{k<k'}(v_k-v_{k'})^{2\rho}
v_l^a(v_l-1)^b(v_l-z)^c\prod_{l<l'}(v_l-v_{l'})^{2\rho}\nn
&&\prod_{k,l}(v_l-v_k)^{2\rho}\prod_{i,k}(u_i-v_k)^{-2}\prod_{i,l}
(u_i-v_l)^{-2}\prod_{j,k}(u_j-v_k)^{-2}\prod_{j,l}(u_j-v_l)^{-2}\nn
&&(u_i-x)(u_j-x)(v_k-x)^{-\rho}(v_l-x)^{-\rho}\nn
&&v^a(1-v)^b(v-z)^c(v-v_k)^{2\rho}(v_l-v)^{2\rho}(v-u_i)^{-2}(v-u_j)^{-2}
(v-x)^{-\rho}
\eea
Here the variables, $u_i,u_j,v_k,v_l$ are taken along approximately real 
(for $z$ real) contours as for fusion 
rule I. The indices indicate: 
$i=1,...,r; j=r+1,... ,R; k=1,...,s; l=s+1,...,S-1$, whereas $v$ runs
along the contour, ${\cal C}_v$ which starts at $x$ and surrounds both $0$ and 
$z$, cf. Fig. \ref{figanII}.  In addition to the s-channel blocks we have 
defined above,
we define additional ones in analogy to the case for minimal models \cite{DF}.
Namely, instead of using complex contours close to 
the real axis (for real $z$), we
may use real ``time ordered'' integrations, with an ordering so that all terms
in the integral expression for ${\cal S}^{(R,S)}_{(r,s,0)}$ become real. 
This block is denoted $S^{(R,S)}_{(r,s,0)}$. Using
arguments similar to Ref. \cite{DF} we find 
\bea
{\cal S}^{(R,S)}_{(r,s,0)}(z,x)&=&
\lambda_r(\rho')\lambda_{R-r}(\rho')\lambda_s(\rho)
\lambda_{S-s}(\rho)S^{(R,S)}_{(r,s,0)}(z,x)\nn
S^{(R,S)}_{(r,s,0)}(z,x)&=&s^{(R,S)}_{(r,s,0)}(z,x)N^{(R,S)}_{(r,s,0)}
\label{sblockI}
\eea
where $s^{(R,S)}_{(r,s,0)}(z,x)$ is normalized in such a way that the
behaviour as $z\rightarrow 0, x\rightarrow 0$ is
\ben
s^{(R,S)}_{(r,s,0)}(z,x)=z^{-h(j_1)-h(j_2)+h(j_I)}(-x)^{j_1+j_2-j_I}
(1+{\cal O}(z,x))
\een
The $\lambda$-functions were defined in \Eq{lambda}.
Similarly for fusion rule II we write
\bea
{\cal S}^{(R,S)}_{(r,s,1)}(z,x)&=&\lambda_r(\rho')\lambda_s(\rho)
\lambda_{R-r}(\rho')\lambda_{S-s-1}(\rho)S^{(R,S)}_{(r,s,1)}(z,x)\nn
S^{(R,S)}_{(r,s,1)}(z,x)&=&s^{(R,S)}_{(r,s,1)}(z,x)N^{(R,S)}_{(r,s,1)}\nn
s^{(R,S)}_{(r,s,1)}(z,x)&=&z^{-h(j_1)-h(j_2)+h(j_{II})}(-x)^{j_1+j_2-j_{II}}
(1+{\cal O}(z,x))
\label{sblockII}
\eea
where
\ben
j_{II}=-j_I-1
\een
The leading behaviour (for $z\rightarrow 0$ followed by $x\rightarrow 0$) 
in this case of fusion rule II is determined by the scalings
\bea
u_i&\rightarrow &zu_i\nn
v_k&\rightarrow &zv_k\nn
v&\rightarrow &(-x)v
\eea
The normalization constants, $N^{(R,S)}_{(r,s,0)}$ and  $N^{(R,S)}_{(r,s,1)}$, 
are found in terms of the famous Dotsenko-Fateev integral (last paper 
Ref. \cite{DF}, appendix A, here with a minor misprint corrected):
\bea
{\cal J}_{nm}(a,b;\rho)&=&\rho^{2nm}\prod_{i=1}^n\frac{\Gamma(i\rho')}
{\Gamma(\rho')}\prod_{i=1}^m\frac{\Gamma(i\rho-n)}{\Gamma(\rho)}\nn
&\times&\prod_{i=0}^{n-1}\frac{\Gamma(1+a'+i\rho')\Gamma(1+b'+i\rho')}
{\Gamma(2-2m+a'+b'+(n-1+i)\rho')}\nn
&\times&\prod_{i=0}^{m-1}\frac{\Gamma(1-n+a+i\rho)\Gamma(1-n+b+i\rho)}
{\Gamma(2-n+a+b+(m-1+i)\rho)}
\label{dfintegral}
\eea
We shall need these normalizations in the calculation of crossing matrices. 
After some calculations we obtain
\bea
N^{(R,S)}_{(r,s,0)}&=&(-)^{R-r+S-s}{\cal J}_{r,s}(a,c;\rho)
{\cal J}_{R-r,S-s}(a+c-2(r-\rho s)-\rho,b;\rho)\nn
&&\prod_{i=0}^{R-r-1}\frac{s(a'+c'-2(s-\rho'r)+1+i\rho')}
{s(a'+b'+c'-2(s-\rho' r)+1+\rho'(R-r-1+i))}\nn
&&\prod_{i=0}^{S-s-1}\frac{s(a+c-2(r-\rho s)-\rho+i\rho)}
{s(a+b+c-2(r-\rho s)-\rho+\rho(S-s-1+i))}
\eea
and
\bea
N^{(R,S)}_{(r,s,1)}&=&N^{(R,S)}_{(r,s+1,0)}
\frac{\Gamma(\rho)\Gamma(1-\rho)\Gamma(2-2r+a+c+2s\rho)}
{\Gamma(-a-c-2\rho s+2r)\Gamma(1-r+a+s\rho)
\Gamma(1-r+c+s\rho)}\nn
&&\frac{1}{\Gamma((s+1)\rho-r)\Gamma(2-r+a+c+(s-1)\rho)}
\eea
For the integral realization considered here \cite{An} it is trivial to obtain
the t-channel forms once the s-channel forms above are given. In fact we have
in an obvious notation ($\epsilon = 0,1$ for fusion rules I and II)
\ben
{\cal T}^{(R,S)}_{(r,s,\epsilon)}(z,x;j_1,j_2,j_3,j_4)=
{\cal S}^{(R,S)}_{(r,s,\epsilon)}(1-z,1-x;j_3,j_2,j_1,j_4)
\een
We notice the following. When in the integral realization, we also transform
all integration variables as $u\rightarrow 1-u, v\rightarrow 1-v$, the 
integrand for the t-channel block is identical to the one for the s-channel 
block, up to phases. In particular, whenever we have $(u-x)$ or $(v-x)^{-\rho}$ 
in the s-channel, we would have $(x-u)$ and $(x-v)^{-\rho}$ in the t-channel.
Also, after transformation of the variables, the integration contours in the 
t-channel are between $z$ and $1$ and between $0$ and $-\infty$, 
and the complex
contour for $v$ in the case of fusion rule II surrounds $z$ and $1$. The above
factors, $(u-x)$ etc. are real provided $x<0$ in the s-channel, or $x>1$ in the
t-channel. These two possibilities map to each other under $x\rightarrow 1-x$.  

\section{Analysis of the equivalence between the integral realizations of PRY 
and of Andreev}
The equality between our form of the conformal blocks and the one provided by 
Andreev turns out to be very closely related to 
an identity in itself remarkable
for minimal models, which was mentioned in Ref. \cite{An}. The proof of this
identity is simpler but very similar to what we shall need. Hence we start by 
discussing the identity for minimal models.

\subsection{An identity for minimal models}

\underline{{\bf Theorem}}
\bea
 &&\int_0^1\prod_{k=1}^Ndv_kv_k^{a'}(1-v_k)^{b'}(1-zv_k)^{c'}\prod_{k<k'}
  (v_k-v_{k'})^{2\rho'}\nn
 &\cdot&\prod_{i=1}^Mdw_iw_i^a(1-w_i)^b(1-zw_i)^c\prod_{i<i'}
  (w_i-w_{i'})^{2\rho}\prod_{k,i}^{N,M}(v_k-w_i)^{-2}\nn
 &=&K_{NM}\int_0^1\prod_{k=1}^Ndv_kv_k^{a'-\delta'}(1-v_k)^{b'+\delta'}
  (1-zv_k)^{c'-\delta'}\prod_{k<k'}(v_k-v_{k'})^{2\rho'}\nn
 &\cdot&\prod_{i=1}^Mdw_iw_i^{a-\delta}(1-w_i)^{b+\delta}
 (1-zw_i)^{c-\delta}
  \prod_{i<i'}(w_i-w_{i'})^{2\rho}\prod_{k,i}^{N,M}(v_k-w_i)^{-2}
\label{minimal}
\eea
where
\bea
 a'=-\rho'a\hspace{1 cm}b'&=&-\rho'b\hspace{1 cm}c'=-\rho'c\hspace{1 cm}
 \delta'=-\rho'\delta\hspace{1 cm}\rho'=1/\rho\nn
 \delta&=&a+c+1-N+(M-1)\rho\nn
 \delta'&=&a'+c'+1-M+(N-1)\rho'
\label{prime}
\eea
and
\bea
  K_{NM}&=&\prod_{i=0}^{N-1}\frac{\Gamma(a'+1+i\rho')\Gamma(b'+1+i\rho')}
    {\Gamma(-c'+M+(-N+1+i)\rho')\Gamma(a'+b'+c'+2-M+(N-1+i)\rho')}\nn
  &\cdot&\prod_{i=0}^{M-1}\frac{\Gamma(a+1-N+i\rho)\Gamma(b+1-N+i\rho)}
   {\Gamma(-c+(-M+1+i)\rho)\Gamma(a+b+c+2-2N+(M-1+i)\rho)}
\eea
The left hand side of \Eq{minimal} has the structure of the standard
integral realization for minimal models \cite{DF}, leaving out some irrelevant 
pre-factors. There are $N,M$ screening charges of the two kinds and they are
at positions $v_k$ and $w_i$ (in Ref. \cite{DF} and in the previous section
they were denoted by $u$ and $v$), except they have been scaled by $z$. 
Also of course here the meaning of the 
letters $a,b,c$ are different from their meaning in the previous section.
The integrations are taken to be (``time''-) ordered, and we only consider
one kind of conformal block, where all the integrations are between $0$ 
and $z$.

The proof we present of this identity is by brute force and takes several 
lengthy calculations, although not so bad as the case we shall be mostly
interested in: the identity between our $SL(2)$ block and that of Andreev.

The idea is simply to consider both sides of \Eq{minimal} as functions of $z$.
Both versions of the conformal blocks  
have singularities only when $z\rightarrow 0,1,\infty$. The 
singularity at $z\rightarrow 0$ has been explicitly removed by the scaling, and
it is rather trivial that both sides had the same power of $z$. The limit
$z\rightarrow 0$ is used for normalization. 
We isolate the other singularities and show that they have the same structure 
and the same strengths.

The limit $z\rightarrow 0$ is simple. In that limit both sides of \Eq{minimal}
are holomorphic in $z$ (since suitable pre-factors have been removed) 
and we may simply put $z=0$. Then both sides may be 
done in terms of the Dotsenko-Fateev integral (\cite{DF}, appendix A). 
This gives
immediately the normalization, $K_{NM}$.

The limits $z\rightarrow 1,\infty$ are much more complicated. Here there will
be several different power singularities of the form $(1-z)^A$ and $z^B$. 
We must
isolate those and compute their strengths and demonstrate that we get the same 
results for both sides of \Eq{minimal}. The basic analytic tool was described
in Ref. \cite{DF}.\\[.2cm]
$\underline{z\rightarrow1}$\\
We split the integration region from $0$ to $1$ using a small positive 
$\epsilon$ as follows (the integrations are time-ordered throughout)
\bea
 \int_0^1&=&\int_0^{1-\epsilon}+\int_{1-\epsilon}^1\nn
  \int_0^1\prod_{k=1}^Ndv_k\prod_{i=1}^Mdw_i&=&\sum_{n,m}
  \int_0^{1-\epsilon}\prod_{k=N-n+1}^Ndv_k\prod_{i=M-m+1}^Mdw_i
\int_{1-\epsilon}^1
   \prod_{l=1}^{N-n}dv_l\prod_{j=1}^{M-m}dw_j
\eea
It is not difficult to check that a particular $n,m$ term will give rise to a
particular power of $(1-z)$. One performs the following scalings of the 
$\int_{1-\epsilon}^1$ integration variables, $w\sim v_l,w_j$,
\bea
  w&\rightarrow&1-(1-z)\frac{1-w}{w}\nn
  dw&\rightarrow&\frac{1-z}{w^2}dw\nn
  1-w&\rightarrow&(1-z)\frac{1-w}{w}\nn
  1-zw&\rightarrow&(1-z)\frac{(1-z)w+z}{w}\sim\frac{1-z}{w}\nn
  \int_{1-\epsilon}^1&\rightarrow&\int_{\frac{1-z}{\epsilon+1-z}}^1\sim
  \int_0^1
\eea
One rather easily finds that the power of $(1-z)$ occurring on both sides of 
\Eq{minimal} is
$$(1-z)^{(N-n)(b'+c'+1)+(N-n)(N-n-1)\rho'+(M-m)(b+c+1)+(M-m)(M-m-1)\rho
  -2(N-n)(M-m)}$$
The coefficient of this singularity may also be evaluated 
on both sides in terms
of Dotsenko-Fateev integrals. It is not, however, immediately obvious that
these coefficients are equal. Both sides involves many products of terms 
involving ratios of $\Gamma$ functions. One employs over and over again the 
simple identity
\ben
\frac{\Gamma(X)}{\Gamma(X-L)}=\prod_{j=0}^{L-1}(X-1-j)
\label{gammaid}
\een
Thus it turns out for example that there are factors on both sides  
involving $\Gamma$ functions with argument involving 
$a'$ only (no $b,'c'$). On the left hand side we have
$$\prod_{i=0}^{n-1}\Gamma(a'+1+i\rho')$$
whereas on the right hand side there are similar factors of the form
$$\frac{\prod_{i=0}^{N-1}\Gamma(a'+1
   +i\rho')}{\prod_{i=0}^{N-n-1}\Gamma(a'+1-M+(n+i)\rho')}$$
Using the identity \Eq{gammaid} we find that the ratio between these two
is
$$\prod_{i=0}^{N-n-1}\prod_{j=0}^{M-1}\frac{1}{(a'-j+(n+i)\rho')}$$ 
By going over the several 
other different factors on both sides and working out the
ratios, one finally shows that the product of all ratios equals $1$.

This completes the proof that the singularities are identical in the limit 
$z\rightarrow 1$.\\[.2cm]
$\underline{z\rightarrow\infty}$\\
The strategy is entirely analogous. We make the split 
(time-ordered integrations)
\bea
  \int_0^1&=&\int_0^{\epsilon}+\int_{\epsilon}^1\nn
 \int_0^1\prod_{k=1}^Ndv_k\prod_{i=1}^Mdw_i&=&\sum_{n,m}\int_0^{\epsilon}
 \prod_{k=N-n+1}^Ndv_k\prod_{i=M-m+1}^Mdw_i\int_{\epsilon}^1\prod_{l=1}^{
  N-n}dv_l\prod_{j=1}^{M-m}dw_j
\label{zsplitinf}
\eea
The $\int_0^{\epsilon}$ integration variables are scaled according to
\bea
 w&\rightarrow&\frac{1-w}{-zw}\nn
 \int_0^{\epsilon}dw&\rightarrow&\int_1^{\frac{1}{1-z\epsilon}}\frac{dw}{zw^2}
  \sim(-z)^{-1}\int_0^1\frac{dw}{w^2}
\label{wwrtinf}
\eea
However, this time we must identify the left hand side with $n,m$ with the 
right
hand side with $N-n,M-m$. It is then simple to verify that the power of $z$
on both sides are
$$(-z)^{-na'+(N-n)c'-n-n(n-1)\rho'-ma+(M-m)c-m-m(m-1)\rho+2nm}$$
To check that the coefficients also agree, as before one carries out explicitly
the integrations in terms of Dotsenko-Fateev integrals resulting in many 
products of ratios of $\Gamma$ functions. Finally one laboriously checks that
the ratios multiply up to $1$.

\subsection{Integral identity in $SL(2)$ current theory}
As previously indicated there is no absolute need for proving the equivalence 
between the PRY 4-point function, \cite{PRY}, and the one by Andreev, 
\cite{An}, since both satisfy the Knizhnik-Zamolodchikov equations. 
Nevertheless, it is of 
some interest to understand better how two such seemingly very different 
expressions can agree, and it is rather nice to be aware of the clarification
provided by the relation to the minimal model case treated in the previous 
subsection. In this subsection we go over  
several of the steps needed for a direct 
analytic proof. In fact, we investigate the singularity structure of the two 
expressions in the double limits, 
$z,x\rightarrow 0, z,x\rightarrow 1, z,x\rightarrow \infty$ and in the single 
limit, $z\rightarrow x$. We restrict ourselves to just one of the s-channel 
conformal blocks.

\underline{{\bf Theorem}}
\bea
 &&\int_0^1du\prod_{k=1}^Ndv_kv_k^{a'}(1-v_k)^{b'}(1-zv_k)^{c'}\left(1-\frac{
  1-v_k}{1-zv_k}\frac{z}{x}u\right)\prod_{k<k'}(v_k-v_{k'})^{2\rho'}\nn  
 &\cdot&\prod_{i=1}^Mdw_iw_i^a(1-w_i)^b(1-zw_i)^c\left(1-\frac{1-w_i}{1-zw_i}
  \frac{z}{x}u\right)^{-\rho}\prod_{i<i'}(w_i-w_{i'})^{2\rho}\nn
 &\cdot&\prod_{k,i}^{N,M}(v_k-w_i)^{-2}u^{-c-1}(1-u)^{b+c-N+(M-1)\rho}\nn
 &=&K_{NM}^x\int_0^1\prod_{k=1}^Ndv_kv_k^{a'-\delta'}(1-v_k)^{b'+\delta'}
  (1-zv_k)^{c'-\delta'}\left(1-\frac{z}{x}v_k\right)\prod_{k<k'}(v_k-v_{k'})^{
  2\rho'}\nn
 &\cdot&\prod_{i=1}^Mdw_iw_i^{a-\delta}(1-w_i)^{b+\delta}(1-zw_i)^{c-\delta}
  \left(1-\frac{z}{x}w_i\right)^{-\rho}\prod_{i<i'}(w_i-w_{i'})^{2\rho}\nn
  &\cdot&\prod_{k,i}^{N,M}(v_k-w_i)^{-2}
\label{pryan}
\eea
where
\ben
 K_{NM}^x=\frac{\Gamma(-c)\Gamma(b+c+1-N+(M-1)\rho)}{\Gamma(b+1-N+(M-1)\rho)}
  K_{NM}
\een
Here, up to irrelevant common pre-factors, 
the left hand side is our form of the
conformal block \Eq{pryblock} for $r=R=N$ and $s=S=M$ in the s-channel. We
denote this by $S^{PRY}$.
Similarly up to the same pre-factors and the new normalization constant, 
$K^x_{NM}$, the right hand side is essentially \Eq{anblockI}. We denote it by
$S^A$.
Notice in particular that now we put 
\ben
a=2j_1, \ b=2j_2+\rho, \ c=2j_3, \ \rho=t
\een
$\delta, \delta'$ are given by the same expressions as for the minimal models:
\bea
\delta&=&a+c+1-N+(M-1)\rho\nn
\delta'&=&a'+c'+1-M+(N-1)\rho'
\eea
Then $a-\delta$ is what was called $a$ in previous sections, $b+\delta$ was 
previously called $c$ and $c-\delta$ was previously called $b$. In subsequent 
sections we shall revert to this notation, but in this section we stick to the
present notation in order to emphasize the similarity with minimal models. Also
notice, that because all integrations are between $0$ and $1$ (after scaling 
by $z$), it is possible to deform the $u$-integration in Fig. \ref{figpryus} to
being along the real axis from $0$ to $1$.

In this case we demonstrate that both the left hand side and the right hand 
side of
the claimed identity have the same singularities in the limits 
$z,x\rightarrow 1$, $z,x\rightarrow \infty$ and $z\rightarrow x$. The proof 
turns out to be rather more laborious than for the minimal models. This
is due to the $x$-dependence and the $u$-integration in the case of $S^{PRY}$.
However, the general strategy is entirely analogous, so we only indicate some
of the steps on the way.

The limit $z,x\rightarrow 0$ is simple to deal with and it gives rise to the
normalization constant, $K^x_{NM}$ differing from the one in the minimal 
models, $K_{NM}$ because of the $u$-integration.\\[.2cm]
$\underline{z,x\rightarrow1}$\\
We first deal with $S^A$. Exactly as in the case of minimal models, we split
the ordered integration ranges for the $v$'s in $n$ $v_k$'s and $m$ $w_i$'s 
in $(0,1-\epsilon)$
and $N-n$ $v_l$'s and $M-m$ $w_j$'s in $(1-\epsilon,1)$. Omitting 
integration signs and products for brevity we find
\bea
S^A_{nm}
 &\sim&(1-z)^{(N-n)(b'+c'+1)+(N-n)(N-n-1)\rho'+(M-m)(b+c+1)
   +(M-m)(M-m-1)\rho-2(N-n)(M-m)}\nn
 &\cdot&(x-1)^{N-n-(M-m)\rho}K_{NM}^x\nn
 &\cdot&v_k^{a'-\delta'}(1-v_k)^{b'+c'+(N-n)2\rho'-2(M-m)}(x-v_k)(v_k-v_{k'}
  )^{2\rho'}\nn
 &\cdot&w_i^{a-\delta}(1-w_i)^{b+c+(M-m)2\rho-2(N-n)}(x-w_i)^{-\rho}(w_i-w_{
   i'})^{2\rho}(v_k-w_i)^{-2}\nn
 &\cdot&v_l^{-b'-c'-2-(N-n-1)2\rho'+2(M-m)}(1-v_l)^{b'+\delta'}(v_l-v_{l'})^{
  2\rho'}\nn
 &\cdot&w_j^{-b-c-2-(M-m-1)2\rho+2(N-n)}(1-w_j)^{b+\delta}(w_j-w_{j'})^{2\rho}
  (v_l-w_j)^{-2}
\label{sanm}
\eea
where we have performed the same scalings as for minimal models. 
The above has to
be summed over $n$ and $m$, but for a fixed value we pick up the pure $(1-z)$ 
and $(x-1)$ singularity indicated. The $l,j$ part of the integration gives 
immediately rise to a standard Dotsenko-Fateev integral. For the $k,i$ part we
perform the further split and scalings
\ben
 \int_0^1=\int_0^{1-\epsilon}+\int_{1-\epsilon}^1
\een
\bea
 v&\rightarrow&1-(1-1/x)\frac{1-v}{v}\nn
 \int_{1-\epsilon}^1dv&\rightarrow&\int_0^1(1-1/x)\frac{dv}{v^2}\nn
 \int_0^1\prod_{k=1}^ndv_k\prod_{i=1}^mdw_i&=&\sum_{n_0,m_0}
  \int_0^{1-\epsilon}\prod_{k_0=n-n_0+1}^{n}dv_{k_0}\prod_{i_0=m-m_0+1}^{m}
dw_{i_0}\nn
&&\int_{1-\epsilon}^1
 \prod_{k=1}^{n-n_0}dv_k\prod_{i=1}^{m-m_0}dw_i
\eea
In the limit $x\rightarrow 1$ we extract the $(1-x)$ power and find the 
coefficient again to be given by the product of two Dotsenko-Fateev integrals.
Analysing the $\Gamma$ functions of these we see that we can only get a non 
vanishing result if $m_0=m$ or if $m_0=m-1$. We denote these cases by 
$S^{AI}_{nm}$ and $S^{AII}_{nm}$. They will turn out to be related to fusion 
rules I and II. Combining everything we find the following singularities
in the limit, $z,x\rightarrow 1$,
\bea
S^{AI}_{nm}&=&(1-z)^{(N-n)(b'+c'+1)+(N-n)(N-n-1)\rho'+(M-m)(b+c+1)
   +(M-m)(M-m-1)\rho-2(N-n)(M-m)}\nn
  &\cdot&(x-1)^{N-n-(M-m)\rho}N(S^{AI}_{nm})\nn
S^{AII}_{nm}&=&(1-z)^{(N-n)(b'+c'+1)+(N-n)(N-n-1)\rho'+(M-m)(b+c+1)
   +(M-m)(M-m-1)\rho-2(N-n)(M-m)}\nn
  &\cdot&(x-1)^{b+c+1-N+n+(M-m-1)\rho}N(S^{AII}_{nm})
\eea
One checks that the singularities exactly correspond to fusion rules I and II.
The normalizations, $N(S^{AI}_{nm})$ and $N(S^{AII}_{nm})$ are found explicitly
in terms of products of Dotsenko-Fateev integrals to be lengthy expressions 
involving many products of ratios of $\Gamma$ functions.

We now turn to a similar analysis of $S^{PRY}$ in the same limit 
$z\rightarrow 1$ followed by $x\rightarrow 1$. We replace $u\rightarrow 1-u$ 
and perform the same split and the same scalings of the $v$ and $w$ variables
as in the case of minimal models. Omitting again integration signs and products
we find
\bea
S^{PRY}_{nm}&\sim&(1-z)^{(N-n)(b'+c'+1)+(N-n)(N-n-1)\rho'+(M-m)(b+c+1)
   +(M-m)(M-m-1)\rho-2(N-n)(M-m)}\nn
 &\cdot&v_k^{a'}(1-v_k)^{b'+c'+(N-n)2\rho'-2(M-m)}(v_k-v_{k'}
  )^{2\rho'}\nn
 &\cdot&w_i^{a}(1-w_i)^{b+c+(M-m)2\rho-2(N-n)}(w_i-w_{
   i'})^{2\rho}(v_k-w_i)^{-2}\nn
 &\cdot&v_l^{-b'-c'-2-(N-n-1)2\rho'+2(M-m)}(1-v_l)^{b'}(x-(1-v_l)(1-u))
 (v_l-v_{l'})^{2\rho'}\nn
 &\cdot&w_j^{-b-c-2-(M-m-1)2\rho+2(N-n)}(1-w_j)^{b}(x-(1-w_j)(1-u))^{-\rho}
  (w_j-w_{j'})^{2\rho}\nn
 &\cdot&(v_l-w_j)^{-2}u^{b+c-N+(M-1)\rho}(1-u)^{-c-1}(x-1+u)^{n-m\rho}
\eea
Here the $k,i$ integrations are independent of $x$ and $u$ and are readily 
evaluated in terms of Dotsenko-Fateev integrals, so we concentrate on the 
$l,j$ part. We perform the split and the scalings
\bea
 \int_0^1\prod_{l=1}^{N-n}dv_l\prod_{j=1}^{M-m}dw_j&=&\sum_{n_0,m_0}
  \int_0^{\epsilon}\prod_{l_0=N-n-n_0+1}^{N-n}dv_{l_0}
\prod_{j_0=M-m-m_0+1}^{M-m}dw_{j_0}\nn
&&\int_{\epsilon}^1
 \prod_{l=1}^{N-n-n_0}dv_l\prod_{j=1}^{M-m-m_0}dw_j\nn
 w&\rightarrow&(x-1)\frac{1-w}{w}\nn
 \int_0^{\epsilon}dw&\rightarrow&\int_1^{\frac{x-1}{\epsilon+x-1}}(1-x)
  \frac{dw}{w^2}\sim(x-1)\int_0^1\frac{dw}{w^2}
\eea
and similarly for $\int_0^1du$. To be able to distinguish we write
\bea
 \int_{\epsilon}^1du&\rightarrow&\int_0^1du\nn
 \int_0^{\epsilon}du&\rightarrow&(x-1)\int_0^1\frac{dy}{y^2}\nn
 u&\rightarrow&(x-1)\frac{1-y}{y}
\eea
and denote them the $u$- and $y$- cases respectively. 
In the $u$-case the arising
Dotsenko-Fateev integrals turn out to vanish unless $n_0=m_0=0$, and we find in
the $u$-case
\bea
S^{PRY,u}_{nm}&\sim&(1-z)^{(N-n)(b'+c'+1)+(N-n)(N-n-1)\rho'+(M-m)(b+c+1)
   +(M-m)(M-m-1)\rho}\nn
 &\cdot&(1-z)^{-2(N-n)(M-m)}\nn
 &\cdot&\prod_{k=1}^n\prod_{i=1}^m
  v_k^{a'}(1-v_k)^{b'+c'+(N-n)2\rho'-2(M-m)}(v_k-v_{k'}
  )^{2\rho'}\nn
 &\cdot&w_i^{a}(1-w_i)^{b+c+(M-m)2\rho-2(N-n)}(w_i-w_{
   i'})^{2\rho}(v_k-w_i)^{-2}\nn
 &\cdot&\prod_{l=1}^{N-n}\prod_{j=1}^{M-m}v_l^{-b'-c'-2-(N-n-1)2\rho'+2(M-m)}
  (1-v_l)^{b'}(v_l-v_{l'})^{2\rho'}\nn
 &\cdot&w_j^{-b-c-2-(M-m-1)2\rho+2(N-n)}(1-w_j)^b(w_j-w_{j'})^{2\rho}
  (v_l-w_j)^{-2}\nn
 &\cdot&(1-(1-v_l)(1-u))(1-(1-w_j)(1-u))^{-\rho}\nn
 &\cdot&u^{b+c-N+n+(M-m-1)\rho}
  (1-u)^{-c-1}
\label{PRYnmu}
\eea
Now we want to establish
\ben
  S^{PRY,u}_{NM}=S^{AI}_{NM}
\label{PRYNMANM1}
\een
and 
\ben
  S^{PRY,u}_{nm}=0\ \ \ ,\ \ \ (n,m)\neq (N,M)
\een
A straightforward analysis shows these to be the satisfied.

In the $y$ case we introduce a similar further splitting resulting in
objects $S^{PRY}_{nm;n_0m_0}$. It turns out to be possible to demonstrate
that
\bea
S^{PRY}_{nm;00}&=&S^{AII}_{nm}\nn
S^{PRY}_{nm;01}&=&S^{AI}_{nm}
\eea
(In principle we should check that higher values of $n_0,m_0$ give zero. We
anticipate no interesting problems here). The analysis contains no new ideas 
over the situation encountered for minimal models, but again the calculations
involved are somewhat lengthy.\\[.2cm]
$\underline{z,z/x,x\rightarrow\infty}$\\
Again we first analyse the $S^A$ case. We introduce the same splitting
of integrations and the same variable transformations as for minimal models, 
and find
\bea
S^A_{nm} &\sim&K_{NM}^x(-z)^{n(-a'+\delta'-1)+(N-n)(c'-\delta'+1)-n(n-1)\rho'}
\nn
 &\cdot&(-z)^{m(-a+\delta-1)+(M-m)(c-\delta-\rho)-m(m-1)\rho+2nm}\nn
 &\cdot&x^{-(N-n)+(M-m)\rho}\nn
 &\cdot&v_k^{-a'-c'+2\delta'-2+2m-(n-1)2\rho'}(1-v_k)^{a'-\delta'}(v_k
  -v_{k'})^{2\rho'}\nn
 &\cdot&w_i^{-a-c+2\delta-2+2n-(m-1)2\rho}(1-w_i)^{a-\delta}(w_i-w_{i'}
  )^{2\rho}(v_k-w_i)^{-2}\nn
 &\cdot&\left(1+\frac{1-v_k}{xv_k}\right)\left(1+\frac{1-w_i}{xw_i}
  \right)^{-\rho}\nn
 &\cdot&v_l^{a'+c'-2\delta'+1-2m+n2\rho'}(1-v_l)^{b'+\delta'}(v_l-v_{l'}
  )^{2\rho'}\nn
 &\cdot&w_j^{a+c-2\delta-2n+(2m-1)\rho}(1-w_j)^{b+\delta}(w_j-w_{j'}
  )^{2\rho}(v_l-w_j)^{-2}
\eea
again omitting integration signs and products, which are just as for the case
of minimal models. The $l,j$ integration is seen to result in Dotsenko-Fateev 
integrals. In the $k,i$ integrals we perform a split of integrals form $0$ to
$\epsilon$ and from $\epsilon$ to $1$. In the $\int_0^\epsilon$ we transform
variables like
\bea
 v&\rightarrow&\frac{1-v}{xv}\nn
 1+\frac{1-v}{xv}&\rightarrow&\frac{1}{1-v}-\frac{1}{x}\sim\frac{1}{1-v}\nn
 \int_0^{\epsilon}dv&\rightarrow&x^{-1}\int_0^1\frac{dv}{v^2}\nn
  \int_0^1\prod_{k=1}^ndv_k\prod_{i=1}^mdw_i&=&\sum_{n_0,m_0}
  \int_0^{\epsilon}\prod_{k_0=n-n_0+1}^{n}dv_{k_0}
  \prod_{i_0=m-m_0+1}^{m}dw_{i_0}\nn
&&\int_{\epsilon}^1\prod_{k=1}^{n-n_0}dv_k\prod_{i=1}^{m-m_0}dw_i
\eea
An analysis of the coefficients of the singularities reveals that this is non 
vanishing only if 
$$(n_0,m_0)=(0,0), (0,1)$$
These two cases we term again (we use the same notation as before, even though
now we consider a different limit), $S^{AI}_{nm}$ and $S^{AII}_{nm}$ for what
turns out to be fusion rules I and II. We find
\bea
S^{AI}_{nm}&=&(-z)^{-(N-n)a'+nc'-(N-n)(N-n-1)\rho'-(M-m)a+mc
  -(M-m)(M-m)\rho+(M-m)(2N-2n-1)}\nn
 &\cdot&x^{-(N-n)+(M-m)\rho}N(S^{AI}_{nm})\nn
S^{AII}_{nm}&=&(-z)^{-(N-n)a'+nc'-(N-n)(N-n-1)\rho'-(M-m)a+mc
  -(M-m)(M-m)\rho}\nn
 &\cdot&(-z)^{(M-m)(2N-2n-1)}\nn
 &\cdot&x^{-a-c-1+N-n+(-M+m)\rho}N(S^{AII}_{nm})
\eea
where the normalizations (different of course to the ones in the previous limit
$z,x\rightarrow 1$), $N(S^{AI}_{nm})$ and $N(S^{AII}_{nm})$ are given (in 
terms of Dotsenko-Fateev integrals) by lengthy products of ratios of 
$\Gamma$ functions. The singularities shown indicate that indeed we are dealing
with fusion rules I and II.

We then treat the $S^{PRY}$ case. Again we first perform the same splittings
and variable transformations as for $S^A$ with the same meaning of 
$v_k,v_l,w_i,w_j$. The $i,k$ part is again simple, whereas the $l,j$ part is 
treated with a split of the ordered integrations as
\bea
 \int_0^1\prod_{l=1}^{N-n}dv_l\prod_{j=1}^{M-m}dw_j&=&\sum_{n_0,m_0}
  \int_0^{\epsilon}\prod_{l_0=N-n-n_0+1}^{N-n}dv_{l_0}
\prod_{j_0=M-m-m_0+1}^{M-m}dw_{j_0}\nn
&&\int_{\epsilon}^1
 \prod_{l=1}^{N-n-n_0}dv_l\prod_{j=1}^{M-m-m_0}dw_j\nn
\eea
followed by the scalings
\bea
v_{l_0}&\rightarrow&\frac{1-v_{l_0}}{xv_{l_0}}\nn
w_{j_0}&\rightarrow&\frac{1-w_{j_0}}{xw_{j_0}}
\eea
We then seek to demonstrate that
\bea
S^{PRY}_{nm;00}&=&S^{AI}_{N-n,M-m}\nn
S^{PRY}_{nm;01}&=&S^{AII}_{N-n,M-m}\nn
S^{PRY}_{nm;n_0m_0}&=&0 ,\ \ \ (n_0,m_0)\neq (0,0),(0,1)
\eea
The proof here is lengthy again, but with no new ideas introduced.\\[.2cm]
$\underline{z\rightarrow x}$\\
This case is the most complicated. We omit nearly all the details, most of
which are similar to what have been described above, and we concentrate on 
the strategy. More details will be presented elsewhere \cite{JR}. 
First one may 
check that the nature of the singularity is a linear combination of just two
different powers of $(z-x)$ namely either $(z-x)^0$ or $(z-x)^{(c+1-\rho)}$.
Second one must investigate whether the coefficients of these two powers are 
the same for the two sides of \Eq{pryan}. That coefficient is a function of $x$
in the limit $z\rightarrow x$, so we must investigate whether the coefficient 
functions defined by the two sides of \Eq{pryan} are equal. As above the 
technique is to investigate the singularity structure in the singular limits
$x=0,1,\infty$. It turns out that the sought equality depends on the 
following identities:
\bea
&&\int_0^1 dwdy w^{-a-2+\rho}(1-w)^a(1-(1-w)(1-(1-x)y))^{-\rho}\nn
&\cdot&y^{-b-c+N-2+(-M+2)\rho}(1-y)^{b+c-N+(M-1)\rho}\nn
&=&(1-x)^{-a-1}\nn
&\cdot&\frac{\Gamma(a+1)\Gamma(b+c+1-N+(M-1)\rho)\Gamma(-a-b-c-2+N+(-M+2)\rho)}
{\Gamma(\rho)}
\eea
and
\ben
\int_0^1 dw\int_0^w dy w^{-a}(1-w)^{a+c-2}(w-y)^{-c}y^{b+c-2}(1-y)^{-b}=0
\een
which are not too difficult to prove.
Next define the Dotsenko-Fateev integrand:
\bea
&&DF(N,M;a,b,\rho;\{v_k\},\{w_i\})\nn
&\equiv&\prod_{k-1}^Nv_k^{a'}(1-v_k)^{b'}\prod_{k<k'}(v_k-v_{k'})^{2\rho'}
\prod_{i=1}^Mw_i^a(1-w_i)^b\prod_{i<i'}(w_i-w_{i'})^{2\rho}
\prod_{k,i}^{N,M}(v_k-w_i)^{-2}\nn
&&
\eea
Then we find that the equality of the two sides of \Eq{pryan} depends on the 
following three identities (generalized Dotsenko-Fateev integrals):
\bea
(I)&&\int_0^1 du\prod_{k=1}^N dv_k\prod_{i=1}^M dw_i 
DF(N,M;a,b,\rho;\{v_k\},\{w_i\})\nn
&\cdot&(1-(1-v_k)u)(1-(1-w_i)u)^{-\rho}u^{-c-1}(1-u)^{b+c-N+(M-1)\rho}\nn
&=&\frac{\Gamma(-c)\Gamma(b+c+1-N+(M-1)\rho)\Gamma(a+b+c+2-N+(M-2)\rho)}
{\Gamma(b+1-\rho)\Gamma(a+b+c+2-2N+(2M-2)\rho)}\nn
&\cdot&\int_0^1\prod_{k=1}^N dv_k\prod_{i=1}^M dw_i 
DF(N,M;a,b-\rho,\rho;\{v_k\},\{w_i\})\nn
(II)&&\int_0^1 du\prod_{k=1}^n dv_k\prod_{i=1}^{m+1} dw_i\nn 
&&DF(n,m+1;a,b+c-2N+2n+(2M-2m-1)\rho,\rho;\{v_k\},\{w_i\})\nn 
&\cdot&(1-v_k-u)(1-w_i-u)^{-\rho}u^{-b-c+N-n-2+(-M+m+2)\rho}
(1-u)^{b+c-N+(M-1)\rho}\nn
&\sim&\frac{\Gamma(b+c+1-N+(M-1)\rho)\Gamma(a+b+c+2-N+(M-2)\rho)}
{\Gamma(a+b+c+2-2N+(2M-2)\rho)}\nn
&\cdot&\frac{\Gamma(-b-c+N-n-1+(-M+m+2)\rho)\Gamma(-N+n+(M-m)\rho)}
{\Gamma(\rho)}\nn
&\cdot&\int_0^1\prod_{k=1}^n dv_k\prod_{i=1}^m dw_i\nn 
&&DF(n,m;a,b+c-2N+2n+(2M-2m-1)\rho,\rho;\{v_k\},\{w_i\})\nn
(III)&&\int_0^1du\prod_{k=1}^n dv_k\prod_{i=1}^m dw_i\nn 
&&DF(n,m;-a-c+2n-2+(-2m+2)\rho,a,\rho;\{v_k\},\{w_i\})\nn
&\cdot&(1-v_ku)(1-w_iu)^{-\rho}u^{-c-1}(1-u)^{b+c-N+(M-1)\rho}\nn
&=&\frac{\Gamma(-c)\Gamma(b+c+1-N+(M-1)\rho)\Gamma(a+b+c+2-N+(M-2)\rho)}
{\Gamma(b+1-N+n+(M-m-1)\rho)\Gamma(a+b+c+2-N-n+(M+m-2)\rho)}\nn
&\cdot&\int_0^1\prod_{k=1}^n dv_k\prod_{i=1}^m dw_i 
DF(n,m;-a-c+2n-2+(-2m+2)\rho,a,\rho;\{v_k\},\{w_i\})\nn
&&
\eea
All the final integrals are of course Dotsenko-Fateev integrals.
In the second identity there is a phase depending on the precise choice of the
integration contour for $u$. These last three identities we have not 
proven directly.
(We have checked for low values of $N,M$). One might take the attitude that the
undoubted identity of our realization  and that of Andreev, i.e. the 
unquestionable correctness of \Eq{pryan}, implies these 
somewhat remarkable integral identities.
\section{Calculation of the crossing matrix}
The crossing matrix, $\alpha^{(R,S)}_{(r,s,\epsilon),(r',s',\epsilon')}$, 
is defined by the equation
\bea
S^{(R,S)}_{(r,s,\epsilon)}(z,x)&=&\sum_{r'=0}^R\sum_{s'=0}^S
\alpha^{(R,S)}_{(r,s,\epsilon),(r',s',0)}T^{(R,S)}_{(r',s',0)}(z,x)\nn
&+&\sum_{r'=0}^R\sum_{s'=0}^{S-1}
\alpha^{(R,S)}_{(r,s,\epsilon),(r',s',1)}T^{(R,S)}_{(r',s',1)}(z,x)
\eea
As explained 
in Ref. \cite{DF}, it is enough to calculate one column and one row
of this matrix in order to determine monodromy invariant greens functions.
We find that a moderate modification of the techniques described 
there suffices for completing the corresponding calculations here.
The main new feature is the fact that we have to observe also the $x$ 
dependence, and the presence of the complex contour in the case of 
conformal blocks for fusion rule II, cf. Fig. \ref{figanII}.

\subsection{The column $\alpha^{(R,S)}_{(r,s,\epsilon),(R,S,0)}$}
Following the idea of Ref. \cite{DF} we define the following object 
(suppressing several variables)
\bea
&&J(r_1,s_1,r_2,s_2,r_3,s_3)=z^{2j_1j_2/t}(1-z)^{2j_2j_3/t}\nn
&&\int_0^z \prod_{i=1}^{r_1}du_i\prod_{k=1}^{s_1}dv_k\int_z^1
\prod_{m=r_1+1}^{r_1+r_2}du_m\prod_{n=s_1+1}^{s_1+s_2}dv_n\int_1^\infty 
\prod_{j=r_1+r_2+1}^{r_1+r_2+r_3}du_j\prod_{l=s_1+s_2+1}^{s_1+s_2+s_3}dv_l\nn
&&\prod_i
u_i^{a'}(1-u_i)^{b'}(z-u_i)^{c'}\prod_{i<i'}(u_i-u_{i'})^{2\rho'}(u_i-x)\nn
&&\prod_m
u_m^{a'}(1-u_m)^{b'}(u_m-z)^{c'}\prod_{m<m'}(u_m-u_{m'})^{2\rho'}(x-u_m)\nn
&&\prod_j
u_j^{a'}(u_j-1)^{b'}(u_j-z)^{c'}\prod_{j<j'}(u_j-u_{j'})^{2\rho'}(u_j-x)\nn
&&\prod_{m,i}(u_m-u_i)^{2\rho'}\prod_{j,i}(u_j-u_i)^{2\rho'}\prod_{j,m}
(u_j-u_m)^{2\rho'}\nn
&&\prod_k 
v_k^a(1-v_k)^b(z-v_k)^c\prod_{k<k'}(v_k-v_{k'})^{2\rho}(v_k-x)^{-\rho}\nn
&&
\prod_nv_n^a(1-v_n)^b(v_n-z)^c\prod_{n<n'}
(v_n-v_{n'})^{2\rho}(x-v_n)^{-\rho}\nn
&&\prod_l
v_l^a(v_l-1)^b(v_l-z)^c\prod_{l<l'}(v_l-v_{l'})^{2\rho}(v_l-x)^{-\rho}\nn
&&\prod_{n,k}(v_n-v_k)^{2\rho}\prod_{l,k}(v_l-v_k)^{2\rho}
\prod_{l,n}(v_l-v_n)^{2\rho}\nn
&&\prod_{\alpha,\beta}(v_\alpha-u_\beta)^{-2}
\eea
In other words, there are $r_1$ and $s_1$ $u$ and $v$ integrations between $0$ 
and $z$, $r_2$ and $s_2$ $u$ and $v$ integrations between $z$ and $1$ and
$r_3$ and $s_3$ $u$ and $v$ integrations between $1$ and $\infty$.
Also the variables, $u_i,v_k,u_j,v_l$ are taken along contours similar to the
ones in Fig. \ref{figanI}, whereas the variables, $u_m,v_n$ are taken along 
similar ones lying between $z$ and $1$.
We notice that 
\bea
J(r,s,0,0,R-r,S-s)&=&{\cal S}^{(R,S)}_{(r,s,0)}\nn
J(0,0,R,S,0,0)&=&{\cal T}^{(R,S)}_{(R,S,0)}
\eea
Therefore we may start from $J(r,s,0,0,R-r,S-s)$ and gradually move integration
contours by contour deformations on to the interval $(z,1)$. In the process
we pick up contributions form integrals between $-\infty$ and $0$, but these 
may be neglected in the calculation of the column, 
$\alpha^{(R,S)}_{(r,s,\epsilon),(R,S,0)}$. The calculational procedure 
\cite{DF} consists in deforming upper and lower $u$ and $v$ contours in 
appropriate ways, and forming suitable linear combinations of the result. 
As explained in Ref. \cite{DF}, one may then derive identities 
for the functions, $J(r_1,s_1,r_2,s_2,r_3,s_3)$,
by carefully keeping track of the phases arising between the result of the 
deformations, and the definitions of the $J$'s. The useful identities turn our
to be after some calculations:
\bea
&&J(r_1,s_1,r_2,s_2,r_3,s_3)\nn
&=&e^{i\pi\rho'(r_2-r_1+1)}\frac{s(b'+\rho'(r_2+r_3))}
{s(b'+c'+\rho'(r_1-1+2r_2+r_3))}J(r_1-1,s_1,r_2+1,s_2,r_3,s_3)+...\nn
&=&-e^{i\pi\rho(s_2-s_1+2)}\frac{s(b+\rho(s_2+s_3))}
{s(b+c+\rho(s_1-1+2s_2+s_3))}J(r_1,s_1-1,r_2,s_2+1,r_3,s_3)+...\nn
&&J(0,0,r_2,s_2,r_3,s_3)\nn
&=&e^{i\pi\rho'(r_2-r_3+1)}\frac{s(c'+\rho'r_2)}
{s(b'+c'+\rho'(2r_2+r_3-1))}J(0,0,r_2+1,s_2,r_3-1,s_3)+...\nn
&=&-e^{i\pi\rho(s_2-s_3+2)}\frac{s(c+\rho s_2)}
{s(b+c+\rho(2s_2+s_3-1))}J(0,0,r_2,s_2+1,r_3,s_3-1)+...
\eea
Here the dots stand for terms that cannot contribute to the crossing matrix 
element. After several further but in principle straightforward calculations
we obtain
\bea
\alpha^{(R,S)}_{(r,s,0),(R,S,0)}&=& (-)^Se^{i\pi S\rho}
\alpha^{(R)}_{r,R}(a',b',c';\rho')\alpha^{(S)}_{s,S}(a,b,c;\rho)\nn
\alpha^{(S)}_{s,S}(a,b,c;\rho)&=&
\frac{\prod_{j=1}^Ss(j\rho)}{\prod_{k=1}^{S-s}s(k\rho)\prod_{m=1}^ss(m\rho)}\nn
&\cdot&\prod_{j=0}^{s-1}\frac{s(b+\rho(S-s+j))}{s(b+c+\rho(S+j-1))}
\prod_{l=0}^{S-s-1}\frac{s(c+\rho(s+l))}{s(b+c+\rho(s+S+l-1))}
\eea
and where $\alpha^{(R)}_{r,R}(a',b',c';\rho')$ is given by a completely similar
expression. The phase is the result of multiplying many phases together.
This completes the calculation of the matrix elements of the relevant column 
as far as fusion rule I is concerned. The result has a form identical to what
is found for minimal models \cite{DF}.

Concerning fusion rule II it turns out that a simple trick allows to obtain
the result rather easily. In fact, a suitable contour deformation of the
complex contour ${\cal C}_v$ allows one to obtain an equation of the form
\ben
{\cal S}^{(R,S)}_{(r,s,1)}=-\frac{1}{\pi}\left \{e^{i\pi\rho s}s(c+\rho s)
{\cal S}^{(R,S)}_{(r,s+1,0)}+s(a+c+2\rho s)\int_x^0...\right \}
\een
The integral from $x$ to $0$ (we imagine $x<0$ in the s-channel) cannot have a 
contribution with a $(1-z)$ singularity appropriate for 
${\cal T}^{(R,S)}_{(R,S,0)}$, and so we do not specify the integrand, and we 
drop the integral in the calculation. Now it is an easy matter to obtain the
missing matrix elements from the ones we have already given. One finds
\ben
\alpha^{(R,S)}_{(r,s,1),(R,S,0)}=-\frac{1}{\pi}\frac{s(c+\rho s)s((s+1)\rho)}
{s(\rho)}\alpha^{(R,S)}_{(r,s+1,0),(R,S,0)}
\een
where
$$s=0,...,S-1$$

\subsection{The row of the transformation matrix, 
$\alpha^{(R,S)}_{(R,S,0),(r,s,\epsilon)}$}
The procedure is to consider the s-channel block, $S^{(R,S)}_{(R,S,0)}$ and
then isolate the t-channel singularities in $(1-z)$ and $(x-1)$. The strengths
of these singularities will tell us which t-channel block is obtained. In this
way we determine modified crossing matrix elements
\bea
S^{(R,S)}_{(r,s,\epsilon)}(z,x)&=&\sum_{r'=0}^R\sum_{s'=0}^S
\alpha^{(R,S)'}_{(r,s,\epsilon),(r',s',0)}t^{(R,S)}_{(r',s',0)}(z,x)\nn
&+&\sum_{r'=0}^R\sum_{s'=0}^{S-1}
\alpha^{(R,S)'}_{(r,s,\epsilon),(r',s',1)}t^{(R,S)}_{(r',s',1)}(z,x)
\eea 
These matrix elements are related to the ones  we have previously considered
by the normalization constants of the last section. Denoting the corresponding 
normalizations in the t-channel by $\tilde{N}^{(R,S)}_{(r,s,0)}(a,b,c;\rho)$,
we have (cf. also Ref. \cite{DF})
\bea
\alpha^{(R,S)}_{(R,S,0),(r,s,0)}&=&\alpha^{(R,S)'}_{(R,S,0),(r,s,0)}/
\tilde{N}^{(R,S)}_{(r,s,0)}(a,b,c;\rho)\nn
&=&\alpha^{(R,S)'}_{(R,S,0),(r,s,0)}/
N^{(R,S)}_{(r,s,0)}(b,a,c;\rho)
\eea
We consider 
the real $T$-ordered form of the integral representation:
\bea
&&e^{i\pi(R-S\rho)}S^{(R,S)}_{(R,S,0)}\nn
&=&z^{2j_1j_2/t}(1-z)^{2j_2j_3/t}\nn
&&z^PT\int_0^1\prod_{I=1}^R du_I\prod_{K=1}^S dv_K u_I^{a'}
(1-zu_I)^{b'}(1-u_I)^{c'}(x-zu_I)
\prod_{I<I'}(u_I-u_{I'})^{2\rho'}\nn
&&v_K^a(1-zv_K)^b(1-v_K)^c(x-zv_K)^{-\rho}\prod_{K<K'}(v_K-v_{K'})^{2\rho}
\prod_{I,K}(u_I-v_K)^{-2}
\label{row}
\eea
The phase on the left hand side takes into account that s- and t-channel blocks
are defined with different phase conventions as far as the factors $(x-u)$ and
$(x-v)^{-\rho}$ are concerned.
The pre-factor $z^P$ is obtained from scaling the integration variables with 
$z$. When
$z\rightarrow 1$ it is regular and we shall ignore it in the following.

Next, we use analytic tricks similar to Ref. \cite{DF} and similar to what
was used in section 5, in order to isolate the
singularities in $(1-z)$. Here we shall need to similarly isolate the 
singularities in $(x-1)$. We consider the integration region, where the first
$r$ $u_i$'s are integrated (ordered) from $1-\epsilon$ to $1$, ($\epsilon$
small $>0$) 
the first $s$ $v_k$'s 
similarly from $(1-\epsilon)$ to $1$, and the remaining variables from $0$ to
$1-\epsilon$. The first $u_i$'s and $v_k$'s are indexed by $i$ and $k$ and the 
remaining ones by $j,l$. The first $u_i$'s are transformed as 
$u_i\rightarrow 1-u_i$, followed by $u_i\rightarrow (1-z) u_i$, 
and likewise $v_k\rightarrow 1-v_k$ followed by 
$v_k\rightarrow (1-z) v_k$. Inserting that in \Eq{row} we find the singular
behaviour as $z\rightarrow 1$:
$$(1-z)^{-h(j_2)-h(j_3)+h(j_I)}$$
where
\ben
j_I=j_2+j_3-r+st
\een
Furthermore one isolates the $x\rightarrow 1$ behaviour
$$(x-1)^{r-st}=(x-1)^{j_2+j_3-j_I}$$
Therefore we may calculate the coefficient of $t^{(R,S)}_{(r,s,0)}$ in the 
expansion of $S^{(R,S)}_{(R,S,0)}$. After some work one finds the result
\bea
\alpha^{(R,S)'}_{(R,S,0),(r,s,0)}&=&e^{i\pi(-R+S\rho)}
{\cal J}_{r,s}(-b-c+2(r-1-(s-1)\rho),c;\rho)\nn
&&{\cal J}_{R-r,S-s}(b+c-\rho-2(r-\rho s),a;\rho)
\eea
Some further calculations give
\bea
\alpha^{(R,S)}_{(R,S,0),(r,s,0)}&=&(-)^Se^{i\pi S\rho}
\alpha^{(R)}_{R,r}(a',b',c',d';\rho')\alpha^{(S)}_{S,s}(a,b,c,d;\rho)\nn
\alpha^{(S)}_{S,s}(a,b,c,d;\rho)&=&
\prod_{i=0}^{s-1}\frac{s(b+i\rho)}{s(b+c+(s-1+i)\rho)}\nn
&&\prod_{i=0}^{S-s-1}\frac{s(a+b+c+d+2(S-1)\rho-i\rho)}
{s(b+c+d+2(S-1)\rho-(S-s-1+i)\rho)}
\eea
Here for convenience of writing we have defined
\bea
d&\equiv&-\rho\nn
d'&\equiv&-d/\rho=1
\eea
The presence of the $d$ dependence 
is the only difference from the corresponding
expression in minimal models \cite{DF}. It originates directly from the factors
$(u-x)$ and $(v-x)^{-\rho}$ in the integral realization. Such factors are not
present in the case of minimal models.

In order to isolate the singularity which corresponds to the t-channel blocks
for fusion rule II, we supplement the above specification of the 
integration region by the requirement that the variable $v_{s+1}$ should be 
integrated between $1-\epsilon$ and $1$, and then transformed as 
$v_{s+1}\rightarrow 1-v_{s+1}$ followed by 
$v_{s+1}\rightarrow (x-1) v_{s+1}$. After some 
calculations we find
\bea
\alpha^{(R,S)'}_{(R,S,0),(r,s,1)}&=&
e^{i\pi(-R+S\rho)}{\cal J}_{r,s}(-b-c+2(r-s\rho)+2(\rho-1),c;\rho)\nn
&&{\cal J}_{R-r,S-s-1}(a,b+c+\rho-2(r-\rho s);\rho)\nn
&&\frac{\Gamma(\rho -1-b-c-2\rho s+2r)\Gamma(b+c+2\rho s-2r+1)}{\Gamma(\rho)}
\eea
and using the normalizations and various $\Gamma$ function identities
\bea
\alpha^{(R,S)}_{(R,S,0),(r,s,1)}&=&\pi\alpha^{(R,S)}_{(R,S,0),(r,s+1,0)}
\frac{s(\rho)}{s(b+\rho s)s(b+c+\rho (s-1))}
\eea

\section{Monodromy invariant greens functions}
Following the discussion in \cite{DF}, monodromy invariant 4-point
greens functions, $$G_{j_1,j_2,j_3,j_4}(z,\overline{z},x,\overline{x})$$ 
can be obtained by writing
\ben
G_{j_1,j_2,j_3,j_4}(z,\overline{z},x,\overline{x})=\sum_{r,s,\epsilon}
|S^{(R,S)}_{(r,s,\epsilon)}(j_1,j_2,j_3,j_4;z,x)|^2X^{(R,S)}_{(r,s,\epsilon)}
\een
This form ensures single valuedness in the limits $z\rightarrow 0$ and 
$x\rightarrow 0$. Single valuedness in the limits $z\rightarrow 1$ and 
$x\rightarrow 1$ is ensured provided the $X$'s are chosen to satisfy \cite{DF}
\ben
X^{(R,S)}_{(r,s,\epsilon)}\propto \frac{\alpha^{(R,S)}_{(R,S,0),(r,s,\epsilon)}
(b,a,c,d;\rho)}{\alpha^{(R,S)}_{(r,s,\epsilon),(R,S,0)}(a,b,c;\rho)}
\een
Using rescaling tricks similar to \cite{DF} we obtain
\bea
X^{(R,S)}_{(r,s,0)}&=&X^{(R)}_r(a',b',c',d';\rho')X^{(S)}_s(a,b,c,d;\rho)\nn
X^{(S)}_s(a,b,c,d;\rho)&=&\prod_{i=1}^s s(i\rho)\prod_{i=1}^{S-s}s(i\rho)
 \prod_{i=0}^{s-1}
\frac{s(a+i\rho)s(c+i\rho)}{s(a+c+(s-1+i)\rho)}\nn
&\cdot&\prod_{i=0}^{S-s-1}\frac{s(b+i\rho)s(1-a-b-c-d-2(S-1)\rho+i\rho)}
{s(1-a-c-d-2(S-1)\rho+(S-s-1+i)\rho)}
\eea
This expression is very similar to the result for minimal models except for
the presence of the $d=-\rho$ and $d'=1$. Finally
\ben
X^{(R,S)}_{(r,s,1)}=-\pi^2\frac{s^2(\rho)}
{s(c+\rho s)s((s+1)\rho)s(a+\rho s)s(a+c+\rho(s-1))}X^{(R,S)}_{(r,s+1,0)}
\een

Consistency requires that the greens functions thus defined automatically are 
single valued also around $z\rightarrow x$. We have checked that indeed for 
one screening charge (of the second kind) this is the case. We expect it to be 
true generally.

It is convenient for studies of the operator algebra to also introduce the
following expansion
\ben
G_{j_1,j_2,j_3,j_4}(z,\overline{z},x,\overline{x})=\sum_{r,s,\epsilon}
|s^{(R,S)}_{(r,s,\epsilon)}(j_1,j_2,j_3,j_4;z,x)|^2f^{(R,S)}_{(r,s,\epsilon)}
\een
where $s^{(R,S)}_{(r,s,\epsilon)}$ 
are defined in \Eq{sblockI} and \Eq{sblockII}.
The coefficients, $f^{(R,S)}_{(r,s,\epsilon)}$, differ from the coefficients,
$X^{(R,S)}_{(r,s,\epsilon)}$ by a factor $(N^{(R,S)}_{(r,s,\epsilon)})^2$. 
It is 
fairly straightforward to collect all the results and obtain the expression
for $f^{(R,S)}_{(r,s,\epsilon)}$. On the way we use the formula 
(\cite{DF} (A.35)) alternative to \Eq{dfintegral}:
\bea
{\cal J}_{nm}(a,b;\rho)&=&\rho^{2nm}\prod_{i,j=1}^{n,m}\frac{1}{(-i+j\rho)}
  \prod_{i=1}^n\frac{\Gamma(i\rho')}{\Gamma(\rho')}\prod_{i=1}^m
  \frac{\Gamma(i\rho)}
  {\Gamma(\rho)}\nn
 &\cdot&\prod_{i,j=0}^{n-1,m-1}\frac{1}
  {(a+j\rho-i)(b+j\rho-i)(a+b+(m-1+j)\rho-(n-1+i))}\nn
 &\cdot&\prod_{i=0}^{n-1}\frac{\Gamma(1+a'+i\rho')\Gamma(1+b'+i\rho')}
  {\Gamma(2-2m+a'+b'+(n-1+i)\rho')}\nn
 &\cdot&\prod_{i=0}^{m-1}\frac{\Gamma(1+a+i\rho)\Gamma(1+b+i\rho)}
  {\Gamma(2-2n+a+b+(m-1+i)\rho)}
\eea
We then obtain
\bea
&&f^{(R,S)}_{(r,s,0)}=\Lambda^{(R,S)}_{r,s}(\rho)\nn
&\cdot&\prod_{i,j=0}^{r-1,s-1}\frac{1}
{(a+j\rho-i)^2(c+j\rho-i)^2(a+c+\rho(s-1+j)-(r-1+i))^2}\nn
&\cdot&\prod_{i,j=0}^{R-r-1,S-s-1}\frac{1}
{(b-i+j\rho)^2(e-i+j\rho)^2(e+b-(R-r-1+i)+(S-s-1+j)\rho)^2}\nn
&\cdot&\prod_{i=0}^{r-1}\frac{G(1+a'+i\rho')G(1+c'+i\rho')}
{G(2-2s+a'+c'+(r-1+i)\rho')}
\prod_{i=0}^{s-1}\frac{G(1+a+i\rho)G(1+c+i\rho)}{G(2-2r+a+c+(s-1+i)\rho)}\nn
&\cdot&\prod_{i=0}^{R-r-1}\frac{G(1+b'+i\rho)G(1+e'+i\rho)}
{G(2+e'+b'-2(S-s)+(R-r-1+i)\rho')}\nn
&\cdot&\prod_{i=0}^{S-s-1}\frac{G(1+b+i\rho)G(1+e+i\rho)}
{G(2+e+b-2(R-r)+(S-s-1+i)\rho)}
\label{fnormI}
\eea
where we have defined
$$G(x)\equiv \frac{\Gamma(x)}{\Gamma(1-x)}=\frac{1}{G(1-x)}$$
and where
\bea
\Lambda^{(R,S)}_{r,s}(\rho)&=&
 \rho^{4rs+4(R-r)(S-s)}\prod_{i=1}^s G(i\rho)\prod_{i=1}^{S-s}G(i\rho)
 \prod_{i=1}^r G(i\rho')\prod_{i=1}^{R-r}G(i\rho')\nn
&\cdot&\prod_{i,j=1}^{r,s}\frac{1}{(i-j\rho)^2}
 \prod_{i,j=1}^{R-r,S-s}\frac{1}{(i-j\rho)^2}
\eea
and where we have defined
\bea
 e&\equiv&-a-b-c-d-2\rho(S-1)+2(R-1)\nn
 e'&\equiv&-e/\rho
\eea
These expressions are like the ones for minimal models except for the 
appearance of the terms $d,d'$ in the definition of $e,e'$. Finally
\bea
 f^{(R,S)}_{(r,s,1)}&=&f^{(R,S)}_{(r,s+1,0)}\nn
 &\cdot&\frac{G(2+a+c+2s\rho-2r)G(1+a+c+2s\rho-2r)G(1-(s+1)\rho+r)}
  {G(1-r+a+s\rho)G(1-r+c+s\rho)G(2-r+a+c+(s-1)\rho)}
\label{fnormII}
\eea

\section{Operator algebra coefficients}
As explained in \cite{DF} the monodromy coefficients in the normalization
\Eq{fnormI} and \Eq{fnormII} determine the operator algebra coefficients
of the theory, $C_{\lambda_1\lambda_2}^\lambda$, defined by
($\lambda \equiv 2j+1$)
\ben
 \phi_{j_2}(z,\overline{z};x,\overline{x})\phi_{j_1}(0,0;0,0)=\sum_j
 \frac{(x\overline{x})^{j_1+j_2-j}}{(z\overline{z})^{h(j_1)+h(j_2)-h(j)}}
 C_{\lambda_1\lambda_2}^\lambda\phi_j(0,0;0,0)
\een
where the contribution from the conformal family is to be understood.
Indeed
\ben
f_{(r,s,\epsilon)}^{(R,S)}(j_1,j_2,j_3,j_4)=
C_{\lambda_1\lambda_2}^{\lambda_\epsilon}
C_{\lambda_3\overline{\lambda_4}}^{\overline{\lambda_\epsilon}}
\een
with $\lambda_\epsilon=\lambda_I,\lambda_{II}$ for $\epsilon =0,1$.
However, this requires that the monodromy coefficients are properly normalized.
The normalization adopted so far follows the
prescription of Dotsenko and Fateev in the case of minimal models, but turns 
out to be inadequate here. Indeed it is completely essential that the above 
factorization takes place in such a way that the operator algebra 
coefficients only depend on the variables indicated and not on anything else.
In particular, $C_{\lambda_1\lambda_2}^{\lambda_\epsilon}$ is allowed to
depend on $r,s$ which are given in terms of the spins (the $\lambda$'s) 
indicated, but it is not allowed to depend on $R,S$ for example. Likewise
$C_{\lambda_3\overline{\lambda_4}}^{\overline{\lambda_\epsilon}}$ is allowed to
depend on $R-r,S-s$ but again, not on $R,S$. However, it is allowed 
(as utilized above) to renormalize the coefficients by arbitrary functions of
$R,S,\lambda_1,\lambda_2,\lambda_3$, ($j_4=j_1+j_2+j_3 -R+St$). It turns out 
to be possible to devise such a normalization with the above criterion 
satisfied. This we have done below. 

We have to use (cf. \Eq{abc})
\bea
a&=&-\lambda_3+R-St+t\nn
b&=&-\lambda_1+R-St+t\nn
e&=&-\lambda_2+R-St+t\nn
c&=&\lambda_4+R-St\nn
e+b&=&-\lambda_I+2(R-r)-1-2t(S-s-1)=+\lambda_{II}+2(R-r)-1-2t(S-s-1)\nn
a+c&=&\lambda_I+2r-2st-1+t=-\lambda_{II}+2r-2st-1+t
\eea
We then find
\bea
C_{\lambda_1\lambda_2}^{\lambda}(r,s;I)&=&
  t^{-2rs}\prod_{i=1}^r G(i/t)\prod_{i=1}^s 
  G(it-r)\nn
&\cdot&\prod_{i=0}^{r-1}\frac{G(1-s+(1-\lambda_1+i)/t)G(1-s+(1-\lambda_2+i)/t)}
  {G(1+s-(1+\lambda+i)/t)}\nn
 &\cdot&\prod_{i=0}^{s-1}
  \frac{G(\lambda_1+it)G(\lambda_2+it)}{G(1+\lambda-(1+i)t)}\nn
C_{\lambda_3\overline{\lambda}_4}^{\overline{\lambda}}(R-r,S-s;I)&=&
  t^{-2(R-r)(S-s)}\prod_{i=1}^{R-r}G(i/t)\prod_{i=1}^{S-s}G(it-(R-r))\nn
 &\cdot&\prod_{i=0}^{R-r-1}\frac{G(1-(S-s)+(1-\lambda_3+i)/t)
  }{G(S-s-(1-\lambda+i)/t)}\nn
 &\cdot&\prod_{i=0}^{R-r-1}
  G(-(S-s)+(1+\lambda_4+i)/t)\nn
 &\cdot&\prod_{i=0}^{S-s-1}\frac{G(\lambda_3+it)G(-\lambda_4+(1+i)t)}
  {G(1-\lambda-it)}\nn
 &=&C_{\lambda_3,-\lambda_4+t}^{-\lambda+t}(R-r,S-s;I)\nn
 &=&C_{\lambda_3\lambda}^{\lambda_4}(R-r,S-s;I)
\label{oac}
\eea
Here, for clarity we have indicated the dependencies on $r,s$ or $R-r,S-s$ 
and on the fusion rule (here I). In fact this is somewhat superfluous, as we 
shall see in particular in the next section. The point is, that from three
spins, it is always clear by which of the two fusion rules they couple. 
And for each case there is a unique possible value of $(r,s)$ (or $(R-r,S-s)$).
The last identity in (\ref{oac}) confirms the fact that treating the left
vertex ($j_3\overline{j}_4\overline{j}$) in terms of conjugate fields
(indicated by bars) is equivalent to simply considering the coupling
($j_3jj_4$). Hence the factorization of the 4-point function into 3-point
functions is made manifest. 
According to the discussion in the next section of 
cases including fusion rule II 
this fact remains true. This indicates that a factorization of 
N-point functions for any combination of fusion rules is similarly manifest. 

Let us consider the case of fusion rule I and
parametrize the intermediate spin, $j$, as
\ben
2j+1=\lambda=\rho-\sigma t
\label{rhosigma}
\een
Then we have\\(I,I)
\bea
(r,s)&=&(\hf (r_1+r_2-\rho-1),\hf(s_1+s_2-\sigma))\nn
(R-r,S-s)&=&(\hf(\rho+r_3-r_4-1),\hf(\sigma+s_3-s_4))
\label{rsI,I}
\eea
where the label (I,I) indicates that we have fusion rule I at both vertices
of the 4-point block, $(j_1j_2j)$ and $(jj_3j_4)$. 
These $(r,s)$ and $(R-r,S-s)$ are integers precisely 
in that case. Modified expressions for $(r,s)$ or $(R-r,S-s)$ have 
to be used for fusion rule II (see next section).

The expression for the monodromy coefficients for the case where fusion rule 
II is operating at both vertices may similarly be obtained from section 7. 
In the
new normalization adopted here it has the correctly factorized form. However,
in the operator algebra coefficients constructed from it the parameters 
$r,s,R-r,S-s$ have very different significance, since it is based on
$$\lambda_{II}=-\lambda_I,\ \ j_1+j_2-j=r-st$$
where $j=j_I$.
Hence in the next section for fusion rule II we shall base our discussion on a 
quite different treatment, but one with a parametrization similar to the
one for fusion rule I.

\section{Blocks with mixed fusion rules and the operator algebra coefficients}
The 4-point blocks considered so far are ones 
where we have either fusion rule I
operating at both vertices, or fusion rule II operating at both vertices. We
now describe how to obtain 4-point blocks for the case where we have either
fusion rule I for $(j_1j_2j)$ and fusion rule II for $(jj_3j_4)$, we denote
that case by (II,I), or fusion rule II for $(j_1j_2j)$ and fusion rule I for
$(jj_3j_4)$, we denote that case by (I,II). We emphasize that for a collection 
of spins considered so far, so that fusion rule I (or fusion rule II) 
is possible at both vertices, neither (I,II) nor (II,I) will be possible. Hence
there will be no mixing in the crossing matrix calculations.

Our technique is based on the discussion of fusion rules I and II for the 
3-point function in section 2. Namely we modify \Eq{rsI,I} as follows:\\
(I,II)
\bea
(r,s)&=&(\hf(r_1+r_2-\rho-1+p),\hf(s_1+s_2-\sigma+q))\nn
(R-r,S-s)&=&(\hf(\rho+r_3-r_4-1),\hf(\sigma +s_3-s_4))
\label{rsI,II}
\eea
(II,I)
\bea
(r,s)&=&(\hf(r_1+r_2-\rho-1),\hf(s_1+s_2-\sigma))\nn
(R-r,S-s)&=&(\hf(\rho+r_3-r_4-1+p),\hf(\sigma +s_3-s_4+q))
\label{rsII,I}
\eea
Notice that these numbers of screenings (numbers of integrations in Andreev's 
case) are integers precisely when the fusion rules indicated are operating.

Now, in all cases, we have for the intermediate spin, $j$, 
($\lambda =2j+1=\rho-\sigma t$):
$$j_1+j_2-j=r-st, \ \ j+j_3-j_4=(R-r)-(S-s)t$$
independent of whether we have fusion rule I or II. This means that our 
parametrization of the internal spin is quite different for fusion rule II 
from what was used throughout the paper so far, but much more symmetric. 
It would in
fact be somewhat natural to consider now an alternative representation of the
case (II,II), namely\\
(II,II)
\bea
(r,s)&=&(\hf(r_1+r_2-\rho-1+p),\hf(s_1+s_2-\sigma+q))\nn
(R-r,S-s)&=&(\hf(\rho+r_3-r_4-1+p),\hf(\sigma +s_3-s_4+q))
\label{rsII,II}
\eea
This would give us an alternative form of that block from the one 
considered so far, but one with more screenings (more integrations in the case 
of Andreev's representation). 
Our previous treatment is the most economic as far
as the numbers of screenings are concerned, but does not lead to a convenient
expression for the operator algebra coefficients. However, we do not consider 
the case (II,II) further, since a discussion of that is analogous
to the following discussion of the cases (I,II) and (II,I).

It is quite clear from the discussion in section 2 that the ``over-screened''
expressions for $(r,s)$ and $(R-r,S-s)$ produce couplings of the 
spins according
to the fusion rules indicated. From the preceding discussion it is then also
almost trivial that the blocks built (in Andreev's representation) with these
numbers of screenings, using the contours in Fig. \ref{figanI}, will have the
correct singular properties. However, closer inspection shows that the block 
built in that way is not well defined due to a $\Gamma(0)$ singularity. 
But there is a very simple remedy, like
the one employed for the 3-point function in section 2. Namely, suppose we need
to build the block corresponding to the case (I,II). 
Then we modify slightly the
contours in Fig. \ref{figanI}, so that we choose Felder contours \cite{F} for 
the variables, $v_1,...v_s$, more precisely, we may take these variables to run
along circle like contours passing through the point $z$, and surrounding the 
contours for $u_1,...,u_r$, i.e. surrounding the point $0$, in such a way that
the $v_1$-contour lies inside the $v_2$-contour, etc. Similarly for the case
of (II,I), we modify the contours for $v_{s+1},...,v_S$ into circle like 
contours surrounding $0$ (actually surrounding $\infty$) and passing through 
the point $1$, in such a way that the contour for $v_{s+1}$ lies inside the 
contour for $v_{s+2}$ etc. With these contours, we have checked in great 
detail that the blocks are well defined and non-vanishing. 
Actually in these mixed cases, we find
cancellations between Gamma functions of negative integer arguments between 
numerator and denominator \cite{JR}.

The construction of crossing matrices and monodromy coefficients is made
essentially trivial by the following observations. We have previously seen that
the integral for the Dotsenko-Fateev contours, Fig. \ref{figanI}, is related to
a corresponding integral for (time-) ordered integrations by the factor
$$\lambda_r(1/t)\lambda_s(t)\lambda_{R-r}(1/t)\lambda_{S-s}(t)$$
Similarly for the present case of some
contours being of Felder type, we get instead a factor 
$$\lambda_r(1/t)\chi^{(2)}_s(S-s_3-1;t)\lambda_{R-r}(1/t)\lambda_{S-s}(t)$$ 
for (I,II) and a factor
$$\lambda_r(1/t)\lambda_s(t)\lambda_{R-r}(1/t)\chi^{(2)}_{S-s}(S-s_2-1;t)$$ 
for (II,I). These rather 
trivial new normalizations allow us to follow completely the treatment for 
fusion rule I (i.e., the case (I,I)) described above and insert appropriate
$\chi/\lambda$ factors as normalizations. It is rather easy to see, that for
the new monodromy coefficients, $f^{(R,S)}_{(r,s,(I,II))}$ and
$f^{(R,S)}_{(r,s,(II,I))}$ (in a self explanatory notation), 
the only $\chi/\lambda$ factors multiplying $f^{(R,S)}_{(r,s,(I,I))}$
which survive are
ones which do not depend on $(r,s)$ or $(R-r,S-s)$, but only on $(R,S)$ and 
hence may be absorbed into renormalizations. 

The crucial conclusion of all these observations is 
the expected one, that the operator 
algebra coefficients {\em in all cases look the same} namely as given by 
\Eq{oac}. 
However, in each case we should investigate whether the indicated spins couple
via fusion rule I or II, and accordingly use the relevant expression for
$(r,s)$.

\section{Conclusions}
In this paper we have investigated in detail the 4-point blocks for conformal
field theories based on $SL(2)$ current algebra 
with fractional levels and based on admissible 
representations. In particular we have devised integration contours appropriate
for suitable conformal blocks, both using our own representation based on free
fields \cite{PRY} and the one by Andreev applicable only to 4-point functions
\cite{An}. We have found both fusion rules already known in the literature
and we have investigated the relation between the two representations
of 4-point functions.
We have then performed the lengthy calculations to obtain crossing matrices 
and monodromy coefficients. Based on the latter we have isolated the operator 
algebra coefficients of the theory for both fusion rules. They may in some 
sense be considered the principal result of our investigation. They are given
by \Eq{oac} for both fusion rules, even though the numbers of screening
integrations are given by different expressions for fusion rule I and II. 
In his work \cite {An}, Andreev also gives operator algebra 
coefficients. They are obtained without many details and appear to differ from
ours. He has avoided the entire discussion of monodromy invariant combinations
by using integrations over the complex plane, rather than the careful and
cumbersome treatment based on complex contours which we have used, following
\cite{DF}. 
\\[.8 cm] 
{\bf Acknowledgement}\\[.2cm]
JLP thanks the high energy theory group at Hokkaido university, 
where some of the present work was carried out, for hospitality; 
JR thanks the theory group at Academia Sinica, where 
part of the present work was carried out, for hospitality;
MY thanks the Danish Research Academy for supporting most of the present work.
This work was carried out in the framework of the project ``Gauge theories,
applied supersymmetry and quantum gravity", contract SC1-CT92-0789 of the 
European Economic Community.


\begin{thebibliography}{9}
\bibitem{KK} V.G. Kac and D.A. Kazhdan, Adv. Math. {\bf 34} (1979) 79
\bibitem{MFF} F.G. Malikov, B.L. Feigin and D.B. Fuks,
 Funkt. Anal. Prilozhen {\bf 20} (1986) 25
\bibitem{HY} H.-L. Hu and M. Yu, \PL{B 289} (1992) 302;\\
 H.-L. Hu and M. Yu, \NP{B 391} (1993) 389
\bibitem{AGSY} O. Aharony, O. Ganor, J. Sonnenschein and S. Yankielowicz,
\NP{B 399} (1993) 527
\bibitem{Bel} A.A. Belavin, in {\em Proc. of the second Yukawa Symposium, 
 Nishinomiya, Japan, Springer Proceedings in Physics, Vol. 31 (1988) 132}
\bibitem{Pol} A.M. Polyakov, in {\em Physics and Mathematics of Strings,}
 Eds. L. Brink, D. Friedan and A.M. Polyakov (World Scientific, 1990)
\bibitem{BO} M. Bershadsky and H. Ooguri, \CMP{126} (1989) 49
\bibitem{bars}I. Bars, hep-th/9503205 preprint
\bibitem{PRY} J.L. Petersen, J. Rasmussen and M. Yu, 
 \NP{B 457} (1995) 309;\\
 J.L. Petersen, J. Rasmussen and M. Yu, \NP{B 457} (1995) 343;\\
 J.L. Petersen, J. Rasmussen and M. Yu, NBI-HE-95-33/hep-th/9510059, preprint,
 to be published in 
 {\em Proc. on EU network meeting on Gauge Theories, Applied 
 Supersymmetry and Quantum Gravity, Leuven, Belgium, July 1995};\\
 J.L. Petersen, J. Rasmussen and M. Yu, NBI-HE-95-42/hep-th/9512175, preprint,
 to be
 published in {\em Proc. on the 29th Symposium Ahrendshop on the Theory of
 Elementary Particles, Buckow, Germany, 1995}
\bibitem{Wak} M. Wakimoto, \CMP{104} (1986) 60
\bibitem{FGPP} P. Furlan, A.Ch. Ganchev, R. Paunov and V.B. Petkova,
 \PL{B 267} (1991) 63;\\
 P. Furlan, A.Ch. Ganchev, R. Paunov and V.B. Petkova,
 \NP{B 394} (1993) 665;\\
 A.Ch. Ganchev and V.B. Petkova, \PL{B 293} (1992) 56
\bibitem{FF} B.L. Feigin and E. Frenkel,  Lett. Math. Phys. {\bf 19} (1990) 307
\bibitem{An}O. Andreev, HUB-IEB-94/9 hep-th/9407180, preprint;\\
 O. Andreev, \PL{B 363} (1995) 166
\bibitem{AY} H. Awata and Y. Yamada, \MPL{A7} (1992) 1185
\bibitem{FM} B. Feigin and F. Malikov, Lett. Math. Phys {\bf 31} (1994) 315
\bibitem{FM95} B. Feigin and F. Malikov, preprint q-alg/9511011
\bibitem{DF} Vl.S. Dotsenko and V.A. Fateev, \NP{B 240}{[}FS12{]} (1984) 312;\\
 Vl.S. Dotsenko and V.A. Fateev, \NP{B 251}{[}FS13{]} (1985) 691
\bibitem{KZ} V. Knizhnik and A. Zamolodchikov, \NP{B 247} (1984) 83
\bibitem{FZ} V.A. Fateev and A.B. Zamolodchikov, \SJNP{43} (1986) 657
\bibitem{F} G. Felder, \NP{B 317} (1989) 215 {[}Erratum:
{\bf B 324} (1989) 548{]}
\bibitem{JR} J. Rasmussen, Ph.D. thesis, to appear
\end{thebibliography}
\end{document}